\documentclass{aa}

\usepackage{amsmath}
\usepackage{amssymb}
\usepackage{mathrsfs}
\usepackage{graphicx}

\begin{document}
   \title{Secular spin-axis dynamics of exoplanets}

   \author{M. Saillenfest \and J. Laskar \and G. Bou{\'e}}

   \institute{IMCCE, Observatoire de Paris, PSL Research University, CNRS, Sorbonne Universit\'e, LAL, Université de Lille, 75014 Paris, France. \email{melaine.saillenfest@obspm.fr} }

   \date{Received 28/09/2018; accepted 07/01/2019}
 
   \abstract
   {Seasonal variations and climate stability of a planet are very sensitive to the planet obliquity and its evolution. This is of particular interest for the emergence and sustainability of land-based life, but orbital and rotational parameters of exoplanets are still poorly constrained. Numerical explorations usually realised in this situation are thus in heavy contrast with the uncertain nature of the available data.}
   {We aim to provide an analytical formulation of the long-term spin-axis dynamics of exoplanets, linking it directly to physical and dynamical parameters, but still giving precise quantitative results if the parameters are well known. Together with bounds for the poorly constrained parameters of exoplanets, this analysis is designed to allow a quick and straightforward exploration of the spin-axis dynamics.}
   {The long-term orbital solution is decomposed in quasi-periodic series and the spin-axis Hamiltonian is expanded in powers of eccentricity and inclination. Chaotic zones are measured by the resonance overlap criterion. Bounds for the poorly known parameters of exoplanets are obtained from physical grounds (rotational breakup) and dynamical considerations (equipartition of AMD).}
   {This method gives accurate results when the orbital evolution is well known. The chaotic zones for planets of the Solar System can be retrieved in details from simple analytical formulas. For less constrained planetary systems, the maximal extent of the chaotic regions can be computed, requiring only the mass, the semi-major axis and the eccentricity of the planets present in the system. Additionally, some estimated bounds of the precession constant allow to classify which observed exoplanets are necessarily out of major spin-orbit secular resonances (unless the precession rate is affected by the presence of massive satellites).}
   {}


   \maketitle

\section{Introduction}
   From the works by \cite{LASKAR-ROBUTEL_1993} and \cite{LASKAR-etal_1993}, we know that the long-term dynamics of the terrestrial planets of the Solar System feature wide chaotic regions allowing large variations of their obliquity. In particular, Mars is currently in a chaotic region extending from $0^\text{o}$ to $60^\text{o}$ obliquity, whereas the Earth is located in a stable region thanks to the presence of the Moon, resulting in obliquity variations of only a few degrees. Subsequent studies detailed both the past and future spin-axis evolution the Earth \citep{NERONDESURGY-LASKAR_1997,LASKAR-etal_2004a,LI-BATYGIN_2014b}, Venus \citep{CORREIA-etal_2003,CORREIA-LASKAR_2003} and Mars \citep{LASKAR-etal_2004b}. On the other hand, paleorecords on Earth show that even the very slight variations of its orbit and obliquity led to major climate changes \citep[e.g.][]{WEERTMAN_1976,HAYS-etal_1976}. This implies that life on Earth would be very different to what it is now if the Earth had evolved in a large chaotic zone, as it would have without the stabilising effect of the Moon. This conclusion is reinforced by the fact that high-obliquity planets undergo severe seasonal variations \citep[e.g.][]{SPIEGEL-etal_2009} even for a stable obliquity (but conditions suitable for the very emergence of life could still be achieved: in some extreme cases, the amount of liquid water on the surface may even be favoured by large obliquity variations, as reported by \citealp{ARMSTRONG-etal_2014}).
   
   As the formation of the Moon is thought to have resulted from an accidental collision event \citep[e.g.][]{HARTMANN-DAVIS_1975,CANUP-ASPHAUG_2001,LOCK-etal_2018}, a ``moonless'' Earth was a possible (or even likely) outcome of the planetary formation process. In the broader context of exoplanets, the dynamics of such a moonless Earth is thus more than of academic interest. This motivated further works about the structure of the chaotic region \citep{LI-BATYGIN_2014a} and some additional numerical studies \citep{LISSAUER-etal_2012}.
   
   When it comes to exoplanets, we must face the problem of the incomplete and imprecise nature of both dynamical and physical data. Except from very favourable cases, like a fast precession and an important flattening inducing detectable transit depth modulations \citep{CARTER-WINN_2010,CORREIA_2014}, the spin orientation and the flattening of exoplanets are far from being reachable by observations. They must therefore be taken as completely free parameters, in the spirit of the work by \citealp{LASKAR-ROBUTEL_1993} for the Solar System planets. Moreover, the orbital properties of exoplanets are not well known either, especially the respective orientations of the orbits (including the mutual inclinations, which play a crucial role in the spin-axis dynamics). Several authors tackled this problem already \citep[e.g.][]{BRASSER-etal_2014,DEITRICK-etal_2018,SHAN-LI_2018}: they used numerical integrations of the planetary system in order to build the time-dependent perturbation of the spin axis. Applied to extrasolar planets, this method requires to choose a nominal value for both the rotational parameters and the unknown orbital elements. The parameter space to be explored is thus very wide, so that even elaborate numerical explorations require some degree of arbitrariness. Consequently, the use of numerical integrations at this stage could appear a bit in contradiction with the very incomplete nature of the data. However, the secular problem (averaged over rotational and orbital motions) is not as complex as it could appear. Some hints about a possible analytical treatment were actually given by \cite{LASKAR_1996} and partially exploited by \cite{ATOBE-etal_2004}, \cite{LI-BATYGIN_2014a} and \cite{SHAN-LI_2018}. In the exoplanetary case, a refined analytical theory would be very convenient, since it would give in a direct way the sensibility of the spin-axis dynamics to the various known and unknown parameters, instead of giving a list of possible outcomes. Associated with some bounds for the unknown parameters, such a theory would give a clear range for these possible outcomes.
   
   This was the approach used by \cite{ATOBE-etal_2004}, with the aim of finding the probability of small obliquity variations for hypothetical terrestrial planets in the habitable zone of known exoplanetary systems. Their analytical developments, though, were limited to the lowest-order approximations (their parameter space was indeed enormous since, in this case, the planet itself was hypothetical). The recent work by \cite{SHAN-LI_2018} also contains an analytical part applied to the particular case of exoplanets Kepler-62f and Kepler-186f. This time, their calculations were mostly designed to precise and explain numerical results, so they did not try to bring any substantial improvement to the theory by \cite{ATOBE-etal_2004}.
   
   In this context, the goal of this article is twofold: \emph{i)} provide a general analytical formalism for studying the long-term spin-axis dynamics of (exo)planets and \emph{ii)} clarify what kind of information about the spin can be obtained from typical observed exoplanetary systems, that is, with numerous unknown physical and dynamical parameters. In particular, it is crucial for future studies to have a simple way to classify the observed exoplanets according to the characteristics of their spin dynamics. This would allow to determine which exoplanets are worth to be studied in more details (in particular if a complex chaotic spin dynamics is expected) and which ones have necessarily a very simple spin dynamics, making unnecessary any further numerical or analytical study. Such a general analysis will give both a qualitative view of the system if it is poorly known (in the continuation of \citealp{ATOBE-etal_2004}), and a quantitative description of the dynamics if it is well known (as an analytical counterpart of \citealp{LASKAR-ROBUTEL_1993}).
   
   This article is organised as follows: Sect.~\ref{sec:anmod} recalls the Hamiltonian of the secular spin-axis dynamics and shows how it can be expanded in terms of the orbital motion parameters. The secular resonances at all orders can then be isolated and used to delimit the chaotic regions. Then, Sect.~\ref{sec:app} shows how an incomplete set of orbital elements can still be used to constrain the orbital solution of an exoplanet. Combined with the rotational breakup limit, it allows to make a preliminary classification of the ``non-resonant'' exoplanets, for which no chaos can appear and the obliquity variations are constrained by an analytical bound.

\section{Analytical model of the long-term spin dynamics}\label{sec:anmod}
   \subsection{Development of the Hamiltonian}
   Let us consider a system composed of a star and several planets. We study the rotational dynamics of one planet among them. For now, we consider that this planet is far from any spin-orbit resonance. Considering only the lowest-order term of the torque from the star expanded in Legendre polynomials, the Hamiltonian of rotation averaged over orbital and rotational motions is given for instance by \cite{LASKAR-ROBUTEL_1993} and detailed by \cite{NERONDESURGY-LASKAR_1997}. It can be written
   \begin{equation}\label{eq:Hinit}
      \begin{aligned}
         \mathcal{H}(X,-\psi,t) &= -\frac{\alpha}{2}\frac{X^2}{\big(1-e(t)^2\big)^{3/2}} \\
         &- \sqrt{1-X^2}\big(\mathcal{A}(t)\sin\psi + \mathcal{B}(t)\cos\psi\big) \\
         &+ 2X\mathcal{C}(t),
      \end{aligned}
   \end{equation}
   where the conjugate coordinates are $X$ (cosine of obliquity) and $-\psi$ (minus the precession angle). The quantity $\alpha$ is called the ``precession constant'' (contrary to previous studies, we prefer to exclude here the eccentricity $e$ appearing in denominator from the definition of $\alpha$). Following the derivation proposed by \cite{NERONDESURGY-LASKAR_1997}, we obtain
   \begin{equation}\label{eq:alpha}
      \alpha = \frac{3\,\mathcal{G}m_0}{2\,\omega a^3}\,\frac{2C-A-B}{2C}.
   \end{equation}
   In this expression, $\mathcal{G}$ is the gravitational constant; $m_0$ is the mass of the star; $a$ is the semi-major axis of the planet in orbit around the star; $\omega$ is its spin angular velocity, and $A\leqslant B\leqslant C$ are its momenta of inertia. The Hamiltonian~\eqref{eq:Hinit} depends explicitly on time $t$ through the eccentricity $e$ and the functions
   \begin{equation}
      \left\{
      \begin{aligned}
         \mathcal{A}(t) &= \frac{2\big(\dot{q}+p\,\mathcal{C}(t)\big)}{\sqrt{1-p^2-q^2}} \\
         \mathcal{B}(t) &= \frac{2\big(\dot{p}-q\,\mathcal{C}(t)\big)}{\sqrt{1-p^2-q^2}} \\
      \end{aligned}
      \right.
      \hspace{0.5cm},\hspace{0.5cm}
      \mathcal{C}(t) = q\dot{p}-p\dot{q},
   \end{equation}
   in which $q=\sin(I/2)\cos\Omega$ and $p=\sin(I/2)\sin\Omega$, where $I$ and $\Omega$ are respectively the orbital inclination and the longitude of ascending node of the planet. In the following, we will write $\eta\equiv\sin(I/2)$. One can note that if there is only one planet in the system (two-body problem), the obliquity is constant and the precession angle circulates with constant angular velocity $\alpha X/(1-e^2)^{3/2}$.
   
   Let us suppose that the eccentricity and the inclination of the planet are small, such that we can develop the Hamiltonian in series of $e$ and $\eta$. In the following, we present the terms up to order $3$, but the method presented here can be generalised to any order (as we will see, the third order is the first one at which the eccentricity begins to play a substantial role). Using the fact that $\mathcal{C} = \eta^2\dot{\Omega} = \mathcal{O}(\eta^2)$, we obtain
   \begin{equation}\label{eq:ABtrunc}
      \begin{aligned}
         \mathcal{A} &= (2+p^2+q^2)\dot{q} + 2p\,\mathcal{C} + \mathcal{O}(\eta^4) \\
         \mathcal{B} &= (2+p^2+q^2)\dot{p} - 2q\,\mathcal{C} + \mathcal{O}(\eta^4)\,.
      \end{aligned}
   \end{equation}
   Let us now suppose that the secular orbital dynamics of the planet, resulting from the perturbations by the other planets, is quasi-periodic. This amounts to considering that the chaos present in the orbital secular system acts on a much larger timescale than the spin dynamics under study. As we will see below, this holds very well for the Solar System (this methodology was first proposed by \citealp{LASKAR_1996}; it is used for instance by \citealp{LI-BATYGIN_2014a}). In this case, we can write
   \begin{equation}\label{eq:qprep}
      \begin{aligned}
         e\exp(i\varpi) &= \sum_{j=1}^N E_j\exp(i\theta_j) \\
         \eta\exp(i\Omega) &= \sum_{j=1}^M S_j\exp(i\phi_j)\,,
      \end{aligned}
   \end{equation}
   where $\varpi$ is the longitude of pericentre of the planet in orbit around the star. The angles $\theta_j$ and $\phi_j$ evolve linearly with frequencies $\mu_j$ and $\nu_j$, that is,
   \begin{equation}
      \theta_j(t) = \mu_j\,t + \theta_j^{(0)}
      \hspace{0.5cm}\text{and}\hspace{0.5cm}
      \phi_j(t) = \nu_j\,t + \phi_j^{(0)},
   \end{equation}
   whereas the amplitudes $E_j$ and $S_j$ are real constants of order $e$ and $\eta$ or smaller. Such series can be obtained either from analytical theories or from frequency analysis of numerical solutions \citep{LASKAR_1988,LASKAR_1990}. In a general integrable case, $\mu_j$ and $\nu_j$ are integer combinations of the fundamental frequencies of the orbital dynamics (usually noted $g_k$ and $s_k$), and the series contain an infinite number of terms. Arranging the terms by decreasing amplitude, we consider here a truncation with $N$ terms for the eccentricity and $M$ terms for the inclination. We get then
   \begin{equation}
      e^2 = \sum_{j=1}^N E_j^2 + 2\sum_{j<k}^N E_jE_k\cos(\theta_j-\theta_k)\,,
   \end{equation}
   \begin{equation}
       \mathcal{C} = \sum_{j=1}^M \nu_jS_j^2 + \sum_{j<k}^M (\nu_j+\nu_k)S_jS_k\cos(\phi_j-\phi_k)\,,
   \end{equation}
   and from~\eqref{eq:ABtrunc},
   \begin{equation}
      \begin{aligned}
         &\mathcal{A}\sin\psi + \mathcal{B}\cos\psi = 2\sum_{j=1}^M \nu_jS_j\cos(\phi_j+\psi) \\
         &+ \sum_{j=1}^M\left[\nu_jS_j^3-2S_j\left(\sum_{k=1}^M \nu_kS_k^2\right)\right]\cos(\phi_j+\psi) \\
         &- \sum_{j=1}^M\sum_{\substack{k=1\\k\neq j}}^M \nu_kS_j^2S_k\cos(2\phi_j-\phi_k+\psi) \\
         &- 2\sum_{i=1}^M\sum_{\substack{j<k\\j,k\neq i}}^M \nu_iS_iS_jS_k\cos(-\phi_i+\phi_j+\phi_k+\psi)\\
         &+ \mathcal{O}(\eta^4)\,.
      \end{aligned}
   \end{equation}
   In order to obtain an autonomous Hamiltonian, let us introduce the momenta $\Theta_j$ and $\Phi_j$ conjugate to $\theta_j$ and $\phi_j$. The system has now $N+M+1$ degrees of freedom. The new Hamiltonian (that we still denote $\mathcal{H}$) can be written
   \begin{equation}\label{eq:Hgen}
      \mathcal{H} = \mathcal{H}_0 + \varepsilon\mathcal{H}_1 + \varepsilon^2\mathcal{H}_2 + \varepsilon^3\mathcal{H}_3 +\mathcal{O}(\varepsilon^4)\,,
   \end{equation}
   where we suppose that $\mathcal{O}(e) = \mathcal{O}(\eta) = \mathcal{O}(\varepsilon)$. The different parts are respectively
   \begin{equation}\label{eq:H0}
      \mathcal{H}_0(X,\Theta,\Phi) = -\frac{\alpha}{2}X^2 + \sum_{j=1}^N \mu_j\Theta_j + \sum_{j=1}^M \nu_j\Phi_j \,,
   \end{equation}
   \begin{equation}\label{eq:H1}
      \varepsilon\mathcal{H}_1(X,-\psi,\phi) = -2\sqrt{1-X^2}\sum_{j=1}^M \nu_jS_j\cos(\phi_j+\psi) \,,
   \end{equation}
   \begin{equation}\label{eq:H2}
      \begin{aligned}
         \varepsilon^2\mathcal{H}_2(X,\theta,\phi) &= -\frac{3\alpha}{4}X^2\sum_{j=1}^N E_j^2 + 2X\sum_{j=1}^M \nu_jS_j^2\\
         &-\frac{3\alpha}{2}X^2\sum_{j<k}^N E_jE_k\cos(\theta_j-\theta_k)\\
         &+ 2X\sum_{j<k}^M (\nu_j+\nu_k)S_jS_k\cos(\phi_j-\phi_k) \,,
      \end{aligned}
   \end{equation}
   and
   \begin{equation}\label{eq:H3}
      \begin{aligned}
         &\varepsilon^3\mathcal{H}_3(X,-\psi,\phi) = \\
         &-\sqrt{1-X^2}\sum_{j=1}^M\left[\nu_jS_j^3-2S_j\left(\sum_{k=1}^M\nu_kS_k^2\right)\right]\cos(\phi_j+\psi) \\
         &+\sqrt{1-X^2}\sum_{j=1}^M\sum_{\substack{k=1\\k\neq j}}^M\nu_kS_j^2S_k\cos(2\phi_j-\phi_k+\psi)\\
         &+2\sqrt{1-X^2}\sum_{i=1}^M\sum_{\substack{j<k\\j,k\neq i}}^M\nu_iS_iS_jS_k\cos(-\phi_i+\phi_j+\phi_k+\psi) \,.
      \end{aligned}
   \end{equation}
   We note that there can be no resonance among the angles $\phi_j$ and $\theta_j$ because they come from the quasi-periodic solution of the orbital dynamics. By definition, they are thus already ``integrated''.
    
   \subsection{One perturbing term: Colombo's top}\label{sec:coltop}
   From~(\ref{eq:H0}-\ref{eq:H1}), we conclude that at lowest-order to the perturbation, resonant angles can only be of the form $\sigma = \phi_j+\psi$. Let us consider a single resonance with the term $j=p$. We introduce the resonant canonical coordinates by the linear transformation
   \begin{equation}
      \left\{
      \begin{aligned}
         \sigma &= \psi + \phi_p \\
         \xi &= -\phi_p
      \end{aligned}
      \right.
      \hspace{0.5cm}\text{and}\hspace{0.5cm}
      \left\{
      \begin{aligned}
         \Sigma &= -X \\
         \Xi &= -X-\Phi_p \,.
      \end{aligned}
      \right.
   \end{equation}
   Assuming that the system is far from any other resonance, the long-term dynamics at first order to the perturbation is given by averaging the Hamiltonian over all angles but $\sigma$. Dropping the constant terms, we get
   \begin{equation}\label{eq:Fabc}
      \mathcal{F}(\Sigma,\sigma) = -\frac{1}{2}\mathfrak{a}\,\alpha\,\Sigma^2 + \mathfrak{b}\,\Sigma + \mathfrak{c}\,\sqrt{1-\Sigma^2}\cos\sigma \,,
   \end{equation}
   with
   \begin{equation}
      \begin{aligned}
         \mathfrak{a} &= 1+\frac{3}{2}\sum_{j=1}^N E_j^2 \\
         \mathfrak{b} &= \nu_p-2\sum_{j=1}^M \nu_jS_j^2 \\
         \mathfrak{c} &= - 2\,\nu_pS_p - \nu_pS_p^3 + 2S_p\sum_{j=1}^M \nu_jS_j^2 \,.
      \end{aligned}
   \end{equation}
   As shown in Appendix~\ref{asec:spinorb}, the Hamiltonian has the same form in the case of a $1$:$1$ spin-orbit resonance, with though a slightly different expression of the coefficients. This would thus only shift a bit the position of the secular resonances considered here. In this case, the tidal damping (responsible for this capture in spin-orbit resonance) is supposed to act on a much larger timescale than the spin dynamics studied here, such that our approach still holds. Finally, applying the modified time $\mathrm{d}\tau = \mathfrak{a}\,\alpha\,\mathrm{d}t$ to~\eqref{eq:Fabc}, we obtain the following Hamiltonian (that we still denote~$\mathcal{F}$):
   \begin{equation}\label{eq:Hres}
      \mathcal{F}(\Sigma,\sigma) = -\frac{1}{2}\Sigma^2 + \gamma\Sigma + \beta\sqrt{1-\Sigma^2}\cos\sigma \,,
   \end{equation}
   where
   \begin{equation}\label{eq:gambet}
      \gamma = \frac{\mathfrak{b}}{\mathfrak{a}\,\alpha}
      \hspace{0.3cm}\text{and}\hspace{0.3cm}
      \beta = \frac{\mathfrak{c}}{\mathfrak{a}\,\alpha} \,.
   \end{equation}
   The dynamical system with Hamiltonian function~\eqref{eq:Hres} is well known. It was thoroughly studied by \cite{HENRARD-MURIGANDE_1987}, who called it ``Colombo's top'' in memory of \cite{COLOMBO_1966}. In the following, we detail its characteristics of interest here, in terms of the two constant parameters $\gamma$ and $\beta$.
    
\begin{figure*}
      \centering
      \includegraphics[width=0.8\textwidth]{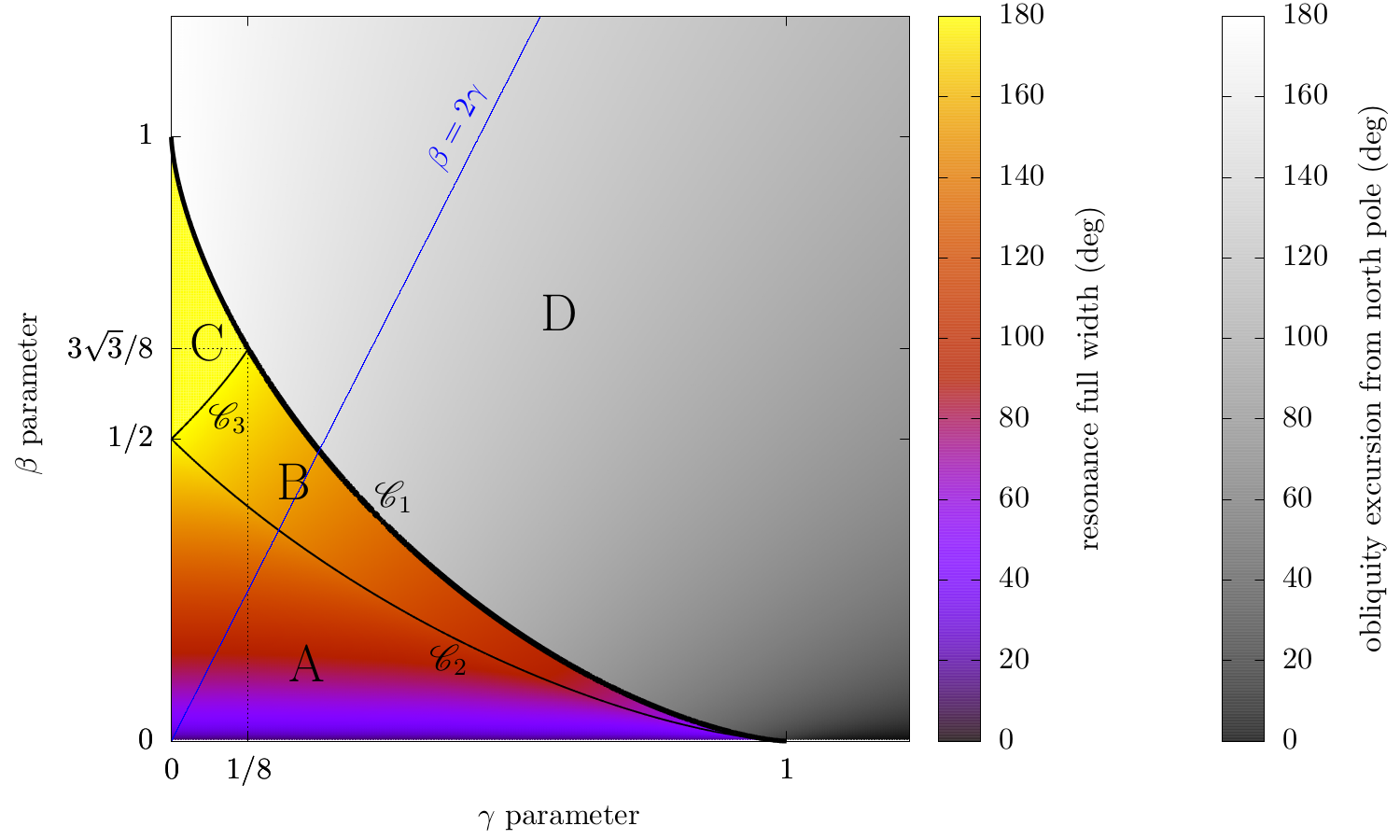}
      \caption{Parameter space of Colombo's top Hamiltonian~\eqref{eq:Hres}. There is no separatrix (and thus no resonance) in region D, delimited by the curve $\mathscr{C}_1$ (Eq.~\ref{eq:C1}, thick black line). The curves $\mathscr{C}_2$ and $\mathscr{C}_3$ (Eq.~\ref{eq:C23}, thin black lines) delimit regions A-B and B-C, respectively. In region A, the resonance island lies between $\Sigma_-$ and $\Sigma_+$ (Eq.~\ref{eq:GSwidth}); in region B, it lies between $\Sigma_-$ and $+1$; in region C, it lies between $-1$ and $+1$ (thus the $180^\text{o}$ full width given by the colour scale). The problem becomes unphysical above $\beta=2\gamma$ (blue line) since it corresponds to an amplitude $S_p>1$ in the inclination series.}
      \label{fig:zones}
   \end{figure*}

\begin{figure*}
      \centering
      \includegraphics[width=0.245\textwidth]{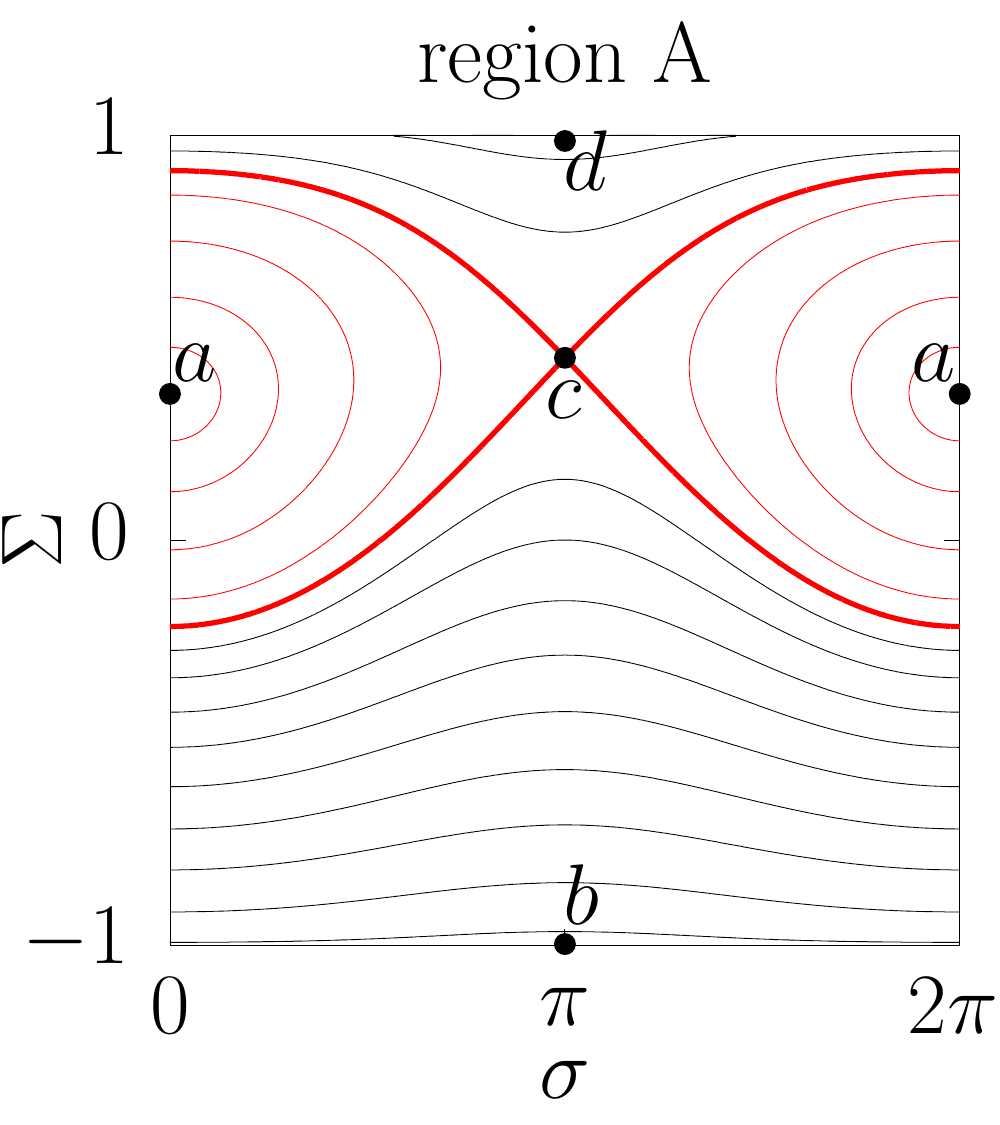}
      \includegraphics[width=0.245\textwidth]{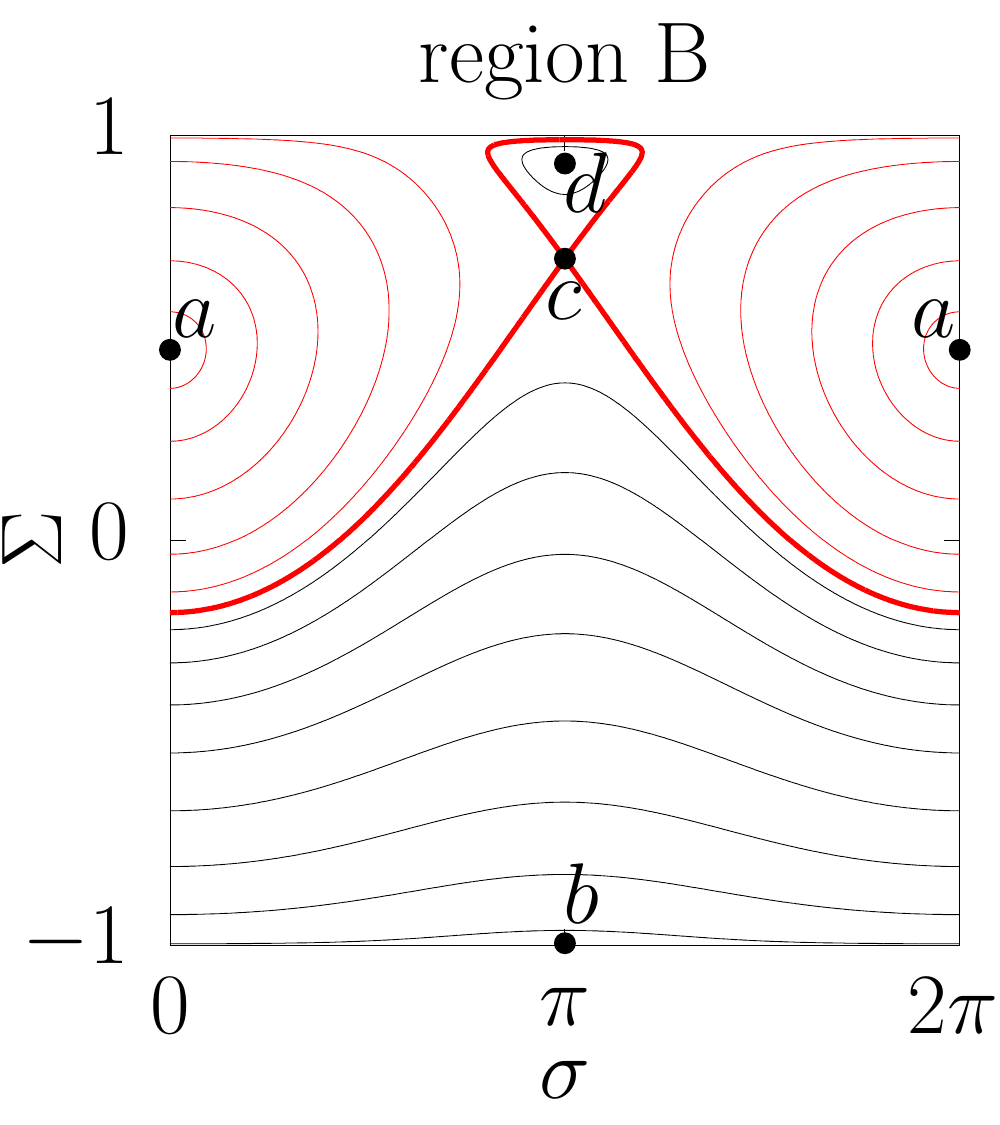}
      \includegraphics[width=0.245\textwidth]{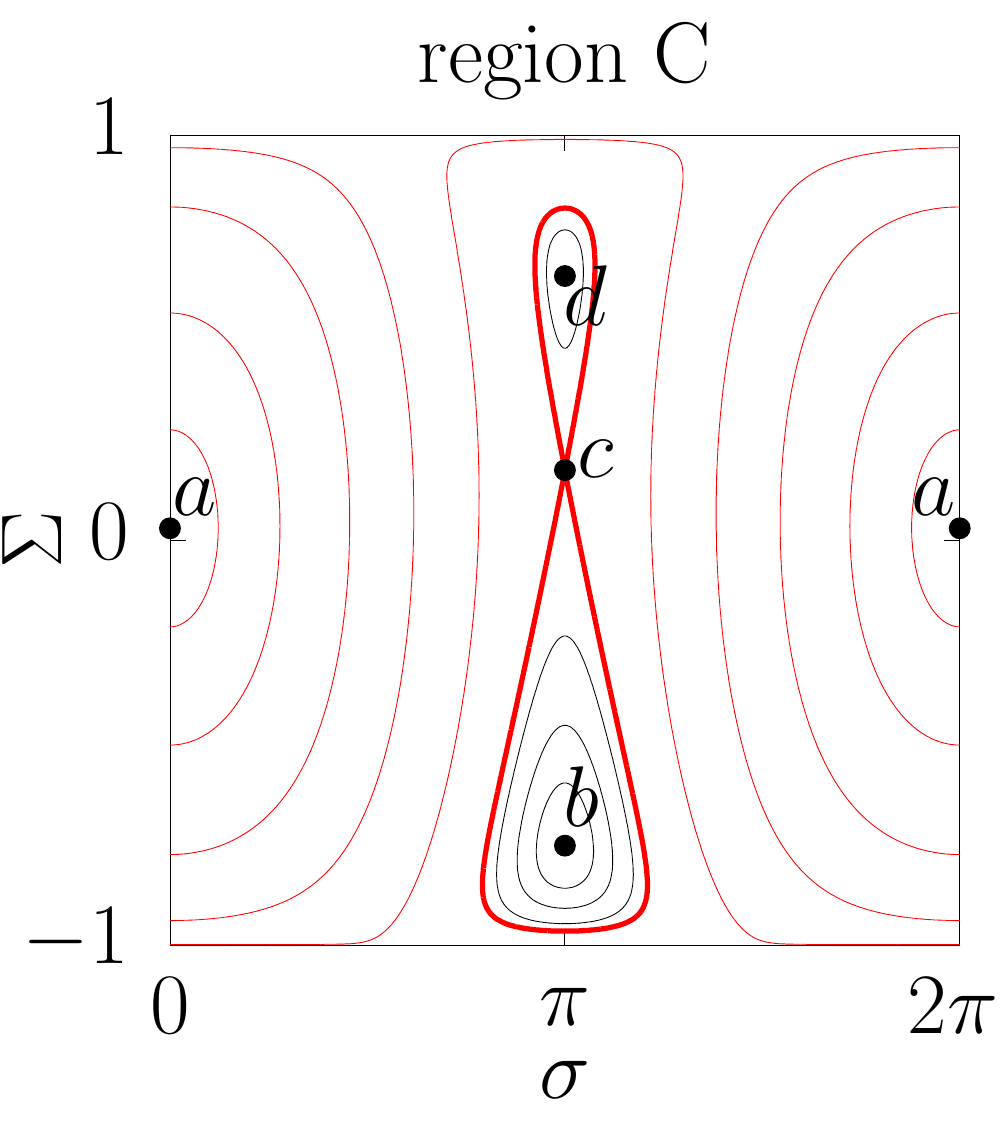}
      \includegraphics[width=0.245\textwidth]{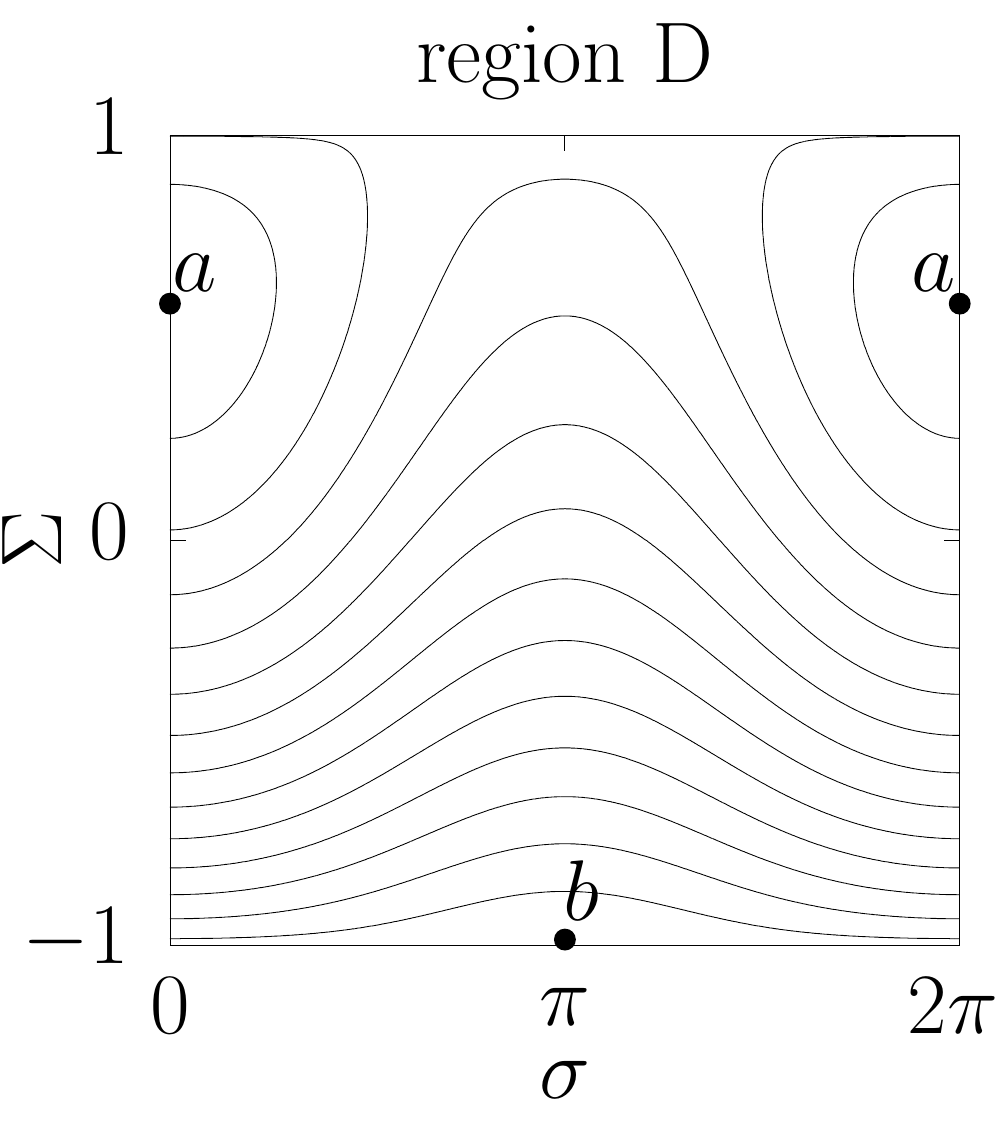}
      \caption{Examples of phase portraits for the four regions of Fig.~\ref{fig:zones}. The equilibrium points are labelled as in Eq.~\eqref{eq:EqP}. The level curves of the Hamiltonian are drawn with black lines out of the resonance island and with red lines inside the resonance. The parameters chosen are, from A to D: $(\gamma,\beta)=(0.4,0.1)$; $(0.55,0.15)$; $(0.05,0.7)$; $(0.8,0.3)$.}
      \label{fig:phase}
   \end{figure*}

   First of all, it is enough to study the case $\gamma\geqslant 0$ and $\beta\geqslant 0$ since we get the negative cases by the transformations $\Sigma\rightarrow -\Sigma$ and $\sigma\rightarrow\sigma+\pi$, respectively. When studying the equilibrium points of the system (\citealp{HENRARD-MURIGANDE_1987}, or Appendix~\ref{assec:eqp} and \ref{assec:1stb}), we find that the phase space can have two different geometries according to the value of $\gamma$ and $\beta$ (see Fig.~\ref{fig:phase} for the phase portraits). The boundary between these two regions of the parameter space is the curve
   \begin{equation}\label{eq:C1}
      \mathscr{C}_1 = \left\{ \gamma,\beta\geqslant 0:\ \ \gamma^{2/3}+\beta^{2/3} = 1\ \right\} \,.
   \end{equation}
   Below $\mathscr{C}_1$ (regions A,B,C of Fig.~\ref{fig:zones}), the dynamical system~\eqref{eq:Hres} has four equilibrium points, that we denote $(a,b,c,d)$. The equilibrium points $c$ and $d$ merge along the curve $\mathscr{C}_1$, and disappear above it (region D of Fig.~\ref{fig:zones}). Their respective positions are
   \begin{equation}\label{eq:EqP}
      \left\{
      \begin{aligned}
         \Sigma_a &\in [0,+1]
         \hspace{0.2cm},\hspace{0.2cm} \sigma_a = 0
         \hspace{0.2cm}\rightarrow\text{ elliptic (A,B,C,D)}\\
         \Sigma_b &\in [-1,0]
         \hspace{0.2cm},\hspace{0.2cm} \sigma_b = \pi
         \hspace{0.2cm}\rightarrow\text{ elliptic (A,B,C,D)}\\
         \Sigma_c &\in [0,+1]
         \hspace{0.2cm},\hspace{0.2cm} \sigma_c = \pi
         \hspace{0.2cm}\rightarrow\text{ hyperbolic (A,B,C)}\\
         \Sigma_d &\in [0,+1]
         \hspace{0.2cm},\hspace{0.2cm} \sigma_d = \pi
         \hspace{0.2cm}\rightarrow\text{ elliptic (A,B,C)} \,.
      \end{aligned}
      \right.
   \end{equation}
   The values $\Sigma_{a,b,c,d}$ have explicit expressions in terms of $\gamma$ and $\beta$, as given in Appendix~\ref{assec:eqp} (they correspond to the different roots of a quartic equation). Since the resonant angle $\psi+\phi_p$ is equal to $0$ or $\pi$ for each of these equilibrium points, they all correspond to configurations where the spin axis, the normal to the orbit (reduced to its $p$th harmonic) and the normal to the reference plane are in the same plane. As such, they are commonly called ``Cassini's states'' and labelled $(1,2,3,4)$ after \cite{PEALE_1969}, corresponding to the equilibrium points $(c,a,b,d)$.
   
   We note the limiting cases
   \begin{equation}\label{eq:beta}
      \begin{aligned}
         &\text{for } \beta \leqslant 1:\left\{
         \begin{aligned}
            &\lim\limits_{\gamma\rightarrow 0}\Sigma_a = \lim\limits_{\gamma\rightarrow 0}\Sigma_c = 0 \\
            &\lim\limits_{\gamma\rightarrow 0}\Sigma_d = -\lim\limits_{\gamma\rightarrow 0}\Sigma_b = \sqrt{1-\beta^2}
         \end{aligned}
         \right.\\
         &\text{for } \beta \geqslant 1:\ \lim\limits_{\gamma\rightarrow 0}\Sigma_a = \lim\limits_{\gamma\rightarrow 0}\Sigma_b = 0 \,,
      \end{aligned}
   \end{equation}
   and
   \begin{equation}\label{eq:gamma}
      \begin{aligned}
         &\text{for } \gamma \leqslant 1:\left\{
         \begin{aligned}
            &\lim\limits_{\beta\rightarrow 0}\Sigma_a = \lim\limits_{\beta\rightarrow 0}\Sigma_c = \gamma \\
            &\lim\limits_{\beta\rightarrow 0}\Sigma_d = -\lim\limits_{\beta\rightarrow 0}\Sigma_b = 1
         \end{aligned}
         \right.\\
         &\text{for } \gamma \geqslant 1;\ \lim\limits_{\beta\rightarrow 0}\Sigma_a = -\lim\limits_{\beta\rightarrow 0}\Sigma_b = 1 \,.
      \end{aligned}
   \end{equation}
   
   We are now interested in the width of the resonance, that is the interval of $\Sigma$ enclosed in the separatrix emerging from the hyperbolic fixed point $c$ and containing the fixed point $a$. A pendulum approximation can be obtained for small values of $\beta$ (as used by \citealp{ATOBE-etal_2004} or \citealp{LI-BATYGIN_2014a}), but this approximation is no longer valid when $\beta$ grows. Since an analytical expression can be derived even in the general case, we will use it here. The computations (Appendix~\ref{assec:sep}) lead to the following extreme values of $\Sigma$ spanned by the resonance:
   \begin{equation}\label{eq:GSwidth}
      \Sigma_{\pm} = 2\gamma - \Sigma_c \pm 2\sqrt{ -\beta^2+\beta\sqrt{1-\Sigma_c^2} } \,.
   \end{equation}
   They are defined whenever $\Sigma_c$ itself is defined (regions A,B,C of Fig.~\ref{fig:zones}). At this point, it is important to note that the coordinates $(\Sigma,\sigma)$ are singular at $\Sigma=\pm 1$ since the problem actually takes place on the sphere \citep{HENRARD-MURIGANDE_1987}. We must thus study carefully the meaning of the limits~\eqref{eq:GSwidth} when one of them crosses $\pm 1$. This leads to two other limits in the parameter space, as the curves
   \begin{equation}\label{eq:C23}
      \begin{aligned}
         \mathscr{C}_2 &= \left\{ \gamma,\beta\geqslant 0:\ \ 8\,\beta^2 = 1-20\gamma-8\gamma^2+(1+8\gamma)^{3/2}\ \right\} \\
         \mathscr{C}_3 &= \Big\{ 0\leqslant\gamma\leqslant1/8\ ,\ \beta\geqslant 0:\\
         &\hspace{2.5cm} 8\,\beta^2 = 1+20\gamma-8\gamma^2+(1-8\gamma)^{3/2}\ \Big\}
      \end{aligned}
   \end{equation}
   (see Appendix~\ref{assec:2ndb}-\ref{assec:3rdb}), delimiting the regions A-B and B-C of Fig.~\ref{fig:zones}, respectively. We note that $\mathscr{C}_1$ and $\mathscr{C}_2$ intersect at $(\gamma;\beta)=(1\,;0)$, $\mathscr{C}_1$ and $\mathscr{C}_3$ intersect at $(1/8\,;3\sqrt{3}/8)$, and $\mathscr{C}_2$ and $\mathscr{C}_3$ intersect at $(0\,;1/2)$. Contrary to $\mathscr{C}_1$, the boundaries $\mathscr{C}_2$ and $\mathscr{C}_3$ do not correspond to actual bifurcations of the dynamical system, but only to the limits where the resonant island contains the north pole of the sphere (region B) and both poles of the sphere (region C). Hence, the resonance lies in $[\Sigma_-;\Sigma_+]$ in region A; in $[\Sigma_-;+1]$ in region B; and in $[-1;+1]$ in region C (see Fig.~\ref{fig:phase}). In region D, there is no more resonance (the separatrix disappears), but the obliquity can still vary substantially. In Fig.~\ref{fig:zones}, the colour shades in region D show the oscillation amplitude of the obliquity as the trajectory passes through $\Sigma=1$.
   
   When $\beta\rightarrow 0$, the width of the island tends to $0$ and all the level curves in the $(\sigma,\Sigma)$ plane tend to be horizontal. In this limit, the resonance width is small and almost independent of $\gamma$ (pendulum approximation). When $\gamma\rightarrow 0$, the phase portrait in the $(\sigma,\Sigma)$ plane tends to be symmetric with respect to the line $\Sigma=0$.
   
   From~\eqref{eq:gambet}, we note that
   \begin{equation}\label{eq:bgsimp}
      \gamma=\frac{\nu_p}{\alpha}+\mathcal{O}(\varepsilon^2)
      \hspace{0.3cm}\text{and}\hspace{0.3cm} \beta=-2S_p\gamma+\mathcal{O}(\varepsilon^2) \,.
   \end{equation}
   Remembering that $S_p$ is the amplitude of a given term in the inclination series~\eqref{eq:qprep}, this means that the parameter region above the line $\beta=2\gamma$ in Fig.~\ref{fig:zones} (that is, $|S_p|=1$) cannot be reached in this problem.
   
   As a simple rule of thumb, one can consider that $\beta$ controls the resonance width, and that $\gamma$ controls the location of the resonance centre (see the Hamiltonian at Eq.~\ref{eq:Hres}). From Eq.~\eqref{eq:bgsimp}, we know that $\beta$ is proportional to the amplitude $S_p$. Therefore, the resonance widths are larger in hot planetary systems, for which the mutual inclinations are large. This was exploited by \cite{BOUE-LASKAR_2010} in their scenario for tilting the spin-axis of Uranus. On the contrary, the secular system cannot produce any obliquity variation if the mutual inclinations are exactly zero. One must keep in mind that $\beta$ depends on the precession constant $\alpha$ (Eq.~\ref{eq:alpha}) as well, so that ``small'' mutual inclinations do not guarantee that the resonances are thin. For the terrestrial planets of the Solar System, some values of $\alpha$ produce first-order resonances larger than $70^\text{o}$, even if the mutual orbital inclinations are modest (see Sect.~\ref{sec:ovl}).
   
   Contrary to $\beta$, the magnitude of $\gamma$ cannot be easily traced back from the planetary architecture: the first-order secular spin-orbit resonances of any planet can be located anywhere between $0^\text{o}$ and $180^\text{o}$ of obliquity. The large majority of them actually lie in $[0^\text{o};90^\text{o}]$ because most of the frequencies $\nu_p$ are negative (the explanation for this property is given in Sect.~\ref{ssec:LL}).
   
   \subsection{Overlap of first-order resonances}\label{sec:ovl}
   Going back to the full Hamiltonian~\eqref{eq:Hgen}, the main chaotic regions of the system can be estimated as the overlap of the first-order resonances taken separately (Chirikov's criterion). With the analytical expression of their respective widths given in Sect.~\ref{sec:coltop}, the computation of the overlapping regions is straightforward for any quasi-periodic representation~\eqref{eq:qprep} of the orbital motion.
   
   When comparing two secular terms, we note that no chaotic zone can form if at least one of them is in region D, because there is no separatrix in D. This is well verified by Poincaré sections. A direct consequence of this property is that if all the terms of the quasi-periodic series are in region D, the secular dynamics of the obliquity cannot be chaotic at first order. Moreover, if all the $S_i$ coefficients are small (low inclination regime), so are the corresponding $\beta$ coefficients, resulting in virtually no secular variation of the obliquity (see the shades of grey in Fig.~\ref{fig:zones}). Actually, a no-chaos criterion can be obtained even if the amplitudes $S_i$ are not known: we just have to check that $\gamma_i\approx \nu_i/\alpha >1\ \forall\ i=1...M$. On the contrary, if $\gamma_i<1$ for at least two $i$, the existence of a chaotic region is possible, but not guaranteed. These properties will be discussed further in Sect.~\ref{ssec:nres}.
   
\begin{figure*}
      \centering
      \includegraphics[width=0.95\textwidth]{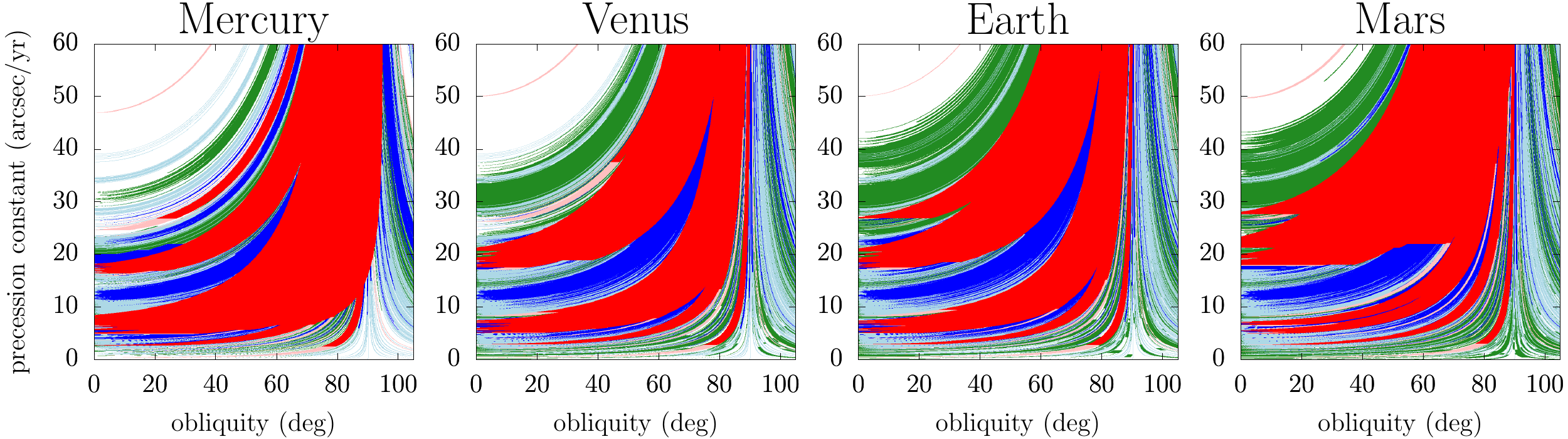}\\
      \vspace{0.2cm}
      \includegraphics[width=0.95\textwidth]{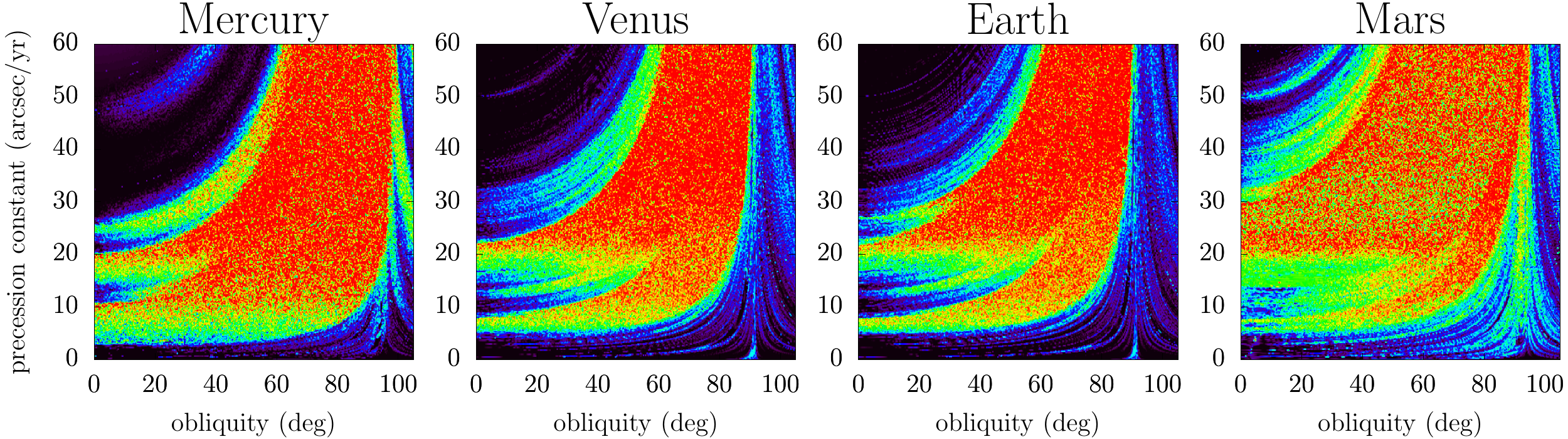}\\
      \vspace{0.2cm}
      \includegraphics[width=0.95\textwidth]{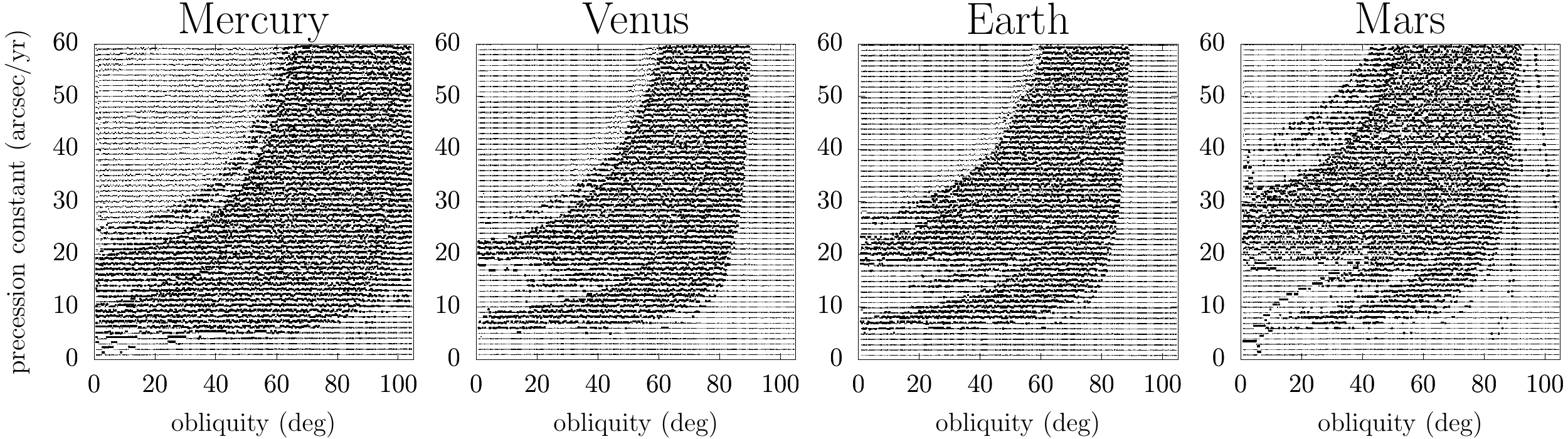}\\
      \vspace{0.2cm}
      \includegraphics[width=0.95\textwidth]{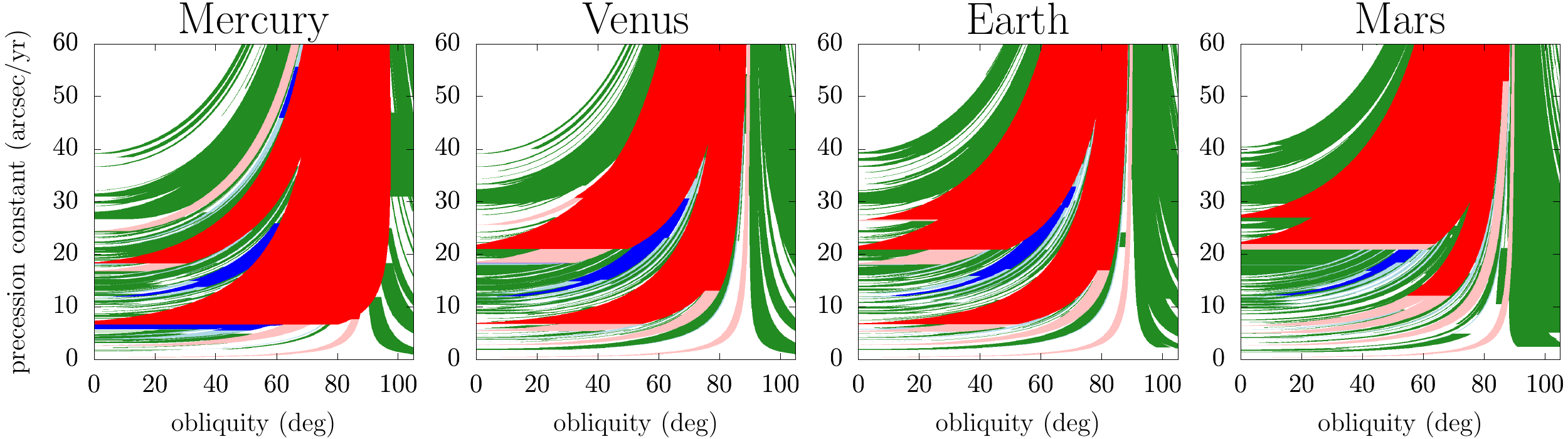}
      \caption{\emph{\textbf{Top row:}} Estimate of the chaotic regions of the spin dynamics as the superposition of secular spin-orbit resonances. The orbital evolution of each planet is approximated by the synthetic representation of \cite{LASKAR_1990}, as detailed in Appendix~\ref{asec:SSsol}. Light-red and dark-red regions represent the first-order resonances and their overlaps, respectively (Colombo's top Hamiltonian, Sect.~\ref{sec:coltop}); light-blue and dark-blue regions represent the second-order resonances and their overlap (Sect.~\ref{sec:highor}); and green regions represent the overlap of third-order resonances. The non-overlapping third-order resonances are not indicated because they are very thin and thus unimportant for a global picture of the dynamics. \emph{\textbf{Second row:}} As a comparison, the system given by Eq.~\eqref{eq:Hinit} is integrated numerically with the same orbital model (quasi-periodic decomposition of \citealp{LASKAR_1990}), and a frequency map analysis is performed to locate the chaotic zones. The colour scale goes from black (no chaos), to red (strong chaos). \emph{\textbf{Third row:}} Same maps obtained from a more detailed model in which the orbital evolution is directly taken from a numerical integration (adapted from \citealp{LASKAR-ROBUTEL_1993}). In the weakly chaotic zones, the dots are shifted vertically according to the level of chaos; in the strongly chaotic zones, they are plotted in boldface. \emph{\textbf{Bottom row:}} Same as top row, but the long-term orbital evolution of each planet is approximated by the Lagrange-Laplace system (Sect.~\ref{ssec:LL}). The eight planets of the Solar System are included, with the initial conditions of \cite{BRETAGNON_1982}.}
      \label{fig:ovlSR}
   \end{figure*}

   This method can be easily checked in the case of the Solar System, since the secular spin dynamics of all planets have been studied in the literature. Moreover, a very accurate quasi-periodic approximation of the orbital dynamics can be obtained, since the properties and initial conditions of the planets are very well known. As an example, the upper row of Fig.~\ref{fig:ovlSR} shows the widths and overlaps of first-order resonances for the terrestrial planets (light and dark-red regions). This figure was produced by applying the previous analytical formulas to the orbital series of \cite{LASKAR_1990}, containing more than 50 terms in both eccentricity and inclination (see Appendix~\ref{asec:SSsol}). We note that most of the first-order resonances overlap (there are almost no light-red regions). This results in wide chaotic zones even if the individual amplitudes $S_i$ are small. However, the full extent of the chaotic regions given by the frequency analysis (Fig.~\ref{fig:ovlSR}, second and third rows) cannot be retrieved by only considering first-order resonances. The following section is thus dedicated to second and third-order resonances.
   
   \subsection{Higher-order resonances}\label{sec:highor}
   Outside of first-order resonances, we can use a canonical change of coordinates close to identity in order to suppress the angular dependency at first order \citep[as already used in a similar context by][]{LI-BATYGIN_2014a}. Let us consider an intermediary Hamiltonian $\mathcal{X} = \varepsilon\mathcal{X}_1$, such that the current coordinates are obtained from the new ones through its flow at time $1$. The Hamiltonian in the new coordinates is then
   \begin{equation}
      \tilde{\mathcal{H}} = \tilde{\mathcal{H}}_0 + \varepsilon\tilde{\mathcal{H}}_1 + \varepsilon^2\tilde{\mathcal{H}}_2 + \mathcal{O}(\varepsilon^3) \,,
   \end{equation}
   where
   \begin{equation}\label{eq:Htilde1}
      \begin{aligned}
         \tilde{\mathcal{H}}_0 &= \mathcal{H}_0 \\
         \tilde{\mathcal{H}}_1 &= \mathcal{H}_1 + \{\mathcal{X}_1,\mathcal{H}_0\} \\
         \tilde{\mathcal{H}}_2 &= \mathcal{H}_2 + \{\mathcal{X}_1,\mathcal{H}_1\} + \frac{1}{2}\{\mathcal{X}_1,\{\mathcal{X}_1,\mathcal{H}_0\}\} \,.
      \end{aligned}
   \end{equation}
   In these expressions, Poisson's brackets are defined as
   \begin{equation}
      \{f,g\} = \sum_{j}\left(\frac{\partial f}{\partial p_j}\frac{\partial g}{\partial q_j} - \frac{\partial f}{\partial q_j}\frac{\partial g}{\partial p_j}\right) \,,
   \end{equation}
   where the pairs $(p_j,q_j)$ are conjugate variables, $p_j$ being the momentum and $q_j$ the coordinate. In order to suppress the angular dependency at order 1, the Hamiltonian $\mathcal{X}$ must fulfil the homological equation
   \begin{equation}
      \tilde{\mathcal{H}}_1 = \mathcal{H}_1 + \{\mathcal{X}_1,\mathcal{H}_0\} = \overline{\mathcal{H}}_1 \,,
   \end{equation}
   in which $\overline{\mathcal{H}}_1$ is the 0th-order term of the multidimensional Fourier decomposition of $\mathcal{H}_1$ (average of $\mathcal{H}_1$ over all angles), which is here equal to zero. By matching the terms of the Fourier decomposition of $\mathcal{H}_1$ and $\mathcal{X}_1$ one by one, the solution to the homological equation is
   \begin{equation}\label{eq:X1}
      \varepsilon\mathcal{X}_1(X,\phi,-\psi) = -2\sqrt{1-X^2}\sum_{j=1}^M\frac{\nu_j\,S_j}{\nu_j+\alpha X}\sin(\phi_j+\psi) \,.
   \end{equation}
   Injecting this function into the expression of the new Hamiltonian~\eqref{eq:Htilde1}, we get $\tilde{\mathcal{H}}_1=0$ as required, and the second order term $\tilde{\mathcal{H}}_2$ is given in Appendix~\ref{asec:2ndor}. The only possible resonant angles at second order have the form $\sigma = \phi_j+\phi_k+2\psi$. The width of second and higher order resonances is quite small, so that their separatrices can be computed assuming that $X$ is near the exact resonance (pendulum approximation). Accordingly, the centre and half width of the second-order resonances are computed in Appendix~\ref{asec:2ndor} and given in the first line of Table~\ref{tab:hres}.
    
   Outside of both first-order and second-order resonances, the same method can be used to compute the location and width of third-order resonances. An intermediary Hamiltonian of the form $\mathcal{X}=\varepsilon\mathcal{X}_1 + \varepsilon^2\mathcal{X}_2$ is used, in which $\mathcal{X}_2$ must satisfy a second homological equation (see Appendix~\ref{asec:3rdor}). The possible resonant angles, as well as the centre and half width of all the third-order resonances are given in Table~\ref{tab:hres}. As before, the same method can be used in the case of synchronous rotation, by adding the term $-\alpha_r(1+X)^2/2$ to $\mathcal{H}_0$ (Appendix~\ref{asec:spinorb}). This would only slightly shift the resonances.
   
   With these values, it is straightforward to compute the overlap regions of every possible second and third-order resonances. Resonances of order 2 or 3 with a centre located inside a resonance of lower order are not considered (in other words, low-order resonances are supposed unaffected by higher-order ones). The result is shown in the top row of Fig.~\ref{fig:ovlSR} for the inner Solar System (blue and green zones). We obtain a much better match with the frequency map analysis, showing the importance of high-order resonances in this context. Indeed, numerous frequencies of the quasi-periodic representation are quite close to each other, which implies that the corresponding resonances overlap massively. Hence, even if the second and third-order resonances are thin, most of them are all located one after another, resulting in large chaotic zones. The chaotic diffusion is though slower for higher-order resonances. In the real Solar System, in which the secular orbital frequencies are actually not fixed but vary slowly \citep{LASKAR_1990}, the diffusion of the obliquity will besides be facilitated by small modulations of the resonances locations.
   
   We note that third-order resonances include terms mixing both eccentricity and inclination (last line of Table~\ref{tab:hres}). The existence of these terms shows that the $s_6+g_5-g_6$ resonance, which is known to play an important role in the future obliquity evolution of the Earth \citep{LASKAR-etal_1993b,LASKAR-etal_2004a}, has two different origins: one is a first-order resonance with $\nu_{23}$, and the other is a third-order resonance between $\mu_1$, $\mu_{10}$ and $\nu_{12}$ (the numbering refers to \citealp{LASKAR-etal_2004a}). The amplitude of the first resonance, which is the one emphasised in the literature, is larger by a factor thirty. This resonance is present in the synthetic representation used in this paper (see the term with frequency $-50.30212\ ''$/yr in Tables~\ref{tab:QPS-Mercury} to \ref{tab:QPS-Mars}). In the top row of Fig.~\ref{fig:ovlSR}, it appears as a thin isolated resonance (upper-left corner of the graphs).
   
   Finally, since tidal dissipations are much more efficient in decreasing the eccentricity than the inclination, one can imagine a planet with an initially chaotic obliquity wandering in a mixed-type-resonance overlap region, becoming frozen out of resonance when the amplitudes $E_i$ decrease due to tidal dissipation. As shown by \citealp{LASKAR-etal_2012}, the eccentricity amplitudes of all the planetary system can be damped even if only one planet dissipates energy with the star. Hence the eccentricity modes of an external planet can be damped even if it is not itself subject to tidal dissipation.
   
   \begin{table*}
      \caption{Critical angle, location and half width of every second and third-order secular spin-orbit resonance.}
      \label{tab:hres}
      \vspace{-0.7cm}
      \begin{equation*}
         \begin{array}{cccc}
            \hline\hline
            \sigma && X_0 & K\text{ (for a half width $2\sqrt{|K|}$)} \\
            \hline
            \phi_j+\phi_k+2\psi && -\frac{1}{2\alpha}(\nu_j+\nu_k) &
            \phantom{\Bigg(}
            8(1-X_0^2)\,\frac{\nu_j\nu_k}{(\nu_j-\nu_k)^2}S_jS_k
            \phantom{\Bigg)} \\
            2\phi_j+\phi_k+3\psi && -\frac{1}{3\alpha}(2\nu_j+\nu_k) &
            \phantom{\Bigg(}
            -\frac{243}{2}(1-X_0^2)^{3/2}\,\frac{\alpha\,\nu_j^2\nu_k}{(\nu_j-\nu_k)^4}S_j^2S_k
            \phantom{\Bigg)} \\
            2\phi_j-\phi_k+\psi && -\frac{1}{\alpha}(2\nu_j-\nu_k) &
            \phantom{\Bigg(}
            \frac{1}{2}\sqrt{1-X_0^2}\,\frac{2(2\nu_j^2+3\nu_j\nu_k-\nu_k^2)(\nu_j-\nu_k)^3+\nu_j^2\nu_k(\nu_k^2-\alpha^2)}{\alpha(\nu_j-\nu_k)^4}S_j^2S_k
            \phantom{\Bigg)} \\
            -\phi_i+\phi_j+\phi_k+\psi && -\frac{1}{\alpha}(-\nu_i+\nu_j+\nu_k) &
            \phantom{\Bigg(}
            2\sqrt{1-X_0^2}\,\frac{P}{\alpha(\nu_i-\nu_j)(\nu_i-\nu_k)(2\nu_i-\nu_j-\nu_k)^2}S_iS_jS_k
            \phantom{\Bigg)} \\
            \phi_i+\phi_j+\phi_k+3\psi && -\frac{1}{3\alpha}(\nu_i+\nu_j+\nu_k) &
            \phantom{\Bigg(}
            -486(1-X_0^2)^{3/2}\,\frac{\alpha\,\nu_i\nu_j\nu_k\big((\nu_i-\nu_j)^2 + (\nu_j-\nu_k)^2 + (\nu_k-\nu_i)^2\big)}{(2\nu_i-\nu_j-\nu_k)^2(2\nu_j-\nu_i-\nu_k)^2(2\nu_k-\nu_i-\nu_j)^2}S_iS_jS_k
            \phantom{\Bigg)} \\
            \phi_i+\theta_j-\theta_k+\psi && -\frac{1}{\alpha}(\nu_i+\mu_j-\mu_k) &
            \phantom{\Bigg(}
            -3X_0\sqrt{1-X_0^2}\,\frac{\nu_i}{\mu_j-\mu_k}S_iE_jE_k
            \phantom{\Bigg)} \\
            \hline
         \end{array}
      \end{equation*}
      \vspace{-0.3cm}
      \tablefoot{In the fourth line, the symbol $P$ stands for: $4\nu_i^5 - 12\nu_i^4(\nu_j+\nu_k) + \nu_i^3(13\nu_j^2+13\nu_k^2+16\nu_j\nu_k) - 2\nu_i^2(3\nu_j^3+3\nu_k^3+\nu_j^2\nu_k+\nu_j\nu_k^2) + \nu_i(\nu_j^4+\nu_k^4-3\nu_j^3\nu_k-3\nu_j\nu_k^3-8\nu_j^2\nu_k^2-2\alpha^2\nu_j\nu_k) + \nu_j\nu_k(\nu_j+\nu_k)^3$.}
   \end{table*}

   \section{Application to exoplanetary systems}\label{sec:app}
   In the previous sections, we saw that in the low-eccentricity and low-inclination regime, the long-term rotational dynamics of planets can be studied very efficiently by a simple analytical model. However, even if numerous exoplanetary systems are known nowadays, most of the information required to characterise the rotation of their planets remains poorly constrained. This information can be split into two groups: \emph{i)} the orbital dynamics (amplitudes and frequencies of the quasi-periodic representation) and \emph{ii)} the rotation parameters ($\alpha$ coefficient). In this section, we will see how these quantities can be estimated from physical and dynamical arguments, even with scarce data.
   
   \subsection{The Lagrange-Laplace system}\label{ssec:LL}
   Regarding the long-term orbital dynamics, one can use nominal orbital elements (either best-fit or assumed ones) and integrate numerically the equations of motion. The so-obtained solution can then be used directly (as did for instance \citealp{BRASSER-etal_2014}, and \citealp{DEITRICK-etal_2018}), or put in the form of quasi-periodic series and used as shown above. However, this method puts a heavy contrast between the very uncertain nature of the orbital elements used and the refined numerical solution applied. Actually, we will see that at this level of precision, the Lagrange-Laplace system is already a good-enough approximation of the orbital dynamics, up to moderate eccentricities and inclinations (and without strong effects coming from mean-motion resonances). It was used for the same purpose by \cite{ATOBE-etal_2004} in the case of a massless hypothetical terrestrial planet.
   
   The Lagrange-Laplace system is the lowest-order model of the long-term orbital dynamics: it uses a development of the Hamiltonian at second order of the eccentricities and inclinations, which is itself averaged over the fast angles (secular model at first order to the mutual perturbations). Let us write
   \begin{equation}
      z_k = e_k\exp(i\varpi_k)
      \hspace{0.5cm}\text{and}\hspace{0.5cm}
      \zeta_k = \sin\frac{I_k}{2}\exp(i\Omega_k) \,,
   \end{equation}
   where the index $k=1,2...N$ represents a given planet of the system. Writing $\mathbf{z}$ and $\boldsymbol{\zeta}$ the vectors of all $z_k$ and $\zeta_k$, the equations of motion in the Lagrange-Laplace approximation are
   \begin{equation}
      \dot{\mathbf{z}} = iA\,\mathbf{z}
      \hspace{0.5cm}\text{and}\hspace{0.5cm}
      \dot{\boldsymbol{\zeta}} = iB\,\boldsymbol{\zeta} \,,
   \end{equation}
   where the real matrices $A$ and $B$ are only functions of the masses and semi-major axes. They can be retrieved from the lowest-order terms in eccentricity and inclination of the orbital Hamiltonian. Organising the planets by increasing semi-major axes, we get from \cite{LASKAR-ROBUTEL_1995}:
   \begin{equation}\label{eq:matEcc}
      \begin{aligned}
         A_{jj} &= n_j\sum_{k=1}^{j-1}\frac{m_k}{m_0}C_3\left(\frac{a_k}{a_j}\right) + n_j\sum_{k=j+1}^{N}\frac{m_k}{m_0}\frac{a_j}{a_k}C_3\left(\frac{a_j}{a_k}\right) \\
         A_{jk} &=
         \left\{
         \begin{aligned}
            &2n_j\frac{m_k}{m_0}C_2\left(\frac{a_k}{a_j}\right)
            \hspace{0.5cm}&\text{if}\ k<j \\
            &2n_j\frac{m_k}{m_0}\frac{a_j}{a_k}C_2\left(\frac{a_j}{a_k}\right)
            \hspace{0.5cm}&\text{if}\ k>j
         \end{aligned}
         \right.
      \end{aligned}
   \end{equation}
   and
   \begin{equation}\label{eq:matInc}
      \begin{aligned}
         B_{jj} &= -n_j\sum_{k=1}^{j-1}\frac{m_k}{m_0}C_3\left(\frac{a_k}{a_j}\right) - n_j\sum_{k=j+1}^{N}\frac{m_k}{m_0}\frac{a_j}{a_k}C_3\left(\frac{a_j}{a_k}\right) \\
         B_{jk} &=
         \left\{
         \begin{aligned}
            &n_j\frac{m_k}{m_0}C_3\left(\frac{a_k}{a_j}\right)
            \hspace{0.5cm}&\text{if}\ k<j \\
            &n_j\frac{m_k}{m_0}\frac{a_j}{a_k}C_3\left(\frac{a_j}{a_k}\right)
            \hspace{0.5cm}&\text{if}\ k>j
         \end{aligned}
         \right.
      \end{aligned}
   \end{equation}
   in which $n_j^2a_j^3=\mathcal{G}(m_0+m_j)$, and the functions $C_2(\alpha)$ and $C_3(\alpha)$ are expressed in terms of the Laplace coefficients $b_s^{(k)}$:
   \begin{equation}
      \begin{aligned}
         C_2(\alpha) &= \frac{3}{8}\alpha\,b_{3/2}^{(0)}(\alpha)-\frac{1}{4}(1+\alpha^2)\,b_{3/2}^{(1)}(\alpha) \\
         C_3(\alpha) &= \frac{1}{4}\alpha\,b_{3/2}^{(1)}(\alpha) \,,
      \end{aligned}
   \end{equation}
   (see \citealp{LASKAR-ROBUTEL_1995}, \citealp{LASKAR-etal_2012} or \citealp{MURRAY-DERMOTT_1999}). The equations of motion for $\mathbf{z}$ and $\boldsymbol{\zeta}$ are decoupled and linear, such that the solution can be obtained by diagonalising the matrices $A$ and $B$. Its expression can be taken directly from \cite{LASKAR-etal_2012}: it has the form of quasi-periodic series~\eqref{eq:qprep} as required by our model. The frequencies $g_k$ and $s_k$ are the eigenvalues of $A$ and $B$, and the amplitude of the term $k$ for the planet $j$ is
   \begin{equation}\label{eq:ES}
      \begin{aligned}
         E_k^{(j)} &= \left|P_{jk}\sum_{i=1}^{N}P^{-1}_{ki}z_i(0)\right|
         \hspace{0.5cm}\text{(excentricity series)}\\
         S_k^{(j)} &= \left|Q_{jk}\sum_{i=1}^{N}Q^{-1}_{ki}\zeta_i(0)\right|
         \hspace{0.5cm}\text{(inclination series)} \,,
      \end{aligned}
   \end{equation}
   where $(P,Q)$ are the matrices composed of the eigenvectors of $(A,B)$, the matrices $(P^{-1},Q^{-1})$ are their inverses, and $z_i(0)$, $\zeta_i(0)$ are the initial conditions of planet~$i$. In this case, we note that there is a single term for each proper frequency of the system; the frequencies $\mu_j$ and $\nu_j$ of the quasi-periodic representation are thus directly equal to one of the $g_k$ and $s_k$, respectively.
   
   From the conservation of total orbital angular momentum, one of the inclination proper frequencies $s_k$ is identically equal to zero (matrix $B$ has rank deficiency of order~1). The inclination series have thus $M=N-1$ terms, whereas the eccentricity series have $N$ terms. Moreover, all the inclination proper frequencies are negative (this can be shown from the Ger\v{s}gorin circles theorem, see Appendix~\ref{asec:gersh}).
   
   The result, in terms of chaotic zones for the spin dynamics, is shown in Fig.~\ref{fig:ovlSR}, bottom row. Although the match with the numerical maps (Fig.~\ref{fig:ovlSR}, second and third rows) is not as good as when we used the synthetic representation of the orbital dynamics (Fig.~\ref{fig:ovlSR}, top row), the estimate obtained is still remarkably good considering the uncertainties of the elements of an exoplanetary system\footnote{Some subtle dynamical effects are not reproduced by the Lagrange-Laplace system, like the $s_6+g_5-g_6$ first-order resonance mentioned in Sect.~\ref{sec:highor}.}. Using this approach with the nominal orbital elements given by \cite{BRASSER-etal_2014} and \cite{DEITRICK-etal_2018}, we retrieve analytically their maps showing the possible obliquity variations of HD\,40307\,g and Kepler-62\,f, in terms of the locations and widths of the secular spin-orbit resonances (see Fig.~8 by \citealp{BRASSER-etal_2014} and Figs.~5, 6, 10, 11 by \citealp{DEITRICK-etal_2018}). The differences of oscillation amplitude that they observe are a natural consequence of the initial position of the planet with respect to the resonance centre (see Fig.~\ref{fig:phase}). This shows that the Lagrange-Laplace system, associated with the development of the Hamiltonian (Sect.~\ref{sec:anmod}), is enough to obtain the level of detail required for studying the long-term rotation of exoplanets up to moderate eccentricities and inclinations. The use of a more elaborate model would add no substantial information, owing to the large uncertainties of the exoplanetary system under study.
   
   \subsection{Maximisation of $E_k$ and $S_k$}\label{ssec:maxES}
   Unfortunately, several orbital elements remain unknown for most of the observed exoplanetary systems. The unknown elements usually include the mutual inclinations and the relative longitudes of ascending node. From now on, we suppose that only the masses, the semi-major axes and the eccentricities are known for all planets of the system. In this case, the Lagrange-Laplace matrices $A$ and $B$ can still be computed since they only depend on the masses and the semi-major axes. We thus obtain the two sets of frequencies $\mu_j$ and $\nu_j$. Because the initial conditions $z_i(0)$ and $\zeta_i(0)$ appearing in Eq.~\eqref{eq:ES} are not fully known, the goal here is to obtain the maximum possible value of the amplitudes $E_k$ and $S_k$ according to the available data.
   
   For the eccentricity, it amounts to maximise the modulus of a sum of complex numbers with unknown phase. The result is thus simply the sum of the moduli:
   \begin{equation}\label{eq:maxE}
      \max\Big[ E_k^{(j)} \Big] = \left|P_{jk}\right|\sum_{i=1}^{N}e_i\left|P^{-1}_{ki}\right| \,,
   \end{equation}
   using the fact that by definition, $|z_i(0)|=e_i$. The problem is more complex for the inclinations, since both the amplitudes and the phases of the initial conditions $\zeta_i(0)$ are unknown. It is thus necessary to introduce additional arguments, either from physical or from dynamical grounds. Guided by statistics on the orbital excitation due to close encounters, \cite{ATOBE-etal_2004}, while dealing mostly with systems with a single observed planet, imposed $I=e/2$ for each of them. This law is also in agreement with statistical distributions of observed exoplanetary systems \citep{XIE-etal_2016}. In our case, though, the application of this statistical result as a strict rule for each planet of a multi-planet system seems a bit simplistic. We will opt here for the hypothesis by \cite{LASKAR-PETIT_2017} of equipartition of the Angular Momentum Deficit (AMD) among the secular degrees of freedom. As they point out, this hypothesis is motivated both by theoretical arguments on chaotic diffusion in the secular dynamics \citep{LASKAR_1994,LASKAR_2008} and by the aforementioned correlations in observed distributions. As shown below, this allows to smooth the statistical law over all the planets contained in the system. Let us introduce the ``coplanar AMD'' of a planetary system, that is, the AMD it would have if it was strictly coplanar:
   \begin{equation}
      C_p = \sum_{j=1}^N \Lambda_j\left(1-\sqrt{1-e_j^2}\right) \,,
   \end{equation}
   where
   \begin{equation}
      \Lambda_j = \frac{m_0m_j}{m_0+m_j}\sqrt{\mathcal{G}(m_0+m_j)\,a_j} \,.
   \end{equation}
   Contrary to \cite{LASKAR-PETIT_2017}, we use here the Hamiltonian decomposition of \cite{LASKAR-ROBUTEL_1995}, where the integrable part is the Sun-planet two-body problem. This is also the one chosen when expressing the matrices $A$ and $B$ of the Lagrange-Laplace system (Eqs.~\ref{eq:matEcc}-\ref{eq:matInc}). The AMD equipartition hypothesis amounts to considering that the total AMD of the system,
   \begin{equation}
      C = \sum_{j=1}^N \Lambda_j\left(1-\sqrt{1-e_j^2}\cos I_j\right) \,,
   \end{equation}
   is equal to
   \begin{equation}\label{eq:equip}
      C = 2\,C_p \,.
   \end{equation}
   Hence, even if the individual orbital inclinations are not known, the so-obtained value of $C$ gives a bound for them. For instance, the maximum possible value of the inclination of the $k$th planet is given by
   \begin{equation}
      \cos\Big[\max I_k\Big] = \max \left[ 1 - \frac{C_p}{\Lambda_k\sqrt{1-e_k^2}}\ ,\ -1 \right] \,.
   \end{equation}
   In our case, we are trying to maximise the quantity
   \begin{equation}
      \max\Big[ S_k^{(j)} \Big] = \left|Q_{jk}\right|\sum_{i=1}^{N}\sin\frac{I_i}{2}\left|Q^{-1}_{ki}\right| \,,
   \end{equation}
   obtained from~\eqref{eq:ES} with unknown $\Omega_i$, using the constraint~\eqref{eq:equip}. This constraint can be rewritten
   \begin{equation}\label{eq:Xparab}
      Z = \sum_{i=1}^N c_i\,\eta_i^2 \,,
   \end{equation}
   where $\eta_i=\sin(I_i/2)$, $c_i = 2\Lambda_i\sqrt{1-e_i^2}$ and $Z = C_p$, whereas the quantity to be maximised can be written
   \begin{equation}\label{eq:Yplane}
      Y = \sum_{i=1}^N b_i\,\eta_i \,,
   \end{equation}
   where $b_i = \left|Q^{-1}_{ki}\right|$. The coefficients $c_i$ and $b_i$ are all positive, and $0\leqslant \eta_i\leqslant 1$. The constraint~\eqref{eq:Xparab} forms an hyper-ellipsoid, whereas the quantity to be maximised~\eqref{eq:Yplane} forms an hyperplane. Except from particular cases that we will dismiss here, there is thus only one solution for the maximisation of $Y$, which corresponds to the tangency of the plane and the ellipsoid. This implies that the two gradients are collinear:
   \begin{equation}\label{eq:xisol}
      \nabla Z = \lambda\nabla Y
      \iff
      \eta_i = \lambda\frac{b_i}{2 c_i} \ \ \forall\ i=1,..,N \,,
   \end{equation}
   where $\lambda > 0$ by definition of $\eta_i$. We get the value of $\lambda$ from the imposed value of $Z$:
   \begin{equation}
      Z = \sum_{i=1}^N \lambda^2\frac{b_i^2}{4 c_i}
      \iff
      \lambda^2 = \frac{Z}{\sum_{i=1}^N\frac{b_i^2}{4 c_i}} \,.
   \end{equation}
   The maximum of $Y$ with the constraint $Z$ is thus
   \begin{equation}
      \max\big[Y\big] = \sqrt{Z\sum_{i=1}^N\frac{b_i^2}{c_i}} \,.
   \end{equation}
   Going back to the original notations, this finally gives
   \begin{equation}\label{eq:maxS}
      \max\Big[ S_k^{(j)} \Big] = \left|Q_{jk}\right|\sqrt{C_p\sum_{i=1}^N\frac{(Q^{-1}_{ki})^2}{2\Lambda_i\sqrt{1-e_i^2}}} \,.
   \end{equation}
   However, Eq.~\eqref{eq:xisol} does not take into account the condition that all the $\eta_i$ are smaller than $1$. In practice, if the value obtained for $\lambda$ implies that one or several $\eta_i$ are larger than~$1$, we just have to fix them to~$1$ and use the same resolution method iteratively\footnote{From the form of the constraint~\eqref{eq:Xparab}, decreasing one $\eta_i$ to $1$ implies that at least one of the remaining $\eta_i$ should increase; from the solution~\eqref{eq:xisol}, this actually means that \emph{all} the remaining $\eta_i$ increase.} with the remaining $\eta_i$ (changing the definition of $Z$ and $Y$ accordingly).
   
\begin{figure}
      \includegraphics[width=0.48\textwidth]{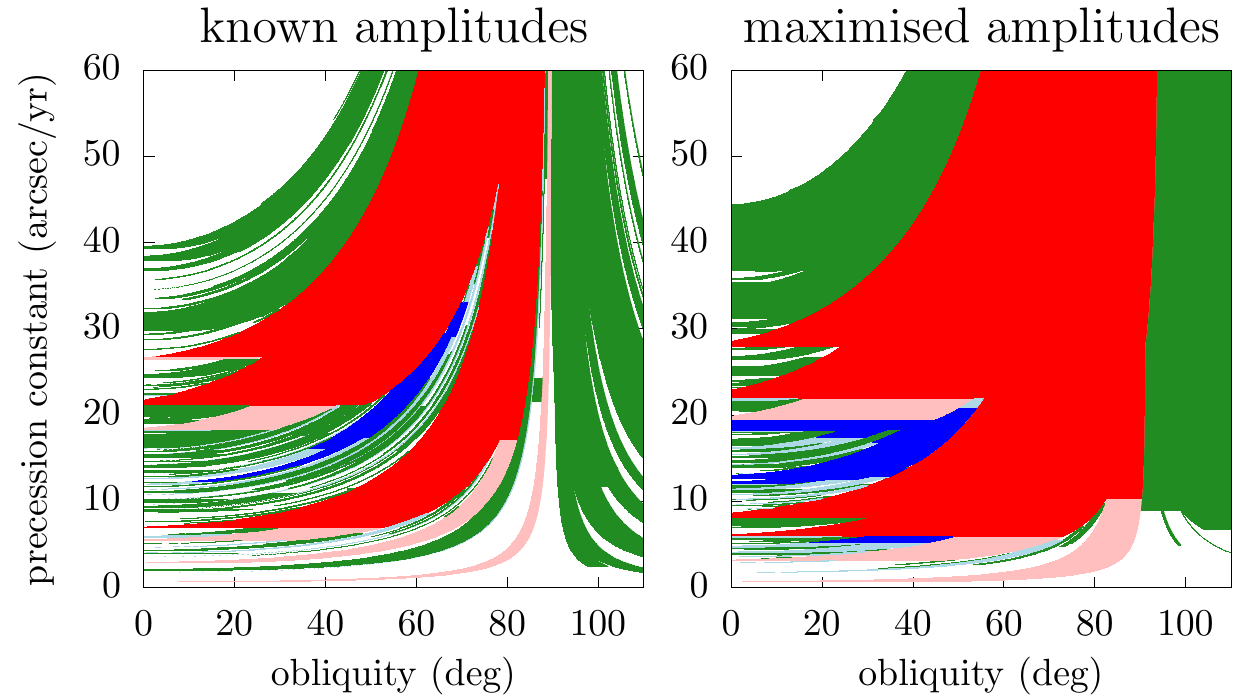}
      \caption{Estimate of the chaotic regions of the spin-axis dynamics for the Earth, where the long-term orbital dynamics is approximated by the Lagrange-Laplace system. The same colour code as Fig.~\ref{fig:ovlSR} is used. On the left, the complete set of initial conditions is used (same as Fig.~\ref{fig:ovlSR}, bottom row). On the right, the orbital elements $(I,\varpi,\Omega)$ are supposed unknown for all planets while the remaining ones are taken from \cite{BRETAGNON_1982}. Accordingly, the coefficients $(E_k,S_k)$ of the quasi-periodic series are maximised according to the estimated AMD value (see Sect.~\ref{ssec:maxES}).}
      \label{fig:ovlLL}
   \end{figure}

   Table~\ref{tab:maxES} shows the comparison of the amplitudes obtained for the Earth with the full Lagrange-Laplace system, and their maximisation supposing that the mutual inclinations and longitudes of node are unknown. As shown in Fig.~\ref{fig:ovlLL}, a chaotic map can be obtained using these maximum values. However, we note that each maximisation is specific to one single amplitude since it implies a distinct set of inclination values. Taking all the maximum amplitudes at once as if they formed one single representation gives thus a large upper bound for the chaotic zones. Moreover, due to the large amplitudes of the series obtained, the small-width approximation for second and third-order resonances does not necessarily hold.
   
   \begin{table}
      \caption{Secular representation of the Earth orbital dynamics given by the Lagrange-Laplace theory.}
      \label{tab:maxES}
      \vspace{-0.7cm}
      \begin{equation*}
         \small
         \begin{array}{rrr}
         \hline\hline
         \mu_i    & E_i\          & \max [E_i]    \\
         (''/\text{yr}) & (\times 10^5) & (\times 10^5) \\
         \hline
          3.7137 & 1628 & 1974 \\
         18.0043 & 1492 & 1917 \\
          7.3460 & 1490 & 3286 \\
         17.3308 & 1057 & 2381 \\
          5.4615 &  404 &  600 \\
         22.2944 &  247 &  385 \\
          2.7015 &   61 &  194 \\
          0.6333 &    1 &    2 \\
         \hline
         \end{array}
         \hfill
         \begin{array}{rrr}
         \hline\hline
         \nu_i     & S_i\          & \max [S_i]    \\
         (''/\text{yr})  & (\times 10^5) & (\times 10^5) \\
         \hline
           0.0000 & 1377 & 2420 \\
         -18.7456 & 1222 & 1989 \\
          -6.5701 &  409 & 3424 \\
          -5.2008 &  425 & 1606 \\
         -17.6358 &  226 &  929 \\
         -25.7514 &  141 &  733 \\
          -2.9039 &   87 &  993 \\
          -0.6778 &   65 &  895 \\
         \hline
         \end{array}
      \end{equation*}
      \vspace{-0.3cm}
      \tablefoot{The eight planets of the Solar System are included. In the third column, each amplitude is maximised in the case where both the mutual inclinations and the longitudes are unknown, assuming the equipartition of AMD between secular degrees of freedom \citep[since the Solar System is hierarchically AMD stable, the AMD of the inner and outer parts were taken separately, see][]{LASKAR-PETIT_2017}. The initial conditions and physical parameters are taken from \cite{BRETAGNON_1982}.}
   \end{table}
   
\begin{figure*}
      \includegraphics[width=\textwidth]{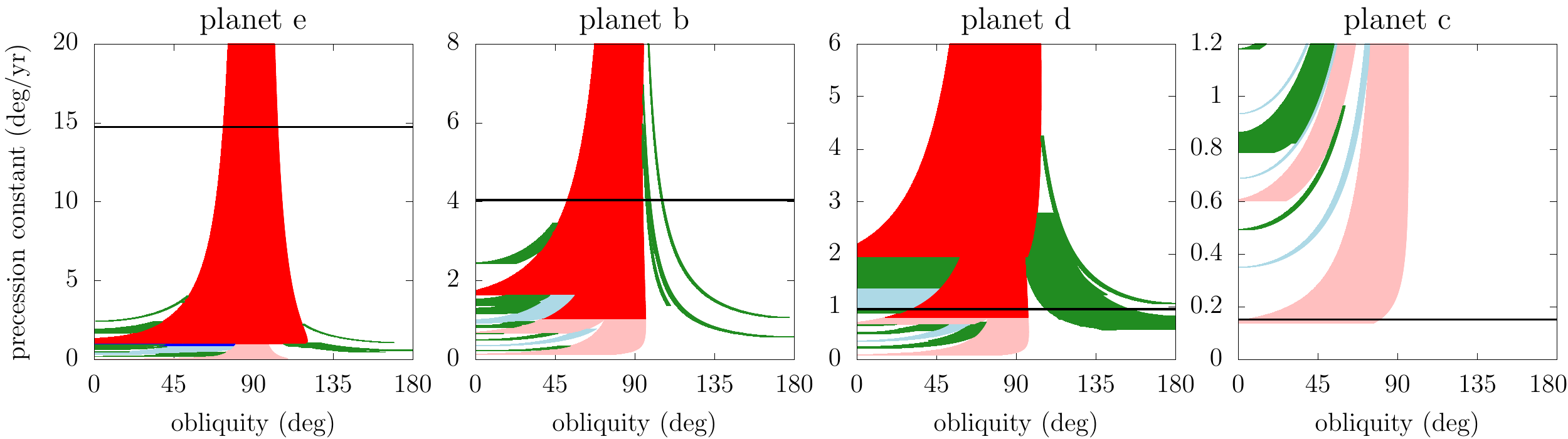}
      \caption{Chaotic regions of the spin-axis dynamics for exoplanets of the GJ\,3293 system, maximised with the method presented in this paper. The colour code is the same as previous figures. The bounds for their precession constants are, from left to right: $\alpha_\text{max}=66$, $20$, $6.2$ and $1.2$~deg/yr (corresponding to maximum rotation periods of a few hours). For each exoplanet, the horizontal line shows the precession constant corresponding to a rotation period equal to the orbital period (obtained from the method of Sect.~\ref{ssec:amax}). They are, from left to right: $13$, $31$, $48$ and $123$~days.}
      \label{fig:GJ3293}
   \end{figure*}

   For simple systems as those studied by \cite{BRASSER-etal_2014}, \cite{DEITRICK-etal_2018} or \cite{SHAN-LI_2018}, the resonances are thin and well separated. A picture of the resonant regions (which do not overlap, in these cases) is thus enough to give clear view of the dynamics. Using the maximised amplitudes gives the largest possible widths of the resonances, which, in turns, show the maximum obliquity variations and their locations. We fully retrieve their results. In contrast to these simple and ordered dynamics, Fig.~\ref{fig:GJ3293} shows as an illustration the maximised chaotic regions for the GJ\,3293 system. The available orbital elements are taken from \cite{ASTUDILLO-DEFRU-etal_2017}, who pointed out that GJ\,3293\,d is in the habitable zone. In the spin-down process toward synchronous rotation due to tidal dissipative effects from the star (thus decreasing the precession constant), planet d is the most likely to suffer from large obliquity changes. One must remember, though, that the chaotic zones are here maximised according to the available orbital data. We also predict a rich obliquity dynamics for the Trappist-1 planets, but due to the confirmed strong effects of mean-motion resonances \citep[see e.g.][]{QUARLES-etal_2017}, the use of the Lagrange-Laplace model is probably inadequate in this case. Building an orbital theory specific to this system would be out the scope of this paper.
   
   
   \subsection{Maximisation of $\alpha$}\label{ssec:amax}
   In Sects.~\ref{ssec:LL} and \ref{ssec:maxES}, we saw how to obtain a quasi-periodic approximation of the long-term orbital motion of a planet, and how to estimate bounds for its coefficients if all the orbital parameters are not known. However, in order to study its long-term spin dynamics, we still lack an estimate of its precession constant $\alpha$. Even in the Solar System, the precession constants of the planets are not very well known. The estimate of $\alpha$ for an extrasolar planet would require observations that are very hard to obtain (its rotation period and a model of interior), and which would be specific to one exoplanet. In order to keep the study as general as possible, we will not try here to obtain a single value for the precession constant: instead, we will look for an upper bound of it from general physical considerations.
   
   After the sphere, the simplest shape model for a rotating planet is given by the Maclaurin ellipsoid \citep{CHANDRASEKHAR_1969}. It describes the equilibrium shape of a self-gravitating homogeneous body in rotation with constant angular velocity. The rotational symmetry is imposed (circular equator), leading to the formula
   \begin{equation}
      \frac{\omega^2}{2\pi\mathcal{G}\rho} = \frac{\sqrt{1-\epsilon^2}}{\epsilon^3}\Big( (3-2\epsilon^2)\arcsin\epsilon - 3\epsilon\sqrt{1-\epsilon^2} \Big) \,,
   \end{equation}
   where $\omega$ and $\rho$ are the rotation velocity and the density of the body and $\epsilon$ is the eccentricity of its ellipsoidal figure (in any plane containing the rotation axis). Studying $f=\omega^2/(2\pi\mathcal{G}\rho)$ as a function of the ellipsoid eccentricity, $f$ is zero for $\epsilon=0$ and $\epsilon=1$, and it has one maximum at $\epsilon_0\approx 0.929956$ with value $f_0\approx 0.224666$. This implies that there is no such equilibrium figure possible for rotation velocities larger than
   \begin{equation}\label{eq:wmax}
      \omega_\text{max} = \sqrt{2\pi\mathcal{G}\rho f_0} \,.
   \end{equation}
   Converting the ellipsoid eccentricity in terms of momenta of inertia, we obtain the relation
   \begin{equation}
      \frac{2C-A-B}{2C} = \frac{1}{2}\epsilon^2 \,.
   \end{equation}
   Injecting it into the expression of the precession constant~\eqref{eq:alpha}, we obtain the maximum value
   \begin{equation}\label{eq:amax}
      \alpha_\text{max} = \frac{3\mathcal{G}m_0}{4a^3}\frac{\epsilon_0^2}{\sqrt{2\pi\mathcal{G}\rho f_0}} \,.
   \end{equation}
   For rotation velocities close to $\omega_\text{max}$, it is known that there exist equilibrium ellipsoidal figures with three unequal axes that have a lower total energy, called Jacobi ellipsoids \citep{CHANDRASEKHAR_1969}. However, we only need an order of magnitude for $\alpha_\text{max}$ and the homogeneous approximation is anyway quite crude, allowing us to stick to the expression~\eqref{eq:amax}. Planets are expected to spin much more slowly than $\omega_\text{max}$ (Eq.~\ref{eq:wmax}), including giant gaseous planets \citep{BATYGIN_2018}. In the remaining part of the article, we allow us to abusively refer to the ``rotational breakup'' velocity.
   
   Using the average density of the Earth, we obtain a minimum rotation period of about $2.4$~hours, leading to a maximum precession constant of about $230~''$/yr. The true value for the Earth is $20~''$/yr, or $50~''$/yr if we include the additional effects of the Moon \citep{LASKAR-ROBUTEL_1993}. This remains well below our bound, but the difference with this ``effective'' precession constant due to the presence of satellites actually constitutes the largest source of uncertainty. Close satellites increase the effective flattening of the planet, whereas far satellites increase the effective torque from the star \citep{BOUE-LASKAR_2006}. In both cases, this increases the precession constant to be used in our model. This effect is particularly problematic for Saturn, because our upper bound gives $\alpha_\text{max}\approx 0.75~''$/yr, while the true value is $0.20~''$/yr, but it increases to $0.83~''$/yr if we take the satellites into account \citep{WARD-HAMILTON_2004}. Hence, the introduction of Saturn's satellites makes the precession constant exceed our upper bound. This problem is unavoidable for exoplanets, because the observation of satellite systems is hard and none has been observed so far. When using our model to study the spin dynamics of exoplanets, we must thus always keep in mind that the presence of numerous or massive satellites could modify our conclusions for borderline cases like Saturn.
   
\begin{figure}
      \includegraphics[width=0.48\textwidth]{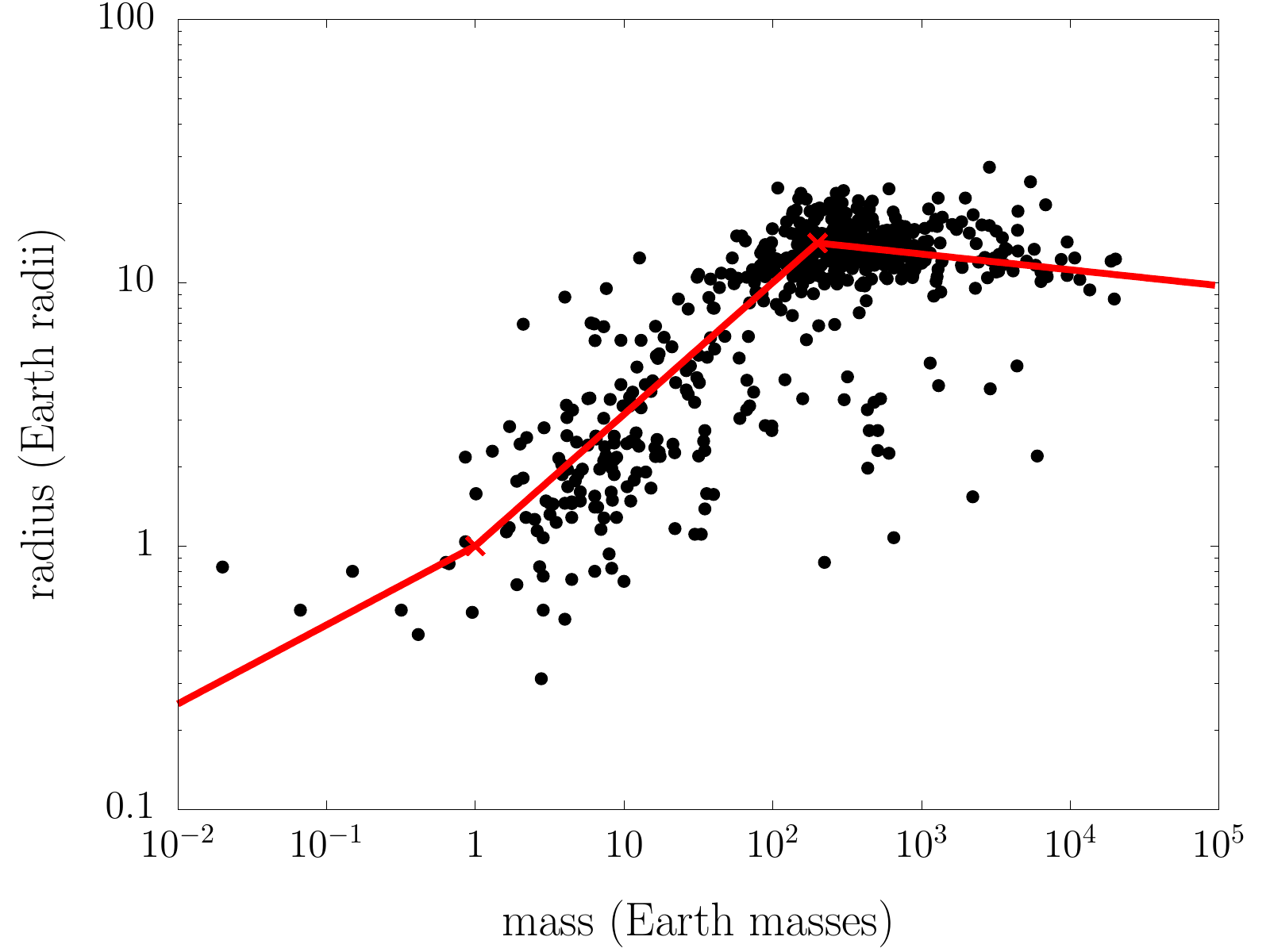}
      \caption{Empirical mass-radius relationship obtained from the exoplanets with known mass and radius (\texttt{http://exoplanet.eu}). The exoplanets are supposed to be rocky up to $1$ Earth mass, gaseous beyond $200$ Earth masses, and of intermediate composition in between. Three power laws are used: $1/3$, $1/2$ and $-0.06$ from left to right, in general accordance with e.g.~\cite{SEAGER-etal_2007} and \cite{WEISS-etal_2013}.}
      \label{fig:MR}
   \end{figure}

   Moreover, we must assume a density $\rho$ for the exoplanets if their radius has not been measured, which adds even more uncertainty. If the radius is unknown, an order of magnitude of the density can be estimated through an empirical law adjusted to the observed mass-radius distribution (see Fig.~\ref{fig:MR}). The density is anyway not the major source of uncertainty of our method.
   
   \subsection{Classification of non-resonant exoplanets}\label{ssec:nres}
   In Sect.~\ref{sec:ovl}, we saw that the overlap of first-order secular resonances, leading to the largest chaotic zones of the spin dynamics, can be produced only if the ratio $\nu_i/\alpha$ is smaller than $1$ at least for two frequencies $\nu_i$ of the inclination quasi-periodic representation. Assuming that $\alpha$ is bounded by $\alpha_\text{max}$, we can deduce that there can be no substantial chaotic zone if $\nu_i/\alpha_\text{max} > 1$ whatever the frequency $\nu_i$. Moreover, if the mutual inclinations are small (as we assume they are), this implies that the obliquity is almost constant. Our bound for $\alpha$ (Eq.~\ref{eq:amax}) can thus be used for a preliminary classification of the exoplanets, while the bounds for the amplitudes (Eqs.~\ref{eq:maxE},\ref{eq:maxS}) are required for a more specific application to one exoplanet (they allow to constrain both $\gamma$ and $\beta$, see Fig.~\ref{fig:zones}).
   
   Table~\ref{tab:class} shows the 94 planets classified strictly non-resonant with this criterion, using the Lagrange-Laplace matrix to estimate the frequencies (Sect.~\ref{ssec:LL}) and Eq.~\eqref{eq:amax} as a bound of $\alpha$. All the exoplanets from \texttt{http://exoplanet.eu} with known mass, semi-major axis and eccentricity were analysed (taking $m\sin I$ instead of the mass if a real-mass estimate was unavailable). At date 2018-03-07, this represents 143 systems with more than one planet, which contain 353 planets in total (plus the Solar System). For some of them, the frequency ratios are so far from $1$ that their classification is quite safe, even when considering the numerous sources of error inherent to our method, and in particular, the possible presence of satellites. This mostly concerns planets that are far from their star, like Uranus and Neptune, and no terrestrial exoplanet has been observed yet is in this category. Such large semi-major axes (third column of Table~\ref{tab:class}) imply that most of the planets listed in Table~\ref{tab:class} are also unaffected by orbital and rotational tidal dissipation resulting from the interaction with their central star.
   
   Most of the exoplanetary systems known so far contain only two planets. In this case, there can be no chaotic region anyway coming from the overlap of first-order resonances (Sect.~\ref{sec:ovl}) because there is only one forced frequency in inclination (Sect.~\ref{ssec:LL}). However, this single term allows to go one step further and compute the variation range of the obliquity at first order. This is obtained by looking at the interval of parameters $\gamma$ and $\beta$ (Fig.~\ref{fig:zones}) allowed for the exoplanet. A very simple formula can be derived if the inclination amplitude $S_1$ (which is the only amplitude of the decomposition) is small. Indeed, this implies that $\beta$ is small as well (Eq.~\ref{eq:bgsimp}), resulting in quite flat level curves for Colombo's top Hamiltonian~\eqref{eq:Hres}. The maximum obliquity variations are achieved around $\Sigma=0$, and at leading order, they are equal to
   \begin{equation}\label{eq:excur}
      \Delta X \approx \frac{2\beta}{\gamma} \approx 4S_1 \,.
   \end{equation}
   This approximation holds very well for small values of $S_1$. For the exoplanet WASP-81\,c, which has the lowest bound for $S_1$ in Table~\ref{tab:class}, we obtain a maximum obliquity excursion of $0.5^\text{o}$. This limit is very close to what is obtained by plotting the level curves of the Hamiltonian~\eqref{eq:Hres}, and it holds as long as $\alpha$ and $S_1$ are below their estimated bounds. Such a good constraint cannot be achieved for every planet in two-planet systems, though, since our bound on $S_1$ from the AMD is sometimes not very informative. It is even dramatic for planets perturbed by a very massive companion: for example the maximum inclination amplitude of HD\,92788\,c is higher than $1$, indicating that all inclinations are possible.
   
   The maximum excursion of the obliquity is harder to obtain if there are more than two planets in the system, since the dynamics is ruled by the superposition of several forcing terms. However, if the maximum maximised amplitude is small (last column of Table~\ref{tab:class}), the superposition of all the terms is unlikely to bring the obliquity over the limit given at Eq.~\eqref{eq:excur}. It can thus be used as well as an order-of-magnitude estimate. This results in a maximum of about $20^\text{o}$ for Uranus and Neptune, showing the crude nature of our maximisation (as shown by previous works, it is very hard to tilt Uranus and Neptune by the mean of planetary perturbations, see e.g. \citealp{BOUE-LASKAR_2010}).
   
   \begin{table*}
      \caption{Exoplanets from \texttt{http://exoplanet.eu} at date 2018-03-07 classified non-resonant with the method detailed in this study.}
      \vspace{-0.7cm}
      \label{tab:class}
      \begin{equation*}
         \tiny
         \begin{array}{lcrrrr}
         \hline\hline
         \text{Name} & j/N &          a & \alpha_\text{max} & \frac{\min|\nu_i|}{\alpha_\text{max}} &     \max[S_i] \\
                     &     & (\text{au})&    (''/\text{yr}) &                                       & (\times 10^4) \\
         \hline
             \text{HD\,113538\,b} & 1/2 & 1.24 &  233.0382 & 1.0 &  1815 \\
             \text{HD\,134987\,c} & 2/2 & 5.80 &    4.0960 & 1.2 &   763 \\
             \text{HD\,163607\,c} & 2/2 & 2.42 &   31.3391 & 1.2 &   626 \\
              \text{HD\,89744\,c} & 1/2 & 0.44 & 5497.1869 & 1.2 &  6954 \\
                \text{WASP-53\,c} & 2/2 & 3.73 &    2.0657 & 1.3 &   221 \\
              \text{HD\,65216\,b} & 2/2 & 1.30 &  242.7438 & 1.4 &   596 \\
              \text{HD\,85390\,c} & 2/2 & 4.23 &    6.5843 & 1.4 &   726 \\
             \text{HD\,183263\,b} & 1/2 & 1.51 &  104.8361 & 1.5 &  2031 \\
             \text{HD\,204313\,b} & 2/3 & 3.17 &    9.2689 & 1.5 &   588 \\
              \text{HD\,27894\,d} & 3/3 & 5.45 &    1.2133 & 1.6 &   268 \\
             \text{HD\,204313\,d} & 3/3 & 3.93 &    8.4220 & 1.7 &  1343 \\
              \text{HD\,37605\,c} & 2/2 & 3.81 &    5.8515 & 1.8 &   850 \\
              \text{HD\,45364\,b} & 1/2 & 0.68 & 1672.3835 & 2.0 &  1165 \\
               \text{HD\,7449\,b} & 1/2 & 2.30 &   53.9106 & 2.1 &  6921 \\
               \text{GJ\,676A\,b} & 3/4 & 1.81 &   25.8112 & 2.2 &  1731 \\
         \text{TYC+1422-614-1\,b} & 1/2 & 0.69 & 1363.7143 & 2.7 &   630 \\
             \text{HD\,155358\,c} & 2/2 & 1.02 &  647.5095 & 2.8 &   754 \\
              \text{HD\,34445\,g} & 6/6 & 6.36 &    3.2020 & 2.8 &   162 \\
              \text{HD\,47366\,b} & 1/2 & 1.21 &  483.1022 & 2.8 &  1270 \\
              \text{HD\,73526\,b} & 1/2 & 0.65 & 1619.2167 & 2.8 &  1647 \\
              \text{HD\,37124\,d} & 3/3 & 2.81 &   30.8760 & 2.9 &   585 \\
                \text{55\,Cnc\,d} & 5/5 & 5.45 &    1.6778 & 3.0 &    34 \\
                \text{24\,Sex\,b} & 1/2 & 1.33 &  287.8167 & 3.0 &   702 \\
            \text{HD\,133131A\,c} & 2/2 & 4.36 &    9.3548 & 3.4 &  2688 \\
              \text{HD\,45364\,c} & 2/2 & 0.90 &  965.6966 & 3.5 &   289 \\
             \text{HD\,102272\,c} & 2/2 & 1.57 &  185.6220 & 3.9 &  3092 \\
             \text{HD\,108874\,c} & 2/2 & 2.68 &   34.1534 & 3.9 &   885 \\
             \text{HD\,147873\,c} & 2/2 & 1.36 &  222.9728 & 3.9 &  1300 \\
             \text{HIP\,67851\,c} & 2/2 & 3.82 &    6.7632 & 4.0 &   234 \\
             \text{HD\,125612\,d} & 3/3 & 4.20 &    3.0777 & 4.1 &   844 \\
             \text{HD\,147018\,c} & 2/2 & 1.92 &   28.5927 & 4.2 &   357 \\
              \text{HD\,12661\,c} & 2/2 & 2.56 &   32.4707 & 4.4 &  1213 \\
              \text{HD\,74156\,c} & 2/2 & 3.82 &    4.3343 & 4.7 &   581 \\
              \text{HD\,11506\,b} & 2/2 & 2.43 &   26.5808 & 5.0 &   462 \\
             \text{HD\,141399\,e} & 4/4 & 5.00 &    7.2669 & 5.0 &  1144 \\
               \text{HD\,4732\,c} & 2/2 & 4.60 &    7.1381 & 5.4 &   730 \\
              \text{HD\,38529\,c} & 2/2 & 3.70 &    3.0111 & 5.7 &    86 \\
                \text{WASP-81\,c} & 2/2 & 2.43 &    4.6451 & 5.9 &    22 \\
             \text{HD\,154857\,c} & 2/2 & 5.36 &    4.2372 & 6.6 &   898 \\
                \text{nu\,Oph\,b} & 1/2 & 1.90 &   45.1541 & 6.7 &  1183 \\
                \text{24\,Sex\,c} & 2/2 & 2.08 &  124.2755 & 6.9 &  1300 \\
             \text{Kepler-419\,c} & 2/2 & 1.68 &   52.9030 & 7.0 &  1024 \\
               \text{eta\,Cet\,b} & 1/2 & 1.27 &  317.3862 & 7.1 &   716 \\
             \text{HD\,159868\,b} & 2/2 & 2.25 &   41.0373 & 7.4 &   159 \\
              \text{HD\,82943\,c} & 1/3 & 0.75 &  391.5783 & 7.9 &  1902 \\
                \text{mu\,Ara\,e} & 4/4 & 5.24 &    3.5196 & 8.2 &   405 \\
             \text{HD\,169830\,c} & 2/2 & 3.60 &    8.7471 & 8.4 &   984 \\
         \hline
         \end{array}
         \hfill
         \begin{array}{lcrrrr}
         \hline\hline
         \text{Name} & j/N &           a & \alpha_\text{max} & \frac{\min|\nu_i|}{\alpha_\text{max}} &     \max[S_i] \\
                     &     & (\text{au}) &    (''/\text{yr}) &                                       & (\times 10^4) \\
         \hline
              \text{HD\,92788\,c} & 1/2 & 0.60 & 3698.5423 &    8.5 & 10293 \\
             \text{HD\,113538\,c} & 2/2 & 2.44 &   27.9248 &    8.6 &   501 \\
              \text{HD\,33844\,b} & 1/2 & 1.60 &  190.8341 &    8.8 &   725 \\
                  \textbf{Uranus} & \mathbf{7/8} & 19.2 & \mathbf{0.0682} & \mathbf{9.9} & \mathbf{1009} \\
              \text{HD\,47366\,c} & 2/2 & 1.85 &  131.0535 &   10.5 &   967 \\
              \text{HD\,60532\,b} & 1/2 & 0.77 &  565.4395 &   11.1 &  1291 \\
              \text{HD\,73526\,c} & 2/2 & 1.03 &  406.9428 &   11.1 &  1309 \\
             \text{HD\,128311\,b} & 1/2 & 1.10 &  265.4641 &   12.2 &  1563 \\
             \text{HD\,110014\,b} & 2/2 & 2.31 &   28.8866 &   13.0 &   604 \\
              \text{HD\,75784\,c} & 2/2 & 6.50 &    1.2344 &   16.4 &   515 \\
              \text{HD\,89744\,b} & 2/2 & 0.88 &  375.7932 &   17.2 &  1771 \\
              \text{HD\,67087\,c} & 2/2 & 3.86 &    6.1887 &   18.9 &  2256 \\
                \text{HD\,142\,c} & 2/2 & 6.80 &    0.8689 &   20.2 &   317 \\
               \text{HD\,1605\,c} & 2/2 & 3.52 &    9.5614 &   20.7 &   202 \\
              \text{HD\,33844\,c} & 2/2 & 2.24 &   74.3551 &   22.5 &   687 \\
                \text{47\,UMa\,d} & 3/3 & 11.6 &    0.3274 &   23.4 &   569 \\
                \text{GJ\,317\,c} & 2/2 & 30.0 &    0.0069 &   23.9 &  2182 \\
               \text{ups\,And\,d} & 3/4 & 2.55 &    7.8848 &   23.7 &   768 \\
             \text{HD\,200964\,b} & 1/2 & 1.60 &  162.1684 &   28.6 &   412 \\
               \text{eta\,Cet\,c} & 2/2 & 1.93 &   77.3953 &   28.9 &   446 \\
               \text{HD\,7449\,c} & 2/2 & 4.96 &    3.7979 &   29.7 &  2617 \\
             \text{HD\,87646A\,c} & 2/2 & 1.58 &   17.3656 &   31.5 &   619 \\
              \text{HD\,82943\,b} & 2/3 & 1.19 &   96.0380 &   32.1 &  1498 \\
               \text{ups\,And\,e} & 4/4 & 5.25 &    5.6512 &   33.1 &  6502 \\
             \text{HD\,200964\,c} & 2/2 & 1.95 &  137.7501 &   33.7 &   771 \\
             \text{HD\,183263\,c} & 2/2 & 4.25 &    4.5921 &   35.0 &  1163 \\
              \text{HD\,82943\,d} & 3/3 & 2.15 &   86.0321 &   35.8 & 10947 \\
                 \textbf{Neptune} & \mathbf{8/8} & 30.1 & \mathbf{0.0156} & \mathbf{43.4} & \mathbf{810} \\
               \text{HD\,5319\,b} & 1/2 & 1.75 &  130.8026 &   47.8 &   536 \\
             \text{HD\,202206\,B} & 1/2 & 0.83 &   90.1987 &   49.0 &  1183 \\
         \text{TYC+1422-614-1\,c} & 2/2 & 1.39 &   72.4459 &   51.0 &   110 \\
             \text{HD\,168443\,c} & 2/2 & 2.84 &    3.9323 &   53.8 &   401 \\
               \text{HD\,5319\,c} & 2/2 & 2.07 &  107.4439 &   58.2 &   831 \\
              \text{HD\,30177\,b} & 1/2 & 3.58 &    4.4350 &   63.2 &   941 \\
             \text{HD\,128311\,c} & 2/2 & 1.76 &   43.9586 &   73.5 &   643 \\
             \text{GJ\,676\,A\,c} & 4/4 & 6.60 &    0.5295 &  107.0 &   901 \\
              \text{HIP\,5158\,c} & 2/2 & 7.70 &    0.2293 &  134.4 &   168 \\
              \text{BD+202457\,b} & 1/2 & 1.45 &  100.0644 &  148.1 &   682 \\
          \text{NN\,Ser\,(AB)\,d} & 1/2 & 3.39 &    6.7742 &  155.6 &   902 \\
              \text{HD\,60532\,c} & 2/2 & 1.58 &   39.3544 &  159.1 &   382 \\
                \text{nu\,Oph\,c} & 2/2 & 6.10 &    1.2729 &  237.0 &   587 \\
              \text{HD\,30177\,c} & 2/2 & 6.99 &    1.0682 &  262.5 &  1808 \\
              \text{HD\,92788\,b} & 2/2 & 0.97 &  115.8082 &  272.2 &   266 \\
              \text{BD+202457\,c} & 2/2 & 2.01 &   51.6913 &  286.7 &   994 \\
             \text{HIP\,57050\,c} & 2/2 & 0.91 &   24.6886 &  407.6 &    24 \\
             \text{HD\,202206\,c} & 2/2 & 2.41 &    9.7781 &  452.1 &  3521 \\
          \text{NN\,Ser\,(AB)\,c} & 2/2 & 5.38 &    0.8811 & 1196.6 &   237 \\
         \hline
         \end{array}
      \end{equation*}
      \vspace{-0.3cm}
      \tablefoot{The first column gives the name of the exoplanet; the second column gives the rank of the exoplanet (sorted by increasing semi-major axis) and the total number of planets in the system; the third column gives the semi-major axis value; the fourth column gives the maximum value of the precession constant estimated from Eq.~\eqref{eq:amax}; the fifth column gives the minimum ratio of the eigenfrequencies of the Lagrange-Laplace system and $\alpha_\text{max}$ in absolute value ($\gamma$ parameter of Colombo's top); the sixth column gives the maximum amplitude of the series decomposition obtained from the maximisation~\eqref{eq:maxS}, allowing to obtain a maximum bound for the $\beta$ parameter of Colombo's top.}
   \end{table*}
   
   \section{Conclusion}
   The spin-axis dynamics of a planet plays a major role in its climate setting, and, by extension, in its suitability for life. However, the rotation properties of exoplanets are still very poorly constrained. In this paper, we presented an analytical formulation of the long-term spin-axis dynamics of a planet, allowing to link known and unknown parameters to its obliquity evolution and to provide a global picture of the dynamics in a straightforward way.
   
   At first, the orbital solution is modelled by quasi-periodic series. This method is thus valid as long as the orbital chaos, if any, takes place on a much larger timescale than the spin-axis evolution. The spin-axis Hamiltonian is then expanded in powers of the eccentricity and inclination amplitudes of the orbital series. We provided all terms up to order~3 but the development can be conducted to higher orders.
   
   A clear picture of the phase space structure is given by the obliquity ranges associated to the various resonant regions. The resonant dynamics at order~1 can be characterised analytically in terms of two parameters, which are linked to the precession constant $\alpha$ (gathering the physical characteristics of the planet under study) and to the quasi-periodic representation of the orbit. The pendulum approximation is only used at order~2 and beyond, for which the resonances are thin enough. The regions of resonance overlap at all orders are identified as chaotic. In some cases (as for the terrestrial planets of the Solar System), these chaotic regions allow wide excursions of the obliquity. The method presented here allows to retrieve analytically the previous numerical results with a good precision. Numerical integrations prove thus to be necessary only if detailed statistics on the obliquity evolution are required. This is very informative for Solar System planets \citep[as shown by][]{NERONDESURGY-LASKAR_1997,CORREIA-LASKAR_2003,LASKAR-etal_2004b} but not yet for exoplanets because their initial conditions and physical parameters are still poorly known. Hence, the uncertainty of our results remains largely dominated by our lack of knowledge of the exoplanetary systems rather than by the approximations inherent to our method. This allows to stick to the simple analytical formulas presented here.
   
   At this level of uncertainty, the Lagrange-Laplace system provides a good-enough representation of the orbital motions (excepted for exoplanetary systems featuring highly excited orbits or strong effects of mean-motion resonances). The formulas obtained allow to set an upper bound for the amplitude of the eccentricity and inclination terms if the mutual orientations of the orbits are unknown. On the other hand, the AMD equipartition hypothesis can be used, if required, to place a bound on the inclination from the eccentricity values. Through our analytical model of the spin-axis dynamics, these maximum amplitudes provide the maximum extent of the chaotic zones. For example, a large chaotic region is expected for exoplanet GJ\,3293\,d for rotational velocities above the synchronous rotation. Systems very affected by mean-motion resonances (like Trappist-1) can still be studied using the method described here, but with the prior construction of a synthetic representation for the orbital motion, written in the form of a quasi-periodic series.
   
   However, this method does not allow to consider tidal dissipations (playing an important role for exoplanets close to their star), which could be modelled as an adiabatic process acting on a much longer timescale than the obliquity variations \citep[see][]{NERONDESURGY-LASKAR_1997}. This amounts to make the precession constant $\alpha$ and/or the amplitudes of the orbital series gradually vary. This method does not include either the effects of libration around spin-orbit resonances, even if a trick allows to take into account a possible locking in synchronous rotation (Appendix~\ref{asec:spinorb}).
   
   Finally, under the hypothesis of hydrostatic equilibrium, we can set a bound to the precession constant $\alpha$. This bound is obtained from the flattening of the planet corresponding to its rotational breakup velocity. Since $\alpha$ governs the width and location of the resonances, this allows to classify the exoplanets that cannot be subject to first-order secular spin-orbit resonances. Among the sufficiently known systems with more than one planet, we found $94$ planets in this category ($26\%$ of our sample). If they belong to exoplanetary systems with low mutual inclinations (as it is expected in most cases for orbital stability), this implies that their obliquity is almost constant. This bound for $\alpha$ is though invalidated by the possible presence of massive satellites (as our Moon), but some exoplanets are so far from resonance that their classification is quite safe. This is the case of Uranus and Neptune.
   
   Considering the high efficiency of the analytical method proposed here, an obliquity stability map could be designed easily in the future for each new exoplanet discovered, and in particular for those classified as ``habitable''. However, such a stability map should always be computed again if any additional planet is found in the system. Indeed, it would shift the existing frequencies (especially if the new planet is massive), and add one frequency in both the inclination and eccentricity series, multiplying the possibilities of resonance. On the other hand, the total AMD of the system would increase, resulting in wider maximised chaotic zones.

\begin{acknowledgements}
   We thank the anonymous referee for her or his detailed review.
\end{acknowledgements}

\bibliographystyle{aa}
\bibliography{secularspin}

\begin{thebibliography}{49}
\expandafter\ifx\csname natexlab\endcsname\relax\def\natexlab#1{#1}\fi

\bibitem[{{Armstrong} {et~al.}(2014){Armstrong}, {Barnes}, {Domagal-Goldman},
  {Breiner}, {Quinn}, \& {Meadows}}]{ARMSTRONG-etal_2014}
{Armstrong}, J.~C., {Barnes}, R., {Domagal-Goldman}, S., {et~al.} 2014,
  Astrobiology, 14, 277

\bibitem[{{Astudillo-Defru} {et~al.}(2017){Astudillo-Defru}, {Forveille},
  {Bonfils}, {S{\'e}gransan}, {Bouchy}, {Delfosse}, {Lovis}, {Mayor}, {Murgas},
  {Pepe}, {Santos}, {Udry}, \& {W{\"u}nsche}}]{ASTUDILLO-DEFRU-etal_2017}
{Astudillo-Defru}, N., {Forveille}, T., {Bonfils}, X., {et~al.} 2017, \aap,
  602, A88

\bibitem[{{Atobe} {et~al.}(2004){Atobe}, {Ida}, \& {Ito}}]{ATOBE-etal_2004}
{Atobe}, K., {Ida}, S., \& {Ito}, T. 2004, \icarus, 168, 223

\bibitem[{{Batygin}(2018)}]{BATYGIN_2018}
{Batygin}, K. 2018, \aj, 155, 178

\bibitem[{{Bou{\'e}} \& {Laskar}(2006)}]{BOUE-LASKAR_2006}
{Bou{\'e}}, G. \& {Laskar}, J. 2006, \icarus, 185, 312

\bibitem[{{Bou{\'e}} \& {Laskar}(2010)}]{BOUE-LASKAR_2010}
{Bou{\'e}}, G. \& {Laskar}, J. 2010, \apjl, 712, L44

\bibitem[{{Brasser} {et~al.}(2014){Brasser}, {Ida}, \&
  {Kokubo}}]{BRASSER-etal_2014}
{Brasser}, R., {Ida}, S., \& {Kokubo}, E. 2014, \mnras, 440, 3685

\bibitem[{{Bretagnon}(1982)}]{BRETAGNON_1982}
{Bretagnon}, P. 1982, \aap, 114, 278

\bibitem[{{Canup} \& {Asphaug}(2001)}]{CANUP-ASPHAUG_2001}
{Canup}, R.~M. \& {Asphaug}, E. 2001, \nat, 412, 708

\bibitem[{{Carter} \& {Winn}(2010)}]{CARTER-WINN_2010}
{Carter}, J.~A. \& {Winn}, J.~N. 2010, \apj, 716, 850

\bibitem[{{Chandrasekhar}(1969)}]{CHANDRASEKHAR_1969}
{Chandrasekhar}, S. 1969, {Ellipsoidal figures of equilibrium} (Yale University
  Press)

\bibitem[{{Colombo}(1966)}]{COLOMBO_1966}
{Colombo}, G. 1966, \aj, 71, 891

\bibitem[{{Correia}(2014)}]{CORREIA_2014}
{Correia}, A.~C.~M. 2014, \aap, 570, L5

\bibitem[{{Correia} \& {Laskar}(2003)}]{CORREIA-LASKAR_2003}
{Correia}, A.~C.~M. \& {Laskar}, J. 2003, \icarus, 163, 24

\bibitem[{{Correia} {et~al.}(2003){Correia}, {Laskar}, \& {de
  Surgy}}]{CORREIA-etal_2003}
{Correia}, A.~C.~M., {Laskar}, J., \& {de Surgy}, O.~N. 2003, \icarus, 163, 1

\bibitem[{{Deitrick} {et~al.}(2018){Deitrick}, {Barnes}, {Quinn}, {Armstrong},
  {Charnay}, \& {Wilhelm}}]{DEITRICK-etal_2018}
{Deitrick}, R., {Barnes}, R., {Quinn}, T.~R., {et~al.} 2018, \aj, 155, 60

\bibitem[{{Ger\v{s}gorin}(1931)}]{GERSHGORIN_1931}
{Ger\v{s}gorin}, S. 1931, Bulletin de l'Académie des Sciences de l'URSS,
  Classe des sciences mathématiques et naturelles, 6, 749

\bibitem[{{Hartmann} \& {Davis}(1975)}]{HARTMANN-DAVIS_1975}
{Hartmann}, W.~K. \& {Davis}, D.~R. 1975, \icarus, 24, 504

\bibitem[{{Hays} {et~al.}(1976){Hays}, {Imbrie}, \&
  {Shackleton}}]{HAYS-etal_1976}
{Hays}, J.~D., {Imbrie}, J., \& {Shackleton}, N.~J. 1976, Science, 194, 1121

\bibitem[{{Henrard} \& {Murigande}(1987)}]{HENRARD-MURIGANDE_1987}
{Henrard}, J. \& {Murigande}, C. 1987, Celestial Mechanics, 40, 345

\bibitem[{{Laskar}(1988)}]{LASKAR_1988}
{Laskar}, J. 1988, \aap, 198, 341

\bibitem[{{Laskar}(1990)}]{LASKAR_1990}
{Laskar}, J. 1990, \icarus, 88, 266

\bibitem[{{Laskar}(1994)}]{LASKAR_1994}
{Laskar}, J. 1994, \aap, 287, L9

\bibitem[{{Laskar}(1996)}]{LASKAR_1996}
{Laskar}, J. 1996, Celestial Mechanics and Dynamical Astronomy, 64, 115

\bibitem[{{Laskar}(2008)}]{LASKAR_2008}
{Laskar}, J. 2008, \icarus, 196, 1

\bibitem[{{Laskar} {et~al.}(2012){Laskar}, {Bou{\'e}}, \&
  {Correia}}]{LASKAR-etal_2012}
{Laskar}, J., {Bou{\'e}}, G., \& {Correia}, A.~C.~M. 2012, \aap, 538, A105

\bibitem[{{Laskar} {et~al.}(2004{\natexlab{a}}){Laskar}, {Correia},
  {Gastineau}, {Joutel}, {Levrard}, \& {Robutel}}]{LASKAR-etal_2004b}
{Laskar}, J., {Correia}, A.~C.~M., {Gastineau}, M., {et~al.}
  2004{\natexlab{a}}, \icarus, 170, 343

\bibitem[{{Laskar} {et~al.}(1993{\natexlab{a}}){Laskar}, {Joutel}, \&
  {Boudin}}]{LASKAR-etal_1993b}
{Laskar}, J., {Joutel}, F., \& {Boudin}, F. 1993{\natexlab{a}}, \aap, 270, 522

\bibitem[{{Laskar} {et~al.}(1993{\natexlab{b}}){Laskar}, {Joutel}, \&
  {Robutel}}]{LASKAR-etal_1993}
{Laskar}, J., {Joutel}, F., \& {Robutel}, P. 1993{\natexlab{b}}, \nat, 361, 615

\bibitem[{{Laskar} \& {Petit}(2017)}]{LASKAR-PETIT_2017}
{Laskar}, J. \& {Petit}, A.~C. 2017, \aap, 605, A72

\bibitem[{{Laskar} \& {Robutel}(1993)}]{LASKAR-ROBUTEL_1993}
{Laskar}, J. \& {Robutel}, P. 1993, \nat, 361, 608

\bibitem[{{Laskar} \& {Robutel}(1995)}]{LASKAR-ROBUTEL_1995}
{Laskar}, J. \& {Robutel}, P. 1995, Celestial Mechanics and Dynamical
  Astronomy, 62, 193

\bibitem[{{Laskar} {et~al.}(2004{\natexlab{b}}){Laskar}, {Robutel}, {Joutel},
  {Gastineau}, {Correia}, \& {Levrard}}]{LASKAR-etal_2004a}
{Laskar}, J., {Robutel}, P., {Joutel}, F., {et~al.} 2004{\natexlab{b}}, \aap,
  428, 261

\bibitem[{{Li} \& {Batygin}(2014{\natexlab{a}})}]{LI-BATYGIN_2014a}
{Li}, G. \& {Batygin}, K. 2014{\natexlab{a}}, \apj, 790, 69

\bibitem[{{Li} \& {Batygin}(2014{\natexlab{b}})}]{LI-BATYGIN_2014b}
{Li}, G. \& {Batygin}, K. 2014{\natexlab{b}}, \apj, 795, 67

\bibitem[{{Lissauer} {et~al.}(2012){Lissauer}, {Barnes}, \&
  {Chambers}}]{LISSAUER-etal_2012}
{Lissauer}, J.~J., {Barnes}, J.~W., \& {Chambers}, J.~E. 2012, \icarus, 217, 77

\bibitem[{{Lock} {et~al.}(2018){Lock}, {Stewart}, {Petaev}, {Leinhardt},
  {Mace}, {Jacobsen}, \& {Cuk}}]{LOCK-etal_2018}
{Lock}, S.~J., {Stewart}, S.~T., {Petaev}, M.~I., {et~al.} 2018, Journal of
  Geophysical Research (Planets), 123, 910

\bibitem[{{Murray} \& {Dermott}(1999)}]{MURRAY-DERMOTT_1999}
{Murray}, C.~D. \& {Dermott}, S.~F. 1999, {Solar system dynamics} (Cambridge
  University Press)

\bibitem[{{N{\'e}ron de Surgy} \& {Laskar}(1997)}]{NERONDESURGY-LASKAR_1997}
{N{\'e}ron de Surgy}, O. \& {Laskar}, J. 1997, \aap, 318, 975

\bibitem[{{Peale}(1969)}]{PEALE_1969}
{Peale}, S.~J. 1969, \aj, 74, 483

\bibitem[{{Quarles} {et~al.}(2017){Quarles}, {Quintana}, {Lopez}, {Schlieder},
  \& {Barclay}}]{QUARLES-etal_2017}
{Quarles}, B., {Quintana}, E.~V., {Lopez}, E., {Schlieder}, J.~E., \&
  {Barclay}, T. 2017, \apjl, 842, L5

\bibitem[{{Seager} {et~al.}(2007){Seager}, {Kuchner}, {Hier-Majumder}, \&
  {Militzer}}]{SEAGER-etal_2007}
{Seager}, S., {Kuchner}, M., {Hier-Majumder}, C.~A., \& {Militzer}, B. 2007,
  \apj, 669, 1279

\bibitem[{{Shan} \& {Li}(2018)}]{SHAN-LI_2018}
{Shan}, Y. \& {Li}, G. 2018, \aj, 155, 237

\bibitem[{{Spiegel} {et~al.}(2009){Spiegel}, {Menou}, \&
  {Scharf}}]{SPIEGEL-etal_2009}
{Spiegel}, D.~S., {Menou}, K., \& {Scharf}, C.~A. 2009, \apj, 691, 596

\bibitem[{{Varga}(2004)}]{VARGA_2004}
{Varga}, R.~S. 2004, {Ger\v{s}gorin and His Circles} (Springer-Verlag, Berlin,
  Heidelberg)

\bibitem[{{Ward} \& {Hamilton}(2004)}]{WARD-HAMILTON_2004}
{Ward}, W.~R. \& {Hamilton}, D.~P. 2004, \aj, 128, 2501

\bibitem[{{Weertman}(1976)}]{WEERTMAN_1976}
{Weertman}, J. 1976, \nat, 261, 17

\bibitem[{{Weiss} {et~al.}(2013){Weiss}, {Marcy}, {Rowe}, {Howard}, {Isaacson},
  {Fortney}, {Miller}, {Demory}, {Fischer}, {Adams}, {Dupree}, {Howell},
  {Kolbl}, {Johnson}, {Horch}, {Everett}, {Fabrycky}, \&
  {Seager}}]{WEISS-etal_2013}
{Weiss}, L.~M., {Marcy}, G.~W., {Rowe}, J.~F., {et~al.} 2013, \apj, 768, 14

\bibitem[{{Xie} {et~al.}(2016){Xie}, {Dong}, {Zhu}, {Huber}, {Zheng}, {De Cat},
  {Fu}, {Liu}, {Luo}, {Wu}, {Zhang}, {Zhang}, {Zhou}, {Cao}, {Hou}, {Wang}, \&
  {Zhang}}]{XIE-etal_2016}
{Xie}, J.-W., {Dong}, S., {Zhu}, Z., {et~al.} 2016, Proceedings of the National
  Academy of Science, 113, 11431

\end{thebibliography}

\appendix

   \section{Case of a $1\!:\!1$ spin-orbit resonance}\label{asec:spinorb}
   Numerous exoplanets are observed very close to their star, in a place where the tidal frictions are strong enough to efficiently lock them in synchronous rotation. In this section, we show that if the librations around the synchronous rotation are much faster than the secular spin-axis dynamics, we can retrieve Colombo's top Hamiltonian (Sect.~\ref{sec:coltop}), allowing to use the same approach as in the non-resonant case. As before, though, we will not consider the effect of the tidal dissipation on the obliquity. This is thus only valid for systems for which the tidal damping of the obliquity acts on a larger timescale than the spin-axis dynamics.
   
   We will use the same method as \cite{CORREIA-etal_2003}. Let us write $\lambda$ the mean longitude of the planet in orbit around the star, and $\ell$ its rotation angle. The mean longitude $\lambda$ is measured from the equinox at a reference epoch (for instance J2000), whereas the rotation angle $\ell$ is measured from the equinox of the date up to a fixed point of the equator (principal axis A). If we keep the angles of the form $\ell-\lambda$ during the average over the mean longitude and the fast rotation angles \citep[see][]{NERONDESURGY-LASKAR_1997}, the corresponding ``semi-averaged'' Hamiltonian is
   \begin{equation}
      \begin{aligned}
         &\mathcal{H}(L,\Lambda,Y,\ell,M,-\psi,t) = \frac{L^2}{2C} + n\Lambda - \frac{\alpha}{2}\frac{Y^2}{L\big(1-e(t)^2\big)^{3/2}} \\
         &- \frac{\alpha_r}{2L}(L+Y)^2\cos\big[2(\ell-\lambda-\psi)\big] \\
         &- \sqrt{L^2-Y^2}\big(\mathcal{A}(t)\sin\psi + \mathcal{B}(t)\cos\psi\big) + 2Y\mathcal{C}(t) \,,
      \end{aligned}
   \end{equation}
   where we neglected terms of order $e(B-A)/C$. The momenta $L=C\omega$ and $Y=LX$ are conjugate to $\ell$ and $-\psi$, respectively. The momentum $\Lambda$, conjugate to $\lambda$, has been added such that $\dot{\lambda}=n$ (mean motion). The resonant precession constant is defined as
   \begin{equation}
     \alpha_r = \frac{3\,\mathcal{G}m_0}{8\,\omega a^3}\,\frac{B-A}{C} \,,
   \end{equation}
   using the same notation as Eq.~\eqref{eq:alpha}. We note that the angle $\lambda+\psi$ appearing in the Hamiltonian corresponds to the mean longitude measured from the equinox of the date. Let us use the canonical change of coordinates
   \begin{equation}
      \left\{
      \begin{aligned}
         \theta &= \ell-\lambda \\
         \gamma &= \lambda
      \end{aligned}
      \right.
      \hspace{0.5cm}\text{and}\hspace{0.5cm}
      \left\{
      \begin{aligned}
         I &= L \\
         \Gamma &= L + \Lambda \,.
      \end{aligned}
      \right.
   \end{equation}
    The momentum $\Gamma$ is an arbitrary constant of motion and the Hamiltonian becomes
    \begin{equation}
       \begin{aligned}
          &\mathcal{H}(I,Y,\theta,-\psi,t) = \frac{I^2}{2C} - nI - \frac{\alpha}{2}\frac{Y^2}{I\big(1-e(t)^2\big)^{3/2}} \\
          &- \frac{\alpha_r}{2I}(I+Y)^2\cos(2\theta-2\psi)\\
          &- \sqrt{I^2-Y^2}\big(\mathcal{A}(t)\sin\psi + \mathcal{B}(t)\cos\psi\big) + 2Y\mathcal{C}(t).
       \end{aligned}
    \end{equation}
    We will now suppose that the dynamics of $\theta$, corresponding to the ``semi-secular'' timescale (either circulation or oscillation), is much faster than the evolution of the other degrees of freedom, corresponding to the secular timescale. We thus consider for now that except $(I,\theta)$, all the variables are fixed (adiabatic approximation). The equations of motion are
    \begin{equation}
       \begin{aligned}
          \dot{I} &= -\frac{\partial\mathcal{H}}{\partial\theta} = -\alpha_r\frac{(I+Y)^2}{I}\sin(2\theta-2\psi) \\
          \dot{\theta} &= \frac{\partial\mathcal{H}}{\partial I} = \frac{I}{C} - n + \frac{\alpha}{2}\frac{Y^2}{I^2\big(1-e(t)^2\big)^{3/2}} \\
          &\hspace{1.1cm}
          - \frac{\alpha_r}{2}\frac{I^2-Y^2}{I^2}\cos(2\theta-2\psi) \\
          &\hspace{1.1cm}
          - \frac{I}{\sqrt{I^2-Y^2}}\big(\mathcal{A}(t)\sin\psi + \mathcal{B}(t)\cos\psi\big) \,.
       \end{aligned}
    \end{equation}
    Using the definition of $I$, $Y$ and $\alpha_r$, the first equation gives
    \begin{equation}
       \dot{\omega} = - \frac{3\,\mathcal{G}m_0}{8 a^3}\,\frac{B-A}{C} (1+X)^2\sin(2\theta-2\psi),
    \end{equation}
    resulting, for any value of $X$, to two equilibrium points: $\theta=\psi$ and $\psi+\pi/2\mod\pi$. We note that $\theta=\psi$ is an elliptic equilibrium while $\theta=\psi+\pi/2$ is hyperbolic. Injecting this into the second equation, we obtain
    \begin{equation}
       \dot{\theta} = \frac{I}{C} - n +\text{small terms},
    \end{equation}
    in which the small terms correspond to the precession of the spin axis ($\alpha$ and $\alpha_r$) and the precession of the orbit ($\mathcal{A}$ and $\mathcal{B}$). The equilibrium condition, corresponding to the exact resonance, is thus $\omega\approx n$. Considering that the planet is locked in synchronous rotation, we have thus $\theta=\psi$ and $\omega\approx n$. According to the adiabatic approximation, this will be verified whatever the value of the slow variables, such that we can inject them into the full Hamiltonian:
   \begin{equation}
      \begin{aligned}
         &\mathcal{H}(X,-\psi,t) = -\frac{\alpha}{2}\frac{X^2}{\big(1-e(t)^2\big)^{3/2}} - \frac{\alpha_r}{2}(1+X)^2 \\
         &- \sqrt{1-X^2}\big(\mathcal{A}(t)\sin\psi + \mathcal{B}(t)\cos\psi\big) + 2X\mathcal{C}(t) \,,
      \end{aligned}
   \end{equation}
    where this time, we use $X$ as conjugate momentum of $-\psi$ (the Hamiltonian is thus divided by the constant $L$). In the expression of $\alpha$ and $\alpha_r$, we must replace $\omega$ by $n$. We get here one extra term with respect to~\eqref{eq:Hinit}, due to the spin-orbit resonance. Using the same method as in Sect.~\ref{sec:coltop}, the Hamiltonian in case of a first-order secular spin-orbit resonance is
    \begin{equation}
       \mathcal{F}(\Sigma,\sigma) = -\frac{1}{2}(\mathfrak{a}\,\alpha+\alpha_r)\Sigma^2 + (\mathfrak{b}+\alpha_r)\Sigma + \mathfrak{c}\,\sqrt{1-\Sigma^2}\cos\sigma,
    \end{equation}
    which must be compared to~\eqref{eq:Fabc}. This Hamiltonian has the same general form and it can be reduced to Colombo's top. We can thus apply the same method of resolution (redefining the constants accordingly).

   \section{Characteristic quantities of Colombo's top}
   
   \subsection{Equilibrium points}\label{assec:eqp}
   From~\eqref{eq:Hres}, the equations of motion are
   \begin{equation}
      \left\{
      \begin{aligned}
         \dot{\Sigma} &= -\frac{\partial\mathcal{F}}{\partial\sigma} = \beta\sqrt{1-\Sigma^2}\sin\sigma \\
         \dot{\sigma} &= +\frac{\partial\mathcal{F}}{\partial\Sigma} = - \Sigma + \gamma - \beta\frac{\Sigma}{\sqrt{1-\Sigma^2}}\cos\sigma \,.
      \end{aligned}
      \right.
   \end{equation}
   Apart from the coordinate singularity at $\Sigma=\pm 1$, the first equation implies that $\dot{\Sigma}=0$ when $\sigma=0$ or $\pi$. Injecting this into the second equation, we get
   \begin{equation}
      (\gamma-\Sigma)\sqrt{1-\Sigma^2} = \pm\beta\Sigma
   \end{equation}
   where $\beta\geqslant 0$ by hypothesis. The resolution of this equation requires to square left and right-hand terms, loosing the information\footnote{After having computed one solution $\Sigma_0$, this information is retrieved by checking the sign of $\Sigma_0/(\gamma-\Sigma_0)$.} about the sign of $\cos\sigma$. We obtain a quartic equation in~$\Sigma$:
   \begin{equation}
      P_4(\Sigma) = \Sigma^4 - 2\gamma\Sigma^3 + (\gamma^2+\beta^2-1)\Sigma^2 + 2\gamma\Sigma - \gamma^2 = 0 \,,
   \end{equation}
   with discriminant
   \begin{equation}\label{eq:Delta}
      \begin{aligned}
         \Delta_4 = 16\gamma^2\beta^2\Big[& - \gamma^6 + 3(1-\beta^2)\gamma^4 \\
         &- 3(1+7\beta^2+\beta^4)\gamma^2 + (1-\beta^2)^3 \Big] \,.
      \end{aligned}
   \end{equation}
   It is zero for the particular cases $\gamma=0$ or $\beta=0$, for which the polynomial can be factored into, respectively,
   \begin{equation}
      \begin{aligned}
         P_4(\Sigma)\big|_{\beta=0} &= (\Sigma-1)(\Sigma+1)(\Sigma-\gamma)^2 \\
         P_4(\Sigma)\big|_{\gamma=0} &= (\Sigma-\sqrt{1-\beta^2})(\Sigma+\sqrt{1-\beta^2})\Sigma^2 \,,
      \end{aligned}
   \end{equation}
   showing the corresponding solutions and their multiplicities. They constitute equilibrium points of the system whenever they are real and in the interval $[-1;1]$.

   For $\gamma > 0$ and $\beta > 0$, the discriminant can be either negative (two equilibrium points), zero (three equilibrium points among which one double root), or positive (four equilibrium points). The corresponding solutions can be written analytically according to the general resolution of quartic equations. They are namely
   \begin{equation}
      \begin{aligned}
         \Sigma_a &= \frac{1}{2}\gamma - V + \frac{1}{2}\sqrt{2C-D+\gamma\frac{1+\beta^2}{V}} \\
         \Sigma_b &= \frac{1}{2}\gamma - V - \frac{1}{2}\sqrt{2C-D+\gamma\frac{1+\beta^2}{V}} \\
         \Sigma_c &= \frac{1}{2}\gamma + V - \frac{1}{2}\sqrt{2C-D-\gamma\frac{1+\beta^2}{V}} \\
         \Sigma_d &= \frac{1}{2}\gamma + V + \frac{1}{2}\sqrt{2C-D-\gamma\frac{1+\beta^2}{V}} \,,
      \end{aligned}
   \end{equation}
   where numerous intermediary variables are required in order to get compact expressions:
   \begin{equation}
      \begin{aligned}
         W &= \gamma^2 + \beta^2 - 1 \\
         Z &= 108\gamma^2\beta^2 + 2W^3 \\
         U &= \sqrt[3]{\frac{1}{2}\left(Z+\sqrt{Z^2-4W^6}\right)}
      \end{aligned}
      \hspace{0.5cm}
      \begin{aligned}
         C &= \gamma^2-\frac{2}{3}W \\
         D &= \frac{1}{3}\left(U+\frac{W^2}{U}\right) \\
         V &= \frac{1}{2}\sqrt{C+D} \,.
      \end{aligned}
   \end{equation}
   We note that $\Sigma_{c,d}$ are real solutions only when $\Delta_4\geqslant 0$ (see below for the limit in terms of $\gamma$ and $\beta$). The corresponding values of $\sigma$ are
   \begin{equation}
      \sigma_a = 0
      \hspace{0.3cm},\hspace{0.3cm}
      \sigma_b = \pi
      \hspace{0.3cm},\hspace{0.3cm}
      \sigma_c = \pi
      \hspace{0.3cm},\hspace{0.3cm}
      \sigma_d = \pi \,.
   \end{equation}
   The points $a$, $b$ and $d$ are elliptic fixed points, whereas the point $c$ is hyperbolic.
   
   \subsection{First boundary (BC/D)}\label{assec:1stb}
   The zero value of~\eqref{eq:Delta} corresponds to a bifurcation. Its position can be computed by solving the equation $\Delta_4=0$, which corresponds to solving a cubic equation either in $\gamma^2$ or $\beta^2$. Choosing to solve it in terms of $\beta$, the discriminant is
   \begin{equation}
      \Delta = -19683\,\gamma^4(1+\gamma^2)^2 < 0 \,,
   \end{equation}
   meaning that there is only one real solution. This solution is
   \begin{equation}
      \beta^2 = \Big(1-\gamma^{2/3}\Big)^3
      \hspace{0.5cm}\text{or}\hspace{0.5cm}
      \gamma^2 = \Big(1-\beta^{2/3}\Big)^3 \,,
   \end{equation}
   which is the boundary $\mathscr{C}_1$~\eqref{eq:C1}.
   
   \subsection{Second boundary (A/B)}\label{assec:2ndb}
   The other two boundaries can be obtained by studying the level curves of the Hamiltonian passing through $\Sigma=\pm 1$ (which is singular using the coordinates $\Sigma$ and $\sigma$, but it does not matter here).
    
   Let us begin with the $+1$ case, for which the Hamiltonian has value $-1/2+\gamma$. We now look for this specific level curve along the axes $\sigma=0$ and $\sigma=\pi$. This leads to the equation
   \begin{equation}
      -\frac{1}{2}\Sigma^2 + \gamma\Sigma \pm \beta\sqrt{1-\Sigma^2} = -\frac{1}{2}+\gamma \,,
   \end{equation}
   for which $\Sigma=+1$ is a solution. By reorganising the terms, taking the square (thus loosing the information about the sign of $\cos\sigma$), and dividing by $(\Sigma-1)$, we get
   \begin{equation}
      \begin{aligned}
         P_3(\Sigma) &= \frac{1}{4}\Sigma^3 + \left(\frac{1}{4}-\gamma\right)\Sigma^2 + \left(-\frac{1}{4}+\gamma^2+\beta^2\right)\Sigma \\
         &+ \left(-\frac{1}{4}+\beta^2+\gamma-\gamma^2\right) = 0 \,,
      \end{aligned}
   \end{equation}
   which is a cubic equation in $\Sigma$. Its determinant is
   \begin{equation}
      \Delta_3 = \beta^2\left[ - \beta^4  + \left(\frac{1}{4}-5\gamma-2\gamma^2\right)\beta^2 + \gamma(1-\gamma)^3 \right] \,.
   \end{equation}
   Once again, it is zero for $\beta=0$. Moreover the solutions for $\gamma=0$ can be easily computed. In these two particular cases, the polynomial can be factored into, respectively,
   \begin{equation}
      \begin{aligned}
         P_3(\Sigma)\big|_{\beta=0} &= \frac{1}{4}(\Sigma-1)(\Sigma+1-2\gamma)^2 \\
         P_3(\Sigma)\big|_{\gamma=0} &= \frac{1}{4}(\Sigma-\sqrt{1-4\beta^2})(\Sigma+\sqrt{1-4\beta^2})(\Sigma+1) \,,
      \end{aligned}
   \end{equation}
   showing the solutions and their multiplicities. For $\gamma > 0$ and $\beta > 0$, the discriminant can be either negative (one solution), zero (three solutions among which one double root), or positive (three solutions). The zero value corresponds to the limit we are looking for. Its position can be computed by solving the equation $\Delta_3=0$, which amounts to solving a quadratic equation in $\beta^2$ or a quartic equation in $\gamma$. Choosing to solve it in terms of $\beta$, the only positive solution is
   \begin{equation}
      \beta^2 = \frac{1}{8}\Big( 1 - 20\gamma - 8\gamma^2 + (1+8\gamma)^{3/2} \Big) \,,
   \end{equation}
   which is the boundary $\mathscr{C}_2$~\eqref{eq:C23}.
   
   \subsection{Third boundary (B/C)}\label{assec:3rdb}
   Let us now study the level curve of the Hamiltonian passing in $\Sigma=-1$, which has value $-1/2-\gamma$. The procedure is the same as for the second boundary, and the new formulas are obtained simply by replacing $\gamma$ by $-\gamma$. There is though an ambiguity because there are two positive solutions $\beta^2$ (as a function of $\gamma$) which cancel the determinant. The one corresponding to the bifurcation is the largest, that is,
   \begin{equation}
      \beta^2 = \frac{1}{8}\Big( 1 + 20\gamma - 8\gamma^2 + (1-8\gamma)^{3/2} \Big) \,,
   \end{equation}
   which is the boundary $\mathscr{C}_3$~\eqref{eq:C23}.
   
   \subsection{Separatrices}\label{assec:sep}
   The position at $\sigma=0$ or $\pi$ of the separatrix emerging from the hyperbolic point $(\Sigma,\sigma)=(\Sigma_c,\pi)$ defines the boundaries of the resonant region (see Fig.~\ref{fig:zones}). Writing $f = \mathcal{F}(\Sigma_c,\pi)$, the equations to solve are
   \begin{equation}
      -\frac{1}{2}\Sigma^2 + \gamma\Sigma \pm\beta\sqrt{1-\Sigma^2} = f \,.
   \end{equation}
   The resolution of this equation requires to square left and right-hand terms, loosing the information\footnote{After having computed one solution $\Sigma_0$, this information is retrieved by checking the sign of $-\Sigma_0^2/2+\gamma\Sigma_0-f$.} about the sign of $\cos\sigma$. We obtain a quartic equation in~$\Sigma$,
   \begin{equation}
      \frac{1}{4}\Sigma^4 - \gamma\Sigma^3 + (\gamma^2+\beta^2+f)\Sigma^2 - 2f\gamma\Sigma + f^2-\beta^2 = 0 \,,
   \end{equation}
   in which $\Sigma_c$ is a double root. It can thus be divided by $(\Sigma-\Sigma_c)^2$, leading to the quadratic equation
   \begin{equation}
      \begin{aligned}
         P_2(\Sigma) &= \frac{1}{4}\Sigma^2 + \left(\frac{1}{2}\Sigma_c-\gamma\right)\Sigma \\
         &+ \left(\frac{3}{4}\Sigma_c^2 + f + \beta^2 - 2\gamma\Sigma_c + \gamma^2\right) = 0 \,.
      \end{aligned}
   \end{equation}
   This equation has always two real solutions, provided that $\Sigma_c$ exists (that is, in zones A, B or C). These solutions are
   \begin{equation}
      \Sigma_{\pm} = 2\gamma - \Sigma_c \pm 2\sqrt{ -\beta^2+\beta\sqrt{1-\Sigma_c^2} } \,,
   \end{equation}
   where we replaced $f$ by its expression~\eqref{eq:Hres} in terms of $\Sigma_c$.
   
   \section{Second-order resonances}\label{asec:2ndor}
   Using the intermediary Hamiltonian $\mathcal{X}=\varepsilon\mathcal{X}_1$~\eqref{eq:X1}, the Hamiltonian in the new coordinates is obtained term by term from Eq.~\eqref{eq:Htilde1}. The two first terms are simple: we have $\tilde{\mathcal{H}}_0=\mathcal{H}_0$~(given at Eq.~\ref{eq:H0}) and $\tilde{\mathcal{H}}_1=0$ by definition of $\mathcal{X}$. The second-order term is more complex since it requires to compute Poisson's brackets. Using of the fact that $\{\mathcal{X}_1,\mathcal{H}_0\} = -\mathcal{H}_1$ and reorganising the terms adequately, we obtain
   \begin{equation}\label{eq:e2H2}
      \begin{aligned}
         \varepsilon^2\tilde{\mathcal{H}}_2 &= -\frac{3}{4}\alpha X^2\sum_{j=1}^N E_j^2 \\
         &+ 2X \sum_{j=1}^M \nu_jS_j^2 \\
         &- 2X \sum_{j=1}^M\frac{\nu_j^2S_j^2}{\nu_j+\alpha X} \\
         &- \alpha (1-X^2)\sum_{j=1}^M\frac{\nu_j^2S_j^2}{(\nu_j+\alpha X)^2} \\
         &- \frac{3}{2}\alpha X^2\sum_{j<k}^N E_jE_k\cos(\theta_j-\theta_k) \\
         &+ \sum_{j<k}^M S_jS_k\Big[2X(\nu_j+\nu_k) - \frac{2X\nu_j\nu_k}{\nu_j+\alpha X} - \frac{2X\nu_j\nu_k}{\nu_k+\alpha X} \\
         &\hspace{0.5cm}
         - \frac{\alpha(1-X^2)\nu_j\nu_k}{(\nu_j+\alpha X)^2} - \frac{\alpha(1-X^2)\nu_j\nu_k}{(\nu_k+\alpha X)^2}\Big]\cos(\phi_j-\phi_k) \\
         &+ \alpha (1-X^2)\sum_{j<k}^M \nu_j\nu_kS_jS_k\Big[\frac{1}{(\nu_j+\alpha X)^2} \\
         &\hspace{2.9cm}
         +\frac{1}{(\nu_k+\alpha X)^2}\Big]\cos(\phi_j+\phi_k+2\psi) \\
         &+ \alpha (1-X^2)\sum_{j=1}^M \frac{\nu_j^2S_j^2}{(\nu_j+\alpha X)^2}\cos(2\phi_j+2\psi) \,.
      \end{aligned}
   \end{equation}
   Since by hypothesis there is no first-order resonance in the system, the only possible resonant angles at second order are of the form $\sigma = \phi_j+\phi_k+2\psi$. Let us perform the canonical change of coordinates
   \begin{equation}
      \begin{pmatrix}
         \sigma \\
         \gamma_1 \\
         \gamma_2
      \end{pmatrix}
      =
      \begin{pmatrix}
         -2 & 1 & 1 \\
          1 & 1 & 0 \\
          0 & 0 & 1
      \end{pmatrix}
      \begin{pmatrix}
         -\psi \\
         \phi_j \\
         \phi_k
      \end{pmatrix}
      \,,
   \end{equation}
   and
   \begin{equation}
      \begin{pmatrix}
         \Sigma \\
         \Gamma_1 \\
         \Gamma_2
      \end{pmatrix}
      =
      \begin{pmatrix}
          1 & -1 &  0 \\
         -1 & -2 &  0 \\
         -1 &  1 & -3
      \end{pmatrix}
      \begin{pmatrix}
         X \\
         \Phi_j \\
         \Phi_k
      \end{pmatrix}
      \,.
   \end{equation}
   Assuming that $\sigma$ is the only resonant angle, the dynamics at second order is given by averaging $\tilde{\mathcal{H}}$ over all other angles (this is another change of coordinates close to identity). The momenta $\Gamma_1$ and $\Gamma_2$ become arbitrary constants of motion that we will conveniently choose equal to zero. Dropping the unnecessary constants, the resonant Hamiltonian is thus
   \begin{equation}
      \begin{aligned}
         &\mathcal{F}(\Sigma,\sigma) = - \frac{\alpha}{2}X^2 - \frac{\nu_j+\nu_k}{2}X \\
         &- \frac{3}{4}\alpha X^2\sum_{i=1}^N E_i^2 + 2X\sum_{i=1}^M \nu_iS_i^2 - 2X\sum_{i=1}^M\frac{\nu_i^2S_i^2}{\nu_i+\alpha X} \\
         &- \alpha (1-X^2)\sum_{i=1}^M\frac{\nu_i^2S_i^2}{(\nu_i+\alpha X)^2} \\
         &+ \alpha (1-X^2)\nu_j\nu_kS_jS_k\left(\frac{1}{(\nu_j+\alpha X)^2}+\frac{1}{(\nu_k+\alpha X)^2}\right)\cos\sigma
      \end{aligned}
   \end{equation}
   in which $X$ must be replaced by $-2\Sigma$. High-order resonances are quite thin, so it is enough to consider the dynamics in the neighbourhood of the resonance centre at first order:
   \begin{equation}
      \begin{aligned}
         &\dot{\sigma}=\frac{\partial\mathcal{F}}{\partial\Sigma} = -4\alpha\Sigma + \nu_j+\nu_k + \mathcal{O}(\varepsilon^2) = 0 \\
         &\iff \Sigma_0 = \frac{\nu_j+\nu_k}{4\alpha} \,,
      \end{aligned}
   \end{equation}
   or equivalently $X_0=-2\Sigma_0=-(\nu_j+\nu_k)/(2\alpha)$. Considering that $|X-X_0|=\mathcal{O}(\varepsilon)$, we have then
   \begin{equation}
      \mathcal{F}(\Sigma,\sigma) = -\frac{\alpha}{2}(X-X_0)^2 + \alpha K\cos\sigma \,,
   \end{equation}
   in which we dropped the unnecessary constants, and where
   \begin{equation}
      K = \frac{8}{(\nu_j-\nu_k)^2}\left(1-\frac{(\nu_j+\nu_k)^2}{4\alpha^2}\right)\nu_j\nu_kS_jS_k \,.
   \end{equation}
   By injecting the momentum $\Sigma$ instead of $X$ and by using the modified time $\mathrm{d}\tau=-4\alpha\mathrm{d}t$, we obtain
   \begin{equation}
      \mathcal{F}(\Sigma,\sigma) = \frac{1}{2}(\Sigma-\Sigma_0)^2 - \frac{K}{4}\cos\sigma \,.
   \end{equation}
   This is the Hamiltonian of a pendulum of centre $\Sigma_0$ and half width $\sqrt{|K|}$. In terms of the obliquity cosine $X$, the resonance has position $X_0$ and half width $2\sqrt{|K|}$.

   \section{Third-order resonances}\label{asec:3rdor}
   If there is no resonance at first and second orders, we can use a canonical change of coordinates close to identity in order to suppress the angular dependency at first and second orders. Let us consider an intermediary Hamiltonian $\mathcal{X} = \varepsilon\mathcal{X}_1 + \varepsilon^2\mathcal{X}_2$, such that the new coordinates are given by its flow at time $1$. The Hamiltonian in the new coordinates is then
   \begin{equation}
      \tilde{\mathcal{H}} = \tilde{\mathcal{H}}_0 + \varepsilon\tilde{\mathcal{H}}_1 + \varepsilon^2\tilde{\mathcal{H}}_2 + \varepsilon^3\tilde{\mathcal{H}}_3 + \mathcal{O}(\varepsilon^4) \,,
   \end{equation}
   where
   \begin{equation}
      \begin{aligned}
         \tilde{\mathcal{H}}_0 &= \mathcal{H}_0 \\
         \tilde{\mathcal{H}}_1 &= \mathcal{H}_1 + \{\mathcal{X}_1,\mathcal{H}_0\} \\
         \tilde{\mathcal{H}}_2 &= \mathcal{H}_2 + \{\mathcal{X}_2,\mathcal{H}_0\} + \{\mathcal{X}_1,\mathcal{H}_1\} + \frac{1}{2}\{\mathcal{X}_1,\{\mathcal{X}_1,\mathcal{H}_0\}\} \\
         \tilde{\mathcal{H}}_3 &= \mathcal{H}_3 + \{\mathcal{X}_1,\mathcal{H}_2\} + \{\mathcal{X}_2,\mathcal{H}_1\} + \frac{1}{2}\{\mathcal{X}_1,\{\mathcal{X}_2,\mathcal{H}_0\}\} \\
         &+ \frac{1}{2}\{\mathcal{X}_2,\{\mathcal{X}_1,\mathcal{H}_0\}\} + \frac{1}{2}\{\mathcal{X}_1,\{\mathcal{X}_1,\mathcal{H}_1\}\} \\
         &+ \frac{1}{6}\{\mathcal{X}_1,\{\mathcal{X}_1,\{\mathcal{X}_1,\mathcal{H}_0\}\}\} \,.
      \end{aligned}
   \end{equation}
   The first-order part of $\mathcal{X}$ required to suppress the angular dependency at order 1 can be directly taken from Eq.~\eqref{eq:X1}. Let us write
   \begin{equation}
      \mathcal{A}_2 = \mathcal{H}_2 + \{\mathcal{X}_1,\mathcal{H}_1\} + \frac{1}{2}\{\mathcal{X}_1,\{\mathcal{X}_1,\mathcal{H}_0\}\} = \overline{\mathcal{A}_2} + \widetilde{\mathcal{A}_2}
   \end{equation}
   (average plus oscillating part) for which the expression is given by~\eqref{eq:e2H2}. The homological equation for order 2 is then
   \begin{equation}
      \begin{aligned}
         \{\mathcal{X}_2,\mathcal{H}_0\} + \mathcal{A}_2 = \overline{\mathcal{A}_2} \,,
      \end{aligned}
   \end{equation}
   which defines the Hamiltonian $\mathcal{X}_2$. This leads to
   \begin{equation}
      \begin{aligned}
         \varepsilon^2\mathcal{X}_2 &= -\frac{3}{2}\alpha X^2\sum_{j<k}^N\frac{E_jE_k}{\mu_j-\mu_k}\sin(\theta_j-\theta_k) \\
         &+ \sum_{j<k}^M\frac{S_jS_k}{\nu_j-\nu_k}\Big[2X(\nu_j+\nu_k) - \frac{2X\nu_j\nu_k}{\nu_j+\alpha X} - \frac{2X\nu_j\nu_k}{\nu_k+\alpha X} \\
         &\hspace{0.5cm}
         - \frac{\alpha(1-X^2)\nu_j\nu_k}{(\nu_j+\alpha X)^2} - \frac{\alpha(1-X^2)\nu_j\nu_k}{(\nu_k+\alpha X)^2}\Big]\sin(\phi_j-\phi_k) \\
         &+ \alpha (1-X^2)\sum_{j<k}^M\frac{\nu_j\nu_kS_jS_k}{\nu_j+\nu_k+2\alpha X}\Big[\frac{1}{(\nu_j+\alpha X)^2} \\
         &\hspace{2.9cm}
         + \frac{1}{(\nu_k+\alpha X)^2}\Big]\sin(\phi_j+\phi_k+2\psi) \\
         &+ \frac{1}{2}\alpha (1-X^2)\sum_{j=1}^M\frac{\nu_j^2S_j^2}{(\nu_j+\alpha X)^3}\sin(2\phi_j+2\psi) \,.
      \end{aligned}
   \end{equation}
   We must now compute the remainders at order 3. First of all, we can simplify their expressions by taking into account that, by definition: $\{\mathcal{X}_1,\mathcal{H}_0\} = -\mathcal{H}_1$, $\{\mathcal{X}_2,\mathcal{H}_0\} = -\widetilde{\mathcal{A}_2}$ and $\{\mathcal{X}_1,\mathcal{H}_1\}=2(\mathcal{A}_2-\mathcal{H}_2)$. We have then
   \begin{equation}
      \begin{aligned}
         \tilde{\mathcal{H}}_3 &= \mathcal{H}_3 + \frac{1}{3}\{\mathcal{X}_1,\mathcal{H}_2\} + \frac{1}{2}\{\mathcal{X}_2,\mathcal{H}_1\} \\
         &+ \frac{1}{6}\{\mathcal{X}_1,\widetilde{\mathcal{A}_2}\} + \frac{2}{3}\{\mathcal{X}_1,\overline{\mathcal{A}_2}\} \,,
      \end{aligned}
   \end{equation}
   which gives
   \begin{equation}
      \begin{aligned}
         \varepsilon\tilde{\mathcal{H}}_3 &= \sum_{j=1}^M\Big[D_j\Big]\cos(\phi_j+\psi) \\
         &- \frac{3}{4}\alpha^2(1-X^2)^{3/2}\sum_{j=1}^M\frac{\nu_j^3S_j^3}{(\nu_j+\alpha X)^4}\cos(3\phi_j+3\psi) \\
         &+ \alpha^2(1-X^2)^{3/2}\sum_{j=1}^M\sum_{\substack{k=1\\k\neq j}}^M \nu_j^2\nu_kS_j^2S_k\Big[
         A_{jk} \\
         &\hspace{2.5cm}
         - \frac{3}{4(\nu_j+\alpha X)^4}
         \Big]\cos(2\phi_j+\phi_k+3\psi) \\
         &+ \sqrt{1-X^2}\sum_{j=1}^M\sum_{\substack{k=1\\k\neq j}}^M S_j^2S_k\Big[
         \nu_j^2\nu_k B_{jk} + \nu_k \\
         &\hspace{1.3cm}
         + \nu_j\frac{(\nu_j+\nu_k)(\nu_k+\alpha X)}{(\nu_j-\nu_k)(\nu_j+\alpha X)}
         \Big]\cos(2\phi_j-\phi_k+\psi)\\
         &+ \sqrt{1-X^2}\sum_{i=1}^M\sum_{\substack{j<k\\j,k\neq i}}^M S_iS_jS_k\Big[
         C_{ijk}
         \Big]\\
         &\hspace{4cm}
         \times\cos(-\phi_i+\phi_j+\phi_k+\psi)\\
         &+ \alpha^2(1-X^2)^{3/2}\sum_{i<j<k}^M \nu_i\nu_j\nu_kS_iS_jS_k\Big[A_{ij}+A_{jk}\\
         &\hspace{3.4cm}
         +A_{ik}\Big]\cos(\phi_i+\phi_j+\phi_k+3\psi)\\
         &+ \frac{3}{2}\alpha X\sqrt{1-X^2}\sum_{i=1}^M\sum_{j=1}^N\sum_{\substack{k=1\\k\neq j}}^N
         \nu_iS_iE_jE_k\Big[\frac{1}{\nu_i+\alpha X}\\
         &\hspace{3cm}
         -\frac{1}{\mu_j-\mu_k}\Big]\cos(\phi_i+\theta_j-\theta_k+\psi) \\
      \end{aligned}
   \end{equation}
   where
   \begin{equation}
      \begin{aligned}
         &x_{jk} = \frac{1}{\nu_j+\alpha X} + \frac{1}{\nu_k+\alpha X} \\
         &y_{jk} = \frac{1}{(\nu_j+\alpha X)^2} + \frac{1}{(\nu_k+\alpha X)^2} \\
         &z_{jk} = \frac{1}{(\nu_j+\alpha X)^3} + \frac{1}{(\nu_k+\alpha X)^3}
      \end{aligned}
   \end{equation}
   \begin{equation}
      \begin{aligned}
         &b_{jk} = \frac{y_{jk}}{\nu_j+\nu_k+2\alpha X} \\
         &c_{jk} = \frac{b_{jk} + z_{jk}}{\nu_j+\nu_k+2\alpha X} \\
         &d_{jk} = \frac{1}{3(\nu_j+\alpha X)(\nu_k+\alpha X)}\Bigg(1 + \frac{2\alpha X}{\nu_j+\alpha X}-\frac{2\alpha X}{\nu_k+\alpha X} \\
         &+ \frac{\alpha^2(1-X^2)}{(\nu_j+\alpha X)^2} - \frac{\alpha^2(1-X^2)}{(\nu_k+\alpha X)^2} + \frac{\alpha^2(1-X^2)}{(\nu_j+\alpha X)(\nu_k+\alpha X)}\Bigg) \\
         &e_{jk} = \frac{- x_{jk} + 2\alpha X\,y_{jk} + \alpha^2(1-X^2)\,z_{jk}}{\nu_j-\nu_k} \\
      \end{aligned}
   \end{equation}
   and:
   \begin{equation}
      \begin{aligned}
         A_{jk} &= \frac{(\nu_j-\nu_k)^2}{3(\nu_j+\alpha X)^3(\nu_k+\alpha X)^3} - c_{jk}
         \\
         B_{jk} &= \frac{1}{3(\nu_j+\alpha X)^2}\Bigg(
         1 + \frac{\alpha X}{\nu_j+\alpha X} + \frac{5}{4}\frac{\alpha^2(1-X^2)}{(\nu_j+\alpha X)^2}
         \Bigg) \\
         &+ d_{jk} + e_{jk}
         \\
         C_{ijk} &= 2\nu_i - \nu_j\frac{(\nu_i+\nu_k)(\nu_i+\nu_j-\nu_k+\alpha X)}{(\nu_i-\nu_k)(\nu_j+\alpha X)} \\
         &- \nu_k\frac{(\nu_i+\nu_j)(\nu_i+\nu_k-\nu_j+\alpha X)}{(\nu_i-\nu_j)(\nu_k+\alpha X)}
         \\
         &+ \nu_i\nu_j\nu_k\Big( d_{ji} + d_{ki} + 2\alpha X b_{jk} + \alpha^2(1-X^2)c_{jk} \\
         &- e_{ij} - e_{ik} + \frac{2-2\alpha X\,x_{jk}-\alpha^2(1-X^2)\,y_{jk}}{3(\nu_j+\alpha X)(\nu_k+\alpha X)}\Big) \,.
      \end{aligned}
   \end{equation}
   The expression of the coefficients $D_j$ is very complex. We will not give it here since they have no interest at this stage (the angles $\phi_j+\psi$ are non resonant by hypothesis).
    
   As shown in Appendix~\ref{asec:2ndor}, in the pendulum approximation, the half-width of any possible resonance is two times the square root of its coefficient divided by $\alpha$, and its position is given by the combination of the unperturbed frequencies. Accordingly, the possible resonances at order 3 are gathered in Table~\ref{tab:hres}.
   
   \section{Ger\v{s}gorin circles}\label{asec:gersh}
   In order to prove that the Lagrange-Laplace matrix for the orbital inclinations has only negative or zero eigenvalues, one can use the Ger\v{s}gorin circle theorem (see \citealp{GERSHGORIN_1931} or \citealp{VARGA_2004}). This theorem is recalled below, and we show how it applies to our matrix.
   
   \paragraph{Definition.} \emph{Let $B$ be a complex $N\times N$ matrix with elements $(b_{ij})$. The $i$th ``Ger\v{s}gorin disc'' $\mathscr{G}_i$ $(i=1,2..N)$ is the closed disc of the complex plane centred at $b_{ii}$ and with radius}
   \begin{equation}
      R_i=\sum_{\substack{j=1\\j\neq i}}^N|b_{ij}| \,.
   \end{equation}
   
   \paragraph{Theorem \citep{GERSHGORIN_1931}.} \emph{Any eigenvalue of $B$ lies inside at least one of the $\mathscr{G}_i$ discs, $i=1,2..N$.}
   
   \paragraph{Corollary.} \emph{All the eigenvalues of $B$ are located inside the union of the $\mathscr{G}_i$ discs, $i=1,2..N$.}
   
\begin{figure}
      \centering
      \includegraphics[width=0.35\textwidth]{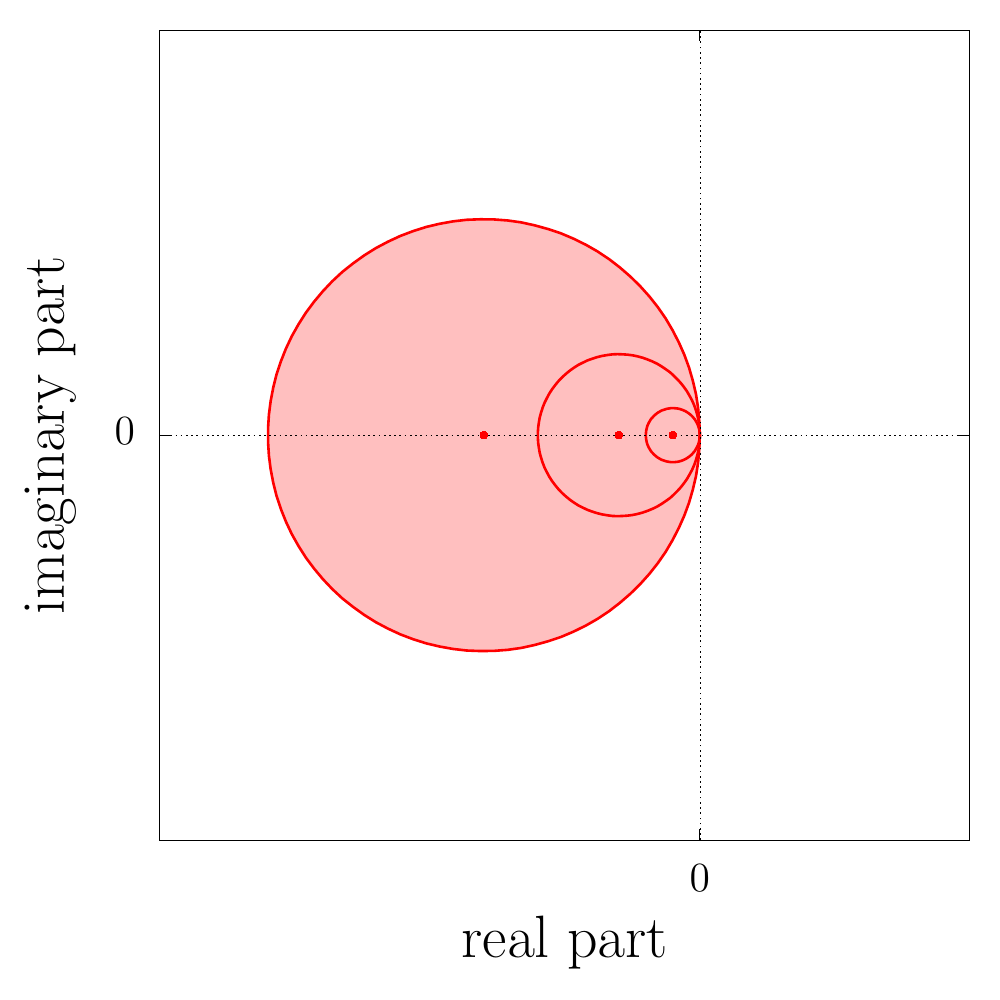}
      \caption{Ger\v{s}gorin discs in the complex plane corresponding to the Lagrange-Laplace matrix $B$ for three planets. The centre of the circles are the diagonal entries of $B$ (red spots). Every eigenvalue of $B$ lies on the real line, inside the union of all the discs.}
      \label{fig:gersh}
   \end{figure}

   \bigskip
   In our case, the matrix $B$ is real (see Eq.~\ref{eq:matInc}). It has only real eigenvalues and one of them is identically equal to zero. Moreover, given the very particular form of this matrix, the centre of each Ger\v{s}gorin disc is located on the real line, with an abscissa equal to the opposite of its radius. Therefore, all the eigenvalues of $B$ are negative or zero, as illustrated in Fig.~\ref{fig:gersh}.
   
   \section{Orbital solution used for the inner Solar System}\label{asec:SSsol}
   
   In order to apply our method to a given planet, we first need a quasi-periodic approximation of its long-term orbital dynamics.
   
   In the case of the Solar System, the search for such series has been a challenge for centuries, eventually leading to very complete solutions (up to the degree of chaos inherent to the system). In the present work, we use the solution of \cite{LASKAR_1990}, obtained by multiplying the normalised proper modes $z_i^\bullet$ and $\zeta_i^\bullet$ (Tables~VI and VII of \citealp{LASKAR_1990}) by the matrix $\tilde{S}$ corresponding to the linear part of the solution (Table~V of \citealp{LASKAR_1990}). In the series obtained, the terms with the same combination of frequencies are then merged together, finally resulting in 56 terms in eccentricity and 60 terms in inclination.
   
   These series are given in Tables~\ref{tab:QPS-Mercury}-\ref{tab:QPS-Mars} for the inner planets, under the form:
   \begin{equation}
      \begin{aligned}
         z = e\exp(i\varpi) &= \sum_{j=1}^N E_j\exp\big[i(\mu_jt+\theta_j^{(0)})\big] \\
         \zeta = \sin\frac{I}{2}\exp(i\Omega) &= \sum_{j=1}^M S_j\exp\big[i(\nu_jt+\phi_j^{(0)})\big] \,,
      \end{aligned}
   \end{equation}
   with $N=56$ and $M=60$. They are used in Fig.~\ref{fig:ovlSR} of the present work.
   
   \begin{table}
      \caption{Quasi-periodic representation of the orbital dynamics of Mercury.}
      \label{tab:QPS-Mercury}
      \vspace{-0.7cm}
      \tiny
      \begin{equation*}
         \begin{array}{rrrrrr}
            \hline\hline
            \multicolumn{3}{c}{z} & \multicolumn{3}{c}{\zeta}\\
            \hline
            \mu_j\,(''/yr) & E_j\!\times\! 10^8 & \theta_j^{(0)}\,(^\text{o}) & \nu_j\,(''/yr) & S_j\!\times\! 10^8 & \phi_j^{(0)}\,(^\text{o}) \\
            \hline
              5.59644 & 18337396 & 110.35    &     -5.61755 &  3995819 & 348.70 \\
              5.47449 &  6902428 & 275.01    &     -7.07963 &  3015900 & 273.77 \\
              5.71670 &  5240271 & 120.52    &     -7.19493 &  1505361 & 105.16 \\
              4.24882 &  3635276 &  30.67    &     -6.96094 &  1429554 &  97.95 \\
              5.35823 &  2815900 &  94.89    &     -5.50098 &  1424811 & 342.89 \\
              7.45592 &  2786428 &  20.24    &      0.00000 &  1372386 & 107.59 \\
              4.36906 &  1312738 & 220.84    &     -6.84091 &  1183049 & 107.89 \\
              5.99227 &  1035633 & 113.56    &     -7.33264 &   872607 & 196.75 \\
              5.65485 &   998897 &  39.22    &     -5.85017 &   481844 & 165.47 \\
              6.93423 &   934569 & 166.16    &     -5.21610 &   360659 &  18.91 \\
              5.23841 &   829067 & 272.97    &     -5.37178 &   358805 &  35.48 \\
              7.05595 &   634974 & 357.62    &     -5.10025 &   351141 & 195.38 \\
              7.34103 &   235292 &  27.85    &     -6.73842 &   285961 &  44.50 \\
             17.91550 &   165568 & 335.25    &     -7.40536 &   264351 & 233.35 \\
              7.57299 &   164186 & 191.47    &     -7.48780 &   245583 &  47.95 \\
             17.36469 &   157893 & 303.95    &     -6.56016 &   230801 & 303.47 \\
              6.82468 &    77097 &  14.53    &     -5.96899 &   205822 & 350.64 \\
             16.81285 &    70120 &  91.98    &     -8.42342 &   192248 & 211.21 \\
              3.08952 &    60554 & 121.36    &     -3.00557 &   159813 & 140.33 \\
             18.46794 &    48514 &   9.97    &    -18.85115 &   156874 & 240.43 \\
              7.20563 &    48115 & 323.91    &     -6.15490 &   149031 &  89.77 \\
             17.08266 &    43934 & 359.38    &    -17.74818 &   119892 & 303.28 \\
             17.63081 &    39761 & 202.03    &     -0.69189 &    70222 &  23.96 \\
              7.71663 &    29983 & 273.52    &     18.14984 &    53922 & 111.19 \\
             28.22069 &    21356 & 307.83    &    -18.30007 &    47541 & 269.86 \\
             17.81084 &    15980 &  58.56    &    -19.40256 &    33322 &  29.01 \\
             19.01870 &    15738 &  39.75    &    -19.13075 &    14506 & 125.90 \\
             17.15752 &    15119 & 145.02    &    -26.33023 &    13964 & 127.29 \\
             18.18553 &    14604 &  57.28    &    -18.01114 &     9092 &  62.09 \\
             17.72293 &    12073 &  48.46    &    -17.66094 &     8620 & 318.93 \\
             18.01611 &    10275 &  44.83    &    -17.83857 &     7011 & 109.13 \\
             16.52731 &     9115 & 311.91    &    -17.54636 &     6248 &  66.71 \\
             17.47683 &     7950 &  80.26    &    -18.97001 &     6027 & 253.36 \\
             16.26122 &     7118 &  58.89    &     -2.35835 &     4687 &  44.73 \\
             17.55234 &     6398 &  17.65    &    -17.94404 &     4398 &  32.26 \\
              5.40817 &     6086 & 120.60    &    -18.59563 &     4035 & 278.11 \\
             18.08627 &     5933 & 356.17    &     -1.84625 &     3435 &  41.72 \\
             52.19257 &     3589 & 225.59    &     -4.16482 &     3303 &  51.62 \\
            -19.72306 &     2363 & 113.24    &    -18.69743 &     3167 &  41.70 \\
              4.89647 &      922 & 292.23    &    -18.77933 &     3104 &  42.83 \\
              0.66708 &      717 &  73.98    &    -18.22681 &     3100 & 226.30 \\
              1.93168 &      618 &  39.55    &    -19.06544 &     2777 & 230.21 \\
              3.60029 &      447 & 121.40    &    -17.19656 &     1298 & 127.26 \\
            -56.90922 &      400 &  44.11    &     -3.11725 &     1067 & 326.97 \\
             53.35188 &      285 & 134.98    &     -0.58033 &      683 &  17.33 \\
             29.37998 &      169 &  37.61    &     -1.19906 &      372 & 133.87 \\
              2.97706 &      158 & 306.81    &     11.50319 &      341 & 281.02 \\
            -20.88236 &       75 & 203.93    &    -26.97744 &      244 &  44.61 \\
             28.86795 &       70 & 212.64    &    -50.30212 &      202 &  29.83 \\
             27.57346 &       62 & 223.74    &      0.46547 &      196 & 286.88 \\
              1.82121 &       50 & 146.09    &     10.34389 &      179 & 191.52 \\
             27.06140 &       45 &  38.56    &     20.96631 &      132 &  57.78 \\
             76.16447 &       16 & 323.03    &      0.57829 &       68 & 103.72 \\
              0.77840 &       12 &  65.10    &     82.77163 &       61 & 128.95 \\
             51.03334 &        9 & 136.30    &      9.18847 &       39 &   1.15 \\
             -0.49216 &        4 & 164.74    &     58.80017 &       36 & 212.90 \\
                      &          &           &     34.82788 &       28 & 294.12 \\
                      &          &           &    -27.48935 &       18 & 218.53 \\
                      &          &           &    -25.17116 &       17 & 215.94 \\
                      &          &           &    -28.13656 &       11 & 314.08 \\
            \hline
         \end{array}
      \end{equation*}
      \vspace{-0.3cm}
      \tablefoot{This representation is used in Fig.~\ref{fig:ovlSR}. It has been directly obtained from \cite{LASKAR_1990}, see text.}
   \end{table}
   
   \begin{table}
      \caption{Quasi-periodic representation of the orbital dynamics of Venus.}
      \label{tab:QPS-Venus}
      \vspace{-0.7cm}
      \tiny
      \begin{equation*}
         \begin{array}{rrrrrr}
            \hline\hline
            \multicolumn{3}{c}{z} & \multicolumn{3}{c}{\zeta}\\
            \hline
            \mu_j\,(''/yr) & E_j\!\times\! 10^8 & \theta_j^{(0)}\,(^\text{o}) & \nu_j\,(''/yr) & S_j\!\times\! 10^8 & \phi_j^{(0)}\,(^\text{o}) \\
            \hline
              7.45592 &  2085594 & 200.24    &      0.00000 &  1377170 & 107.59 \\
              4.24882 &  1963621 &  30.67    &    -18.85115 &   953835 &  60.43 \\
             17.91550 &  1346128 & 335.25    &     -5.61755 &   671575 & 348.70 \\
             17.36469 &  1164633 & 123.95    &    -17.74818 &   575205 & 123.28 \\
              5.59644 &   659312 & 110.35    &     -7.07963 &   404368 &  93.77 \\
             17.08266 &   324058 & 179.38    &    -18.30007 &   298364 &  89.80 \\
              5.47449 &   248173 & 275.01    &     -5.50098 &   239467 & 342.89 \\
             16.81285 &   239435 & 274.43    &     -6.84091 &   208804 & 286.39 \\
              6.93423 &   216696 & 169.77    &     -7.19493 &   201837 & 285.16 \\
             17.63081 &   191660 & 193.67    &     -6.96094 &   191673 & 277.95 \\
              5.71670 &   188411 & 120.52    &    -19.40256 &   188959 & 209.10 \\
              7.05595 &   178786 & 358.98    &     -7.33264 &   116984 &  16.75 \\
              7.34103 &   176112 & 207.85    &     -3.00557 &    99208 & 140.33 \\
             17.81084 &   129713 &  58.56    &    -19.13075 &    88217 & 305.90 \\
              7.57299 &   122891 &  11.47    &     -5.85017 &    80983 & 165.47 \\
             18.18553 &   118540 &  57.28    &     -0.69189 &    65885 &  23.96 \\
             19.01870 &   116085 & 219.75    &     -5.21610 &    60616 &  18.91 \\
             17.15752 &   111520 & 325.02    &     -5.37178 &    60304 &  35.48 \\
             18.46794 &   110955 &   6.95    &     -5.10025 &    59016 & 195.38 \\
              5.35823 &   101244 &  94.89    &    -18.01114 &    43620 & 242.09 \\
             17.72293 &    97999 &  48.46    &    -17.66094 &    41358 & 138.93 \\
             18.01611 &    83405 &  44.83    &     -6.73842 &    38337 & 224.50 \\
             16.52731 &    67232 & 131.91    &    -18.97001 &    36654 &  73.36 \\
             17.47683 &    58638 & 260.26    &     -7.40536 &    35440 &  53.35 \\
             16.26122 &    57777 &  58.89    &     -5.96899 &    34592 & 350.64 \\
              6.82468 &    57706 & 194.53    &    -17.83857 &    33636 & 289.13 \\
              3.08952 &    54138 & 121.36    &     -7.48780 &    32924 & 227.95 \\
             18.08627 &    48156 & 356.17    &     -6.56016 &    30942 & 123.47 \\
             17.55234 &    47190 & 197.65    &    -17.54636 &    29977 & 246.71 \\
              5.99227 &    37236 & 113.56    &     -8.42342 &    25773 &  31.21 \\
              7.20563 &    36013 & 143.91    &     -6.15490 &    25048 &  89.77 \\
              5.65485 &    35915 &  39.22    &    -17.94404 &    21099 & 212.26 \\
              5.23841 &    29809 & 272.97    &    -18.59563 &    19361 &  98.11 \\
              7.71663 &    22441 &  93.52    &    -18.69743 &    19259 & 221.70 \\
              4.36906 &    20218 & 220.79    &    -18.77933 &    18879 & 222.83 \\
             28.22069 &    16949 & 308.38    &    -18.22681 &    18853 &  46.30 \\
              5.40817 &     3036 & 120.48    &    -19.06544 &    16888 &  50.21 \\
            -19.72306 &     1161 & 113.24    &    -17.19656 &    11902 & 171.81 \\
              0.66708 &     1088 &  73.98    &     18.14984 &     9063 & 111.19 \\
             27.06140 &      536 & 218.72    &    -26.33023 &     5577 & 127.29 \\
              4.89647 &      470 & 291.97    &     -2.35835 &     2677 &  44.72 \\
             29.37998 &      416 & 217.51    &     -1.84625 &     2187 &  40.13 \\
             52.19257 &      339 & 225.73    &     -4.16482 &     2022 &  51.60 \\
             28.86795 &      277 &  32.64    &     -3.11725 &      663 & 326.97 \\
             27.57346 &      244 &  43.74    &     -0.58033 &      641 &  17.33 \\
              3.60029 &      242 & 121.40    &    -50.30212 &      215 &  29.83 \\
            -56.90922 &      216 &  44.11    &     11.50319 &      212 & 281.02 \\
              2.97706 &      141 & 306.81    &     -1.19906 &      194 & 133.77 \\
              1.93168 &       69 &  93.94    &      0.46547 &      184 & 286.88 \\
            -20.88236 &       67 & 203.93    &    -26.97744 &      128 &  44.89 \\
             76.16447 &       63 & 143.03    &     10.34389 &       98 & 191.39 \\
              1.82121 &       46 & 148.00    &     20.96631 &       82 &  57.78 \\
             51.03334 &       35 & 316.30    &      0.57829 &       64 & 103.72 \\
              0.77840 &       19 &  65.10    &      9.18847 &       36 &   1.15 \\
             53.35188 &       16 & 135.62    &     82.77163 &       24 & 128.95 \\
             -0.49216 &        5 & 164.74    &     58.80017 &       14 & 212.90 \\
                      &          &           &     34.82788 &       11 & 294.12 \\
                      &          &           &    -27.48935 &        7 & 218.53 \\
                      &          &           &    -25.17116 &        7 & 215.94 \\
                      &          &           &    -28.13656 &        5 & 314.08 \\
            \hline
         \end{array}
      \end{equation*}
      \vspace{-0.3cm}
      \tablefoot{This representation is used in Fig.~\ref{fig:ovlSR}. It has been directly obtained from \cite{LASKAR_1990}, see text.}
   \end{table}
   
   \begin{table}
      \caption{Quasi-periodic representation of the orbital dynamics of the Earth.}
      \label{tab:QPS-Earth}
      \vspace{-0.7cm}
      \tiny
      \begin{equation*}
         \begin{array}{rrrrrr}
            \hline\hline
            \multicolumn{3}{c}{z} & \multicolumn{3}{c}{\zeta}\\
            \hline
            \mu_j\,(''/yr) & E_j\!\times\! 10^8 & \theta_j^{(0)}\,(^\text{o}) & \nu_j\,(''/yr) & S_j\!\times\! 10^8 & \phi_j^{(0)}\,(^\text{o}) \\
            \hline
              4.24882 &  1891285 &  30.67    &      0.00000 &  1377263 & 107.59 \\
              7.45592 &  1614222 & 200.24    &    -18.85115 &   875509 & 240.43 \\
             17.91550 &  1315949 & 155.25    &     -5.61755 &   496020 & 348.70 \\
             17.36469 &   938579 & 303.95    &    -17.74818 &   401987 & 303.28 \\
              5.59644 &   420011 & 110.35    &     -7.07963 &   343071 &  93.77 \\
             17.08266 &   261159 & 359.38    &    -18.30007 &   281401 & 269.74 \\
             17.63081 &   197777 &  14.78    &     -5.50098 &   176869 & 342.89 \\
             28.22069 &   168931 & 128.09    &     -6.84091 &   174079 & 286.47 \\
             16.81285 &   168064 &  95.11    &     -7.19493 &   171242 & 285.16 \\
              6.93423 &   161978 & 169.87    &     -6.96094 &   162618 & 277.95 \\
              5.47449 &   158097 & 275.01    &    -19.40256 &   162229 &  29.19 \\
              7.34103 &   136309 & 207.85    &    -26.33023 &   133519 & 127.29 \\
              7.05595 &   134274 & 359.01    &     -7.33264 &    99249 &  16.75 \\
             18.46794 &   131495 & 187.69    &     -3.00557 &    89258 & 140.33 \\
             17.81084 &   126776 & 238.56    &    -19.13075 &    80968 & 125.90 \\
              5.71670 &   120026 & 120.52    &     -0.69189 &    64554 &  23.96 \\
             18.18553 &   115855 & 237.28    &     -5.85017 &    59814 & 165.47 \\
             17.72293 &    95780 & 228.46    &     -5.21610 &    44770 &  18.91 \\
              7.57299 &    95116 &  11.47    &     -5.37178 &    44540 &  35.48 \\
             19.01870 &    93553 &  39.75    &     -5.10025 &    43589 & 195.38 \\
             17.15752 &    89874 & 145.02    &    -18.97001 &    33642 & 253.36 \\
             18.01611 &    81516 & 224.83    &     -6.73842 &    32525 & 224.50 \\
              5.35823 &    64497 &  94.89    &    -18.01114 &    30484 &  62.09 \\
              3.08952 &    56656 & 121.36    &     -7.40536 &    30067 &  53.35 \\
             16.26122 &    56468 & 238.89    &    -17.66094 &    28903 & 318.93 \\
             16.52731 &    54183 & 311.91    &     -7.48780 &    27932 & 227.95 \\
             17.47683 &    47256 &  80.26    &     -6.56016 &    26251 & 123.47 \\
             18.08627 &    47065 & 176.17    &    -17.19656 &    25667 & 341.65 \\
              6.82468 &    44664 & 194.53    &     -5.96899 &    25550 & 350.64 \\
             17.55234 &    38030 &  17.65    &    -17.83857 &    23507 & 109.13 \\
              7.20563 &    27874 & 143.91    &     -8.42342 &    21866 &  31.21 \\
              5.99227 &    23721 & 113.56    &    -17.54636 &    20949 &  66.71 \\
              5.65485 &    22879 &  39.22    &     -6.15490 &    18500 &  89.77 \\
              5.23841 &    18989 & 272.97    &    -18.69743 &    17676 &  41.70 \\
              7.71663 &    17369 &  93.52    &    -18.77933 &    17327 &  42.83 \\
              4.36906 &     9354 & 220.76    &    -18.22681 &    17304 & 226.30 \\
             52.19257 &     7041 & 225.56    &    -19.06544 &    15500 & 230.21 \\
              5.40817 &     2871 & 120.45    &    -17.94404 &    14745 &  32.26 \\
             29.37998 &     1761 &  37.54    &    -18.59563 &    13530 & 278.11 \\
             27.06140 &     1669 &  38.70    &     18.14984 &     6694 & 111.19 \\
            -19.72306 &     1591 & 113.24    &     -2.35835 &     2098 &  44.69 \\
              0.66708 &     1259 &  73.98    &     -1.84625 &     1981 &  39.73 \\
             28.86795 &     1027 & 212.64    &     -4.16482 &     1812 &  51.59 \\
             27.57346 &      902 & 223.74    &    -26.97744 &     1074 &  43.23 \\
             53.35188 &      584 & 134.92    &     -0.58033 &      628 &  17.33 \\
              4.89647 &      447 & 291.91    &     -3.11725 &      596 & 326.97 \\
             76.16447 &      233 & 323.03    &     82.77163 &      581 & 128.95 \\
              3.60029 &      233 & 121.40    &     58.80017 &      341 & 212.90 \\
            -56.90922 &      208 &  44.11    &     34.82788 &      269 & 294.12 \\
              2.97706 &      148 & 306.81    &     11.50319 &      191 & 281.02 \\
             51.03334 &      129 & 136.30    &      0.46547 &      181 & 286.88 \\
              1.93168 &       70 & 148.98    &    -27.48935 &      173 & 218.53 \\
            -20.88236 &       70 & 203.93    &     -1.19906 &      167 & 133.74 \\
              1.82121 &       49 & 148.46    &    -25.17116 &      163 & 215.94 \\
              0.77840 &       22 &  65.10    &    -28.13656 &      108 & 314.08 \\
             -0.49216 &        6 & 164.74    &     10.34389 &       85 & 191.35 \\
                      &          &           &     20.96631 &       74 &  57.78 \\
                      &          &           &      0.57829 &       62 & 103.72 \\
                      &          &           &      9.18847 &       36 &   1.15 \\
                      &          &           &    -50.30212 &       25 &  29.78 \\
            \hline
         \end{array}
      \end{equation*}
      \vspace{-0.3cm}
      \tablefoot{This representation is used in Fig.~\ref{fig:ovlSR}. It has been directly obtained from \cite{LASKAR_1990}, see text.}
   \end{table}
   
   \begin{table}
      \caption{Quasi-periodic representation of the orbital dynamics of Mars.}
      \label{tab:QPS-Mars}
      \vspace{-0.7cm}
      \tiny
      \begin{equation*}
         \begin{array}{rrrrrr}
            \hline\hline
            \multicolumn{3}{c}{z} & \multicolumn{3}{c}{\zeta}\\
            \hline
            \mu_j\,(''/yr) & E_j\!\times\! 10^8 & \theta_j^{(0)}\,(^\text{o}) & \nu_j\,(''/yr) & S_j\!\times\! 10^8 & \phi_j^{(0)}\,(^\text{o}) \\
            \hline
             17.91550 &  4902750 & 335.25    &    -17.74818 &  3464962 & 303.28 \\
             17.36469 &  4004873 & 303.95    &    -18.85115 &  1541097 &  60.43 \\
              4.24882 &  2030021 &  30.67    &      0.00000 &  1375324 & 107.59 \\
             16.81285 &  1853846 &  91.90    &    -18.30007 &   745752 &  89.07 \\
             18.46794 &  1357476 &   9.91    &    -17.19656 &   543058 & 154.89 \\
             17.63081 &  1139332 & 201.65    &    -26.33023 &   457927 & 127.29 \\
             17.08266 &  1114353 & 359.38    &    -18.01114 &   262761 &  62.09 \\
             28.22069 &   706337 & 128.11    &    -17.66094 &   249135 & 318.93 \\
             17.81084 &   472390 &  58.56    &    -17.83857 &   202620 & 109.13 \\
             18.18553 &   431698 &  57.28    &    -17.54636 &   180575 &  66.71 \\
             19.01870 &   399188 &  39.75    &    -19.13075 &   142530 & 305.90 \\
             17.15752 &   383488 & 145.02    &    -17.94404 &   127098 &  32.26 \\
             17.72293 &   356895 &  48.46    &    -18.59563 &   116625 & 278.11 \\
             18.01611 &   303744 &  44.83    &     -5.61755 &   105161 & 348.70 \\
              7.45592 &   295700 & 200.24    &    -19.40256 &    85957 &  23.33 \\
             16.52731 &   231195 & 311.91    &     -7.07963 &    77142 &  93.77 \\
             16.26122 &   210411 &  58.89    &     -3.00557 &    63897 & 140.33 \\
             17.47683 &   201641 &  80.26    &     -0.69189 &    60870 &  23.96 \\
             18.08627 &   175373 & 356.17    &    -18.97001 &    59221 &  73.36 \\
             17.55234 &   162274 &  17.65    &     -6.84091 &    38863 & 286.50 \\
              3.08952 &    73627 & 121.36    &     -7.19493 &    38505 & 285.16 \\
             52.19257 &    26717 & 225.55    &     -5.50098 &    37498 & 342.89 \\
              6.93423 &    25209 & 170.37    &     -6.96094 &    36566 & 277.95 \\
              7.34103 &    24970 & 207.85    &    -18.69743 &    31116 & 221.70 \\
              7.05595 &    21406 & 359.16    &    -18.77933 &    30502 & 222.83 \\
              7.57299 &    17424 &  11.47    &    -18.22681 &    30460 &  46.30 \\
              6.82468 &     8182 & 194.53    &    -19.06544 &    27285 &  50.21 \\
             29.37998 &     8059 &  37.54    &     -7.33264 &    22307 &  16.75 \\
             27.06140 &     8027 &  38.70    &     -5.85017 &    12681 & 165.47 \\
              5.59644 &     6750 & 110.35    &     -5.21610 &     9492 &  18.91 \\
              7.20563 &     5106 & 143.91    &     -5.37178 &     9443 &  35.48 \\
             28.86795 &     4795 & 212.64    &     -5.10025 &     9241 & 195.38 \\
             27.57346 &     4212 & 223.74    &     -6.73842 &     7310 & 224.50 \\
              7.71663 &     3182 &  93.52    &     -7.40536 &     6758 &  53.35 \\
              4.36906 &     3180 &  40.88    &     -7.48780 &     6278 & 227.95 \\
            -19.72306 &     3055 & 113.24    &     -6.56016 &     5900 & 123.47 \\
              5.40817 &     2931 & 120.37    &     -5.96899 &     5417 & 350.64 \\
              5.47449 &     2541 & 275.01    &     -8.42342 &     4915 &  31.21 \\
             53.35188 &     2250 & 134.91    &     -6.15490 &     3922 &  89.77 \\
              0.66708 &     2008 &  73.98    &    -26.97744 &     3471 &  43.07 \\
              5.71670 &     1929 & 120.52    &     82.77163 &     1993 & 128.95 \\
             76.16447 &     1090 & 323.03    &     -1.84625 &     1457 &  38.22 \\
              5.35823 &     1037 &  94.89    &     18.14984 &     1419 & 111.19 \\
             51.03334 &      605 & 136.30    &     -4.16482 &     1278 &  51.57 \\
              4.89647 &      464 & 291.74    &     58.80017 &     1169 & 212.90 \\
              5.99227 &      381 & 113.56    &     34.82788 &      924 & 294.12 \\
              5.65485 &      368 &  39.22    &     -2.35835 &      628 &  44.46 \\
              5.23841 &      305 & 272.97    &     -0.58033 &      592 &  17.33 \\
              3.60029 &      250 & 121.40    &    -27.48935 &      592 & 218.53 \\
              1.93168 &      227 & 192.09    &    -25.17116 &      559 & 215.94 \\
            -56.90922 &      223 &  44.11    &    -50.30212 &      458 & 209.84 \\
              2.97706 &      192 & 306.81    &     -3.11725 &      427 & 326.97 \\
            -20.88236 &       91 & 203.93    &    -28.13656 &      371 & 314.08 \\
              1.82121 &       65 & 149.54    &      0.46547 &      170 & 286.88 \\
              0.77840 &       35 &  65.10    &     11.50319 &      136 & 281.02 \\
             -0.49216 &       10 & 164.74    &     -1.19906 &       96 & 133.57 \\
                      &          &           &      0.57829 &       59 & 103.72 \\
                      &          &           &     20.96631 &       53 &  57.78 \\
                      &          &           &     10.34389 &       52 & 191.16 \\
                      &          &           &      9.18847 &       33 &   1.15 \\
            \hline
         \end{array}
      \end{equation*}
      \vspace{-0.3cm}
      \tablefoot{This representation is used in Fig.~\ref{fig:ovlSR}. It has been directly obtained from \cite{LASKAR_1990}, see text.}
   \end{table}
   
\end{document}